Diego Di Battista

# AD-HOC CONTROL OF SCATTERING FOR ADAPTIVE OPAQUE LENSES

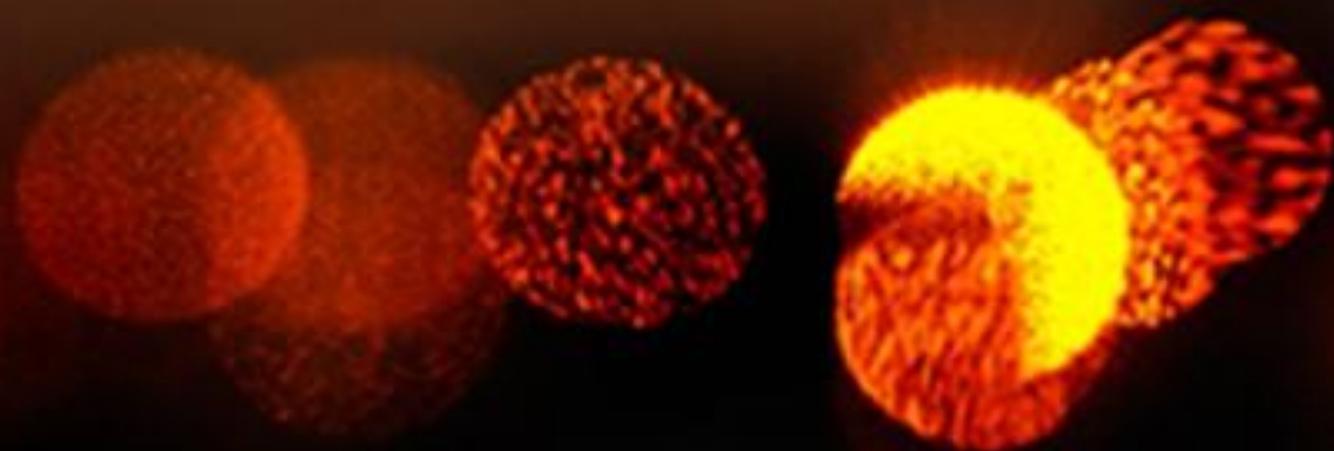

*A PhD Thesis*

# Ad-hoc control of scattering for adaptive opaque lenses

A Doctor of Philosophy (PhD) Thesis
By

**Diego Di Battista**

University of Crete
Department of Material Science

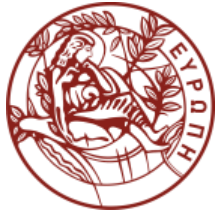

Institute of Electronic Structure and Laser (IESL)
Foundation for Research and Technology - Hellas

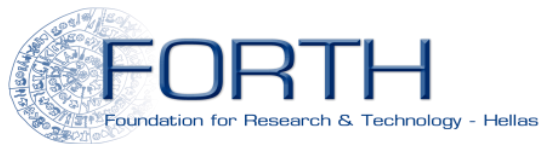

**Heraklion, 15<sup>th</sup> December 2016**

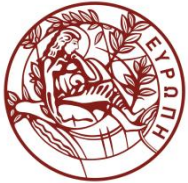

**ΠΑΝΕΠΙΣΤΗΜΙΟ ΚΡΗΤΗΣ**
**ΣΧΟΛΗ ΘΕΤΙΚΩΝ ΚΑΙ ΤΕΧΝΟΛΟΓΙΚΩΝ ΕΠΙΣΤΗΜΩΝ**
**ΤΜΗΜΑ ΕΠΙΣΤΗΜΗΣ ΚΑΙ ΤΕΧΝΟΛΟΓΙΑΣ ΥΛΙΚΩΝ**

**ΠΡΑΚΤΙΚΟ ΔΗΜΟΣΙΑΣ ΠΑΡΟΥΣΙΑΣΗΣ ΚΑΙ ΕΞΕΤΑΣΗΣ**
**ΤΗΣ ΔΙΔΑΚΤΟΡΙΚΗΣ ΔΙΑΤΡΙΒΗΣ ΤΟΥ**
*κ. Di Battista Diego*
**ΥΠΟΨΗΦΙΟΥ ΔΙΔΑΚΤΟΡΑ ΤΟΥ ΤΜΗΜΑΤΟΣ**
**ΕΠΙΣΤΗΜΗΣ ΚΑΙ ΤΕΧΝΟΛΟΓΙΑΣ ΥΛΙΚΩΝ**

Η Επταμελής Επιτροπή της Διδακτορικής Διατριβής του κ Di Battista Diego η οποία ορίσθηκε στην 74η Γ.Σ.Ε.Σ. στις 04/11/2016, εκλήθη την Πέμπτη 15 Δεκεμβρίου 2016 να εξετάσει την σύμφωνα με το Νόμο υποστήριξη της διατριβής του υποψηφίου με τίτλο:

## «Ad Hoc Control of Scattering for Adaptive Opaque Lenses»

Τα παρόντα μέλη της επταμελούς Επιτροπής εκφράζουν ομόφωνα την πλήρη ικανοποίησή τους για την υψηλή ποιότητα του περιεχομένου και της υποστήριξης της διατριβής.
Τα ερευνητικά αποτελέσματα της εργασίας του **κ. Di Battista Diego** είναι σημαντικά, πρωτότυπα και διευρύνουν το πεδίο της έρευνας στην Επιστήμη και Τεχνολογία Υλικών.
**Ως εκ τούτου η Εξεταστική Επιτροπή προτείνει ομόφωνα την απονομή του Διδακτορικού Διπλώματος στον κ. Di Battista Diego.**

**Τα μέλη της επταμελούς επιτροπής για την αξιολόγηση της Διδακτορικής Διατριβής του κ. Di Battista Diego.**

### Η Επταμελής Επιτροπή:

**Τζωρτζάκης Στέλιος, (Επιβλέπων)**
Αναπληρωτής καθηγητής, ΤΕΤΥ, Παν/μιο Κρήτης

___________________________________________________________________

**Κοπιδάκης Γεώργιος,**
Αναπληρωτής καθηγητής ΤΕΤΥ, Παν/μιο Κρήτης

**Τσιρώνης Γεώργιος,**
Καθηγητής Τμήματος Φυσικής, Παν/μιο Κρήτης

**Καφεσάκη Μαρία**
Αναπληρώτρια καθηγήτρια ΤΕΤΥ, Παν/μιο Κρήτης

**Παπάζογλου Δημήτριος**
Μόνιμος επίκουρος καθηγητής ΤΕΤΥ, Παν/μιο Κρήτης

**Νικολόπουλος Γεώργιος**
Ερευνητής Β΄, ΙΗΔΛ, Ίδρυμα Τεχνολογίας και Έρευνας, Ηράκλειο Κρήτης

**Ρακιτζής Πέτρος**
Καθηγητής Τμήματος Φυσικής, Παν/μιο Κρήτης

# ABSTRACT

Microscopy and optical imaging are drastically limited by the inhomogeneities encountered by the light while propagating from the object of interest to the detection system. In this context, adaptive optics and wavefront manipulation are able to improve the contrast (visibility) of systems embedded in turbid and noisy environments. By wavefront shaping, the fluence of the light propagating through complex systems can be controlled, thus, confining the light in a defined microscopic region in the volume or at the back of scattering structures. We can imagine to counterintuitively exploit the optical barriers, turning them into scattering lenses. In this Thesis we consider these new generation of lenses: we demonstrate them to be configurable optical devices able to produce tailored light structures, hence resulting extremely advantageous if integrated in optical systems.

We have initially studied their focusing spatial resolution at different scattering strengths, thereafter we have developed a method that allows to overcome the present limitations and to scale down their resolution of a factor 3 if compared with the previous standard methods. This result has been obtained selecting transmitting optical modes that are directly responsible for the resolution improvement.

Therefore, we show that properly filtering the light transmitted at the back of a scatterer we can also obtain numerous novel optical features. Specifically, we demonstrate the full control of non-diffractive light structures named amorphous speckle patterns. Under wavefront shaping manipulation those complex interference patterns can be converted to Bessel-like beams. The method discloses the possibility to produce Bessel beams at the selected location at the back of scattering barriers. The results presented have potential for providing arranged structured illumination and for enhancing the penetration depth in microscopy providing optimal imaging resolution of thick samples.

Thence, complex scattering systems have been recently used as programmable stand-alone optical devices. We have foreseen their operation in optical circuits, for this reason we designed micro-fabricated scattering systems in the bulk of cover-slip glasses. Our photonic structures are proved to be unreproducible and resistant to the deterioration, hence resulting suitable for authentication control operations. Moreover, fashioning the position and the geometry of the single scatterers via laser ablation we can engineer the process of multiple scattering in order to produce structured illuminations. In particular, we test an anisotropic geometry for the formation of light sheets. When we compare our system with conventional cylindrical lenses we observe a significant spatial resolution enhancement in the light sheet generated. In addition, our opaque lenses result fully tunable, a property that can allow for bypassing mechanical adjustment for chromatic correction or axial scanning in selected plane illumination microscopy. Ultimately, we present a lab-on-a-chip: a programmable platform that can support and inspect biological systems of interest in a controlled environment.

# ACKNOWLEDGMENTS

The "international Year of Light" in 2015 has coincided with the second year of my PhD project. In this stimulating contest, I was delighted for participating in the escalation of the trailblazing technology as onlooker and insider. In this respect, I have always felt in the front row thank to the Marie Curie action that has fully supported my training during the three years of my PhD.

I gratefully acknowledge my scientific supervisor Dr. Giannis Zacharakis for his guidance and support. Giannis, with you I comprehended the profitable art of the intellectual socialism. You consider everybody working in your group on a level field, colleagues sharing the same targets. A patient attitude that is unfortunately quite rare, but you have demonstrated to be strategically advantageous. Your guidelines based on your experience have always boosted my ideas and triggered me on.

I would like to thank all the members of the IVIL group whom contributed to this Thesis with significant discussions: Stella, Stelios, Thanassis, Vagelis L., Vagelis M., Krystalia, Marilena and Giorgos. Mainly, sweet and special acknowledge has to be addressed to my friend, colleague, priest and housemate Daniele Ancora. Thank you my friend for criticizing my ideas, dismounting my arguments, doubting my preliminary results and digging into the weak points of our experiments, seriously you cannot believe how much helpful was it. Above all, I want to thank you for being always honest with me. Two things you should never forget: $1^{st}$ the great team we are, $2^{nd}$ if something goes wrong is your fault.

I would like to express my gratitude to Dr Marco Leonetti, a solid scientist that has been a rock, a reference and an example. You taught me a lot and I will always consider you as my scientific career initiator (well, I don't know if this is good or not though!).

I would like to thank Dr George Tsibidis for being the best officemate ever. Thank you, George, for all your suggestions and the interesting discussions. Your subtle irony has livened up many working days along these three years. You are a good boy and a lovely father. Moreover, I want to thank Maria Dimitriadi for your invaluable help and patient with all the bureaucratic paperwork.

I would like to thank also my friends and confidants Tryfon, Kostis and Stefano. Thank you guys for your support.

Finally, I would like to acknowledge Prof. Horacio Lamela, the coordinator of the "Oiltebia" Marie Curie Training Network project, I have been proud to be an Oiltebia ESR; and Prof. Stelios Tzortzakis for accepting to be my academic supervisor at the University of Crete and supporting my PhD project.

From all those people cited here I have learned a lot and I thank you all again!

The work within this Thesis was supported from the EU Marie Curie Initial Training Network "OILTEBIA" PITN-GA--2012-317526. In addition to the Grants "Skin-DOCTor" and "Neureka!" implemented under the "ARISTEIA" and "Supporting Postdoctoral Researchers" Actions respectively, of the "OPERATIONAL PROGRAMME EDUCATION AND LIFELONG LEARNING", co-funded by the European Social Fund (ESF), National Resources and General Secretariat for Research and Technology (GSRT) (1778). Moreover, it was partially supported by the European Union's Seventh Framework Program (FP7-REGPOT-2012- 2013-1) under Grant Agreement No 316165 and H2020 Laserlab Europe (EC-GA 654148).

# INTRODUCTION

*Light* based technologies, are modernizing the way the universe communicates, observes and operates; thus, "Optics and Photonics" have become of world-wide fundamental interest. Lasers and optical devices guarantee the highest precision and speed, becoming indispensable in many fields for exploring undetected phenomenon [1]. Nowadays, the ensemble of Biomedical optics, Optical communication and Quantum computing is subject to growing demand from the high-tech market, but the innovative solutions are not always at our fingertips, so we are forced to explore new ways into the unknown. In such a way, what had been considered random, can be converted in something complex, then from complex to subcorrelated, up to singularly resolvable. A very present example may be found considering the effort that today is put for understanding the human brain function from the single unit, "the neuron", to the network and vice versa.

In my specific case, during my PhD I have been facing the problem of "scattering", the light distortion encountered by light trespassing an inhomogeneous material. Scattering is generally considered detrimental due to its dispersive nature: it blurs the vision, loses energy and de-correlates signals. My aim has been to invert this concept, exploiting light scattering in something rather beneficial. Together with my Colleagues from the "In Vivo Imaging Lab" supervised by Dr. Giannis Zacharakis, at the "Institute of Electronic Structure and Laser" (IESL), of the "Foundation for Research and Technology – Hellas" (FORTH), in Greece, we have investigated on new systems able to enhance visibility and resolution of optical imaging modalities adopted for interrogating large and thick specimens.

Reviewing the efficiency of the current optical imaging techniques [2] we understand that they are still not suitable for whole-body imaging because of the limited penetration depth: as soon as scattering increases (semi-transparent media) resolution is lost. In fact, many system of interest [3, 4] are transparent in their first embryonic and larva stages only, thereafter they become "opaque" [5]. At 1 mm (1 *transport-mean-free-path* for biological samples) light suffers strong diffusion and the imaging quality is dramatically compromised. On the other hand, sample manipulations such as optical clearing [6] allow deeper scanning, but only after sacrificing the sample or organism.

In general, the problem is tackled resorting Adaptive Optics [7] approaches for reducing the effect of wavefront distortions occurred in presence of scattering, but their use is currently limited to the imaging corrections, meaning that they are not efficient enough for enhancing *in vivo* imaging in deep.

In this scenario, the adaptive technique named "Wavefront Shaping" is finding vast response in the field. Since the pioneering experiment in 2007 [8], Wavefront shaping has been considered the most promising method for light scattering manipulation and it has immediately found direct applications in many fields.

The active control of light propagation through turbid media [9] is becoming an essential tool in microscopy [10], biological and biomedical imaging [11, 2], communication technology [12, 13], and astrophysics [14]. Wavefront shaping [8] is a powerful technique that allows to manipulate the optical paths through scattering media and currently it is possible to generate behind a scattering material multiple light spots, actively driven at user controlled positions, spatiotemporal focusing [9, 15], sub-wavelength foci (which may be employed for high resolution microscopy below the diffraction limit) [16, 17], to transmit image around corners [18], to control nonlinear systems [19] such as random lasers [20] and to permit novel forms of secure communication [21].

We started from the state-of-the-art of Wavefront shaping consisting in the ability to use a static opaque material as a focusing lens ("opaque lenses") by modifying the wavefront of the light impinging onto its volume [9, 22].

Even though the potential of Wavefront shaping is easily deducible, it is a very recent technique and it may have been only partially exploited. There are, indeed, many open questions pending and most of them are based on fundamental concepts that are remaining unclear due to their complexity. In order to reduce the entropy of the system described we decided to introduce specific constraints able to gather the scattered light in sub-regions of easier understanding. From those sub-regions we have extrapolated outstanding optical outputs that inspired new forms of applications. Therefore, we took advantage of the scattering complexity to explore extraordinary properties of unconventional optical features coming out of engineered disorder.

In such a way, we have improved the resolution of opaque lenses in "semi-transparent" environments, which is the nominal regime of interesting biological systems in their developed state [4, 5, 3].

Moreover, we have demonstrated the generation of non-diffractive beams (Bessel-like) through scattering walls, structures able to drastically improve the light penetration into inhomogeneous material. The direct consequence is the enhanced depth-of-focus in mesoscopic imaging applications.

Finally, exploiting lithographic techniques for imprinting designed scattering geometries, we engineered scattering optical elements that result completely configurable by wavefront shaping manipulation. Due to this property, these platforms consist of stand-alone optical elements that are found to be suitable for on-chip integration. We prove that they disclose optical features, such as light sheets that exhibit resolutions inaccessible with conventional lenses.

This Thesis is organized in two sections consisting in an introductory part in the first three Chapters:

- In Chapter 1 we revise the fundamentals of light scattering and describe the formation of speckle patterns. The equations defining the statistic of speckle pattern will be highlighted and commented.

- In Chapter 2 the breakthrough of Wavefront shaping method and the modus-operandi of the Opaque lenses are described.
- In Chapter 3 we introduce the Fourier Optics formalism. We treat a generic optical focusing system emphasizing the physical quantities that play a relevant role in the final focus profile determination.

And a second part containing our results and the related articles published in different scientific journal:

- In Chapter 4 we demonstrate that sorting the scattered light with specific spatial filters makes possible the final focus shape manipulation. We were able to enhance the state-of-the-art of opaque lenses resolution and to realize non-diffractive microstructures at the back of scattering systems.
- In Chapter 5 we show the way for designing scattering structures. For the first time *ad-hoc* disorder is permanently printed in integrated platforms able to produce configurable outputs.

Outlook and future prospective are commented in Chapter 6.

# Tables of contents

# 1 Light Scattering

In this first Chapter we provide a theoretical introduction into light scattering. The concepts presented herein are sufficient to introduce our results, nevertheless the reader may find a more detailed description in the cited references. First of all, we briefly review the standard diffusion model used to quantitatively describe light transport in an isotropic random dielectric medium. From the model, the relevant quantities will be highlighted. Then, we define the *speckle pattern* and examine its statistics. Lastly, we describe the light *memory effect* and the related benefits in the context of light scattering manipulation.

## 1.1 Multiple light scattering

When light propagates from a source to a detector through a homogenous non-scattering medium, it does it in a straight direction. In that case, light propagation is said to be ballistic. On the contrary, when light travels through a very disordered system it is multiply scattered. The multiple scattering process can be intuitively distinguished from the ballistic transport as described in the following example. If we consider a glass of beer, looking into the lower portion of the drink we distinguish the objects behind the glass, while on the contrary, we cannot look through froth. This is due to the continuum interchange of liquid and air (bubbles): the light entering in such medium is scattered numerous times and when it emerges from the material is completely scrambled. Since ambient light is visible (contains all the wavelengths of the visible spectrum), the diffusive media appear white (just as froth). Indeed, in most of the cases, the white materials owe their color to multiple light scattering.

In this regime ballistic propagation and ray tracing cannot accurately describe the transport of light. The multiple scattering of light has a very complicated solution in terms of Maxwell equations when many scatterers have to be taken into account [23]. An approximate approach to tackle this problem is the radiative transfer equation of the dilute medium where phase and light interference are neglected. The solution of the radiative transfer equation can be considerably simplified by introducing further approximations [24]. For instance, the "Diffusion approximation" considers a random walk of photons and imposes a continuity equation for the light intensity $I(r,t)$ disregarding interference effects [25]. Propagation of light can, therefore, be viewed as a diffusion process such as gasses diffuse in a partial pressure gradient. In this approximation we consider a spherical wave attenuating through an infinite homogeneous media, well described by the so called "Constant Illumination Green's Function" [25]:

$$G_{CW} = \frac{e^{ik_0 r}}{4\pi D r} \qquad (1.1.1)$$

where the constant *D* represents the diffusion coefficient.

In the light scattering processes the most important parameter to take into account is the *scattering mean free path*, $\ell_s$, which is the average distance travelled by the light between two consecutive scattering events. This parameter sets the limits of the diffusive approximation as: $\lambda \ll \ell_s \ll L$ (many scattering events occur before the light leaves the system, where $L$ is the system size, hereafter sample thickness) and $k \cdot \ell_s \gg 1$ (limits the approximation to a dilute medium, where $k$ is the light wave vector). After several scattering events, the light propagation is completely randomized. In such a scenario, we can define the *transport mean free path*, $\ell_t$, as the average distance after which the intensity distribution becomes isotropic (the memory from the initial direction of propagation is lost) and is the characteristic length in the regime of multiple scattering.

Indeed, the transport of ballistic or unscattered light in such a medium in space and time is dictated by the Lambert-Beer equation: $I(r,t) = I(0,t) \cdot \exp(-r/\ell_s)$, while diffuse light propagates according to the diffusion equation as follows:

$$\frac{\partial I(r,t)}{\partial t} = D \nabla^2 I(r,t) - \frac{v_e}{\ell_i} I(r,t) + S(r,t) \qquad (1.1.2)$$

where $S(r,t)$ is the light source, $v_e$ is the velocity of the energy [24, 26] and $\ell_i$ is the inelastic absorption length. The diffusion equation is a very general and practical description of numerous transport processes in physics. In the particular case of light diffusion, it describes how light intensity spreads through the system with a rate of transport dictated by the diffusion coefficient, $D$. The larger the diffusion coefficient, the faster the transport process. The exact solution of equation (1.1.2) or for any other distribution of time-indipendent sources will be based on the Green's function of equation (1.1.1).

The inelastic absorption length, $\ell_i$, is the average depth which light propagates ballistically in a homogeneous medium before being attenuated by a factor e. The diffusive absorption length, $\ell_a$, is the distance that the light propagates diffusively before being absorbed. Inside a diffusive and absorbing material, $\ell_a$ is the penetration depth of the diffuse light. Diffuse light propagates a greater distance than ballistic light in a homogeneous non-scattering material to reach the same depth. For this reason, $\ell_a$ is shorter than $\ell_i$. However, both are not independent functions rather they are related to $\ell_t$ as [23]: $\ell_a = \sqrt{(\ell_i \cdot \ell_t)/3}$. In any case, usually, disordered media are considered as opaque and white, i.e. non dispersive.

The scattering systems treated in our experiments have a slab geometry which imposes certain boundary conditions on the diffusion equation: the system can be considered infinite for *x* and *y* directions and limited between $z = 0$ and $z = L$.

Within the diffusion approximation, only diffusive light can be handled and therefore, an incident plane wave cannot be inserted as source in the diffusion equation: it decays exponentially inside the system according to Lambert-Beer equation. Therefore, we assume the source is a diffusive radiation placed at the plane $z = z_p$, where $z_p$ is the so called penetration length. A common phenomenological way to introduce this source is to consider an exponentially decaying one, $S(z) = S(0) \cdot exp(-z/z_p)$ [27].

The solution of the stationary diffusion equation with boundary conditions mentioned above leads to the solution to the stationary diffusion equation, $I(z)$ and the total transmission. In practice, in absence of absorption and taking into account $L \gg \ell_t$, the total transmission of light integrated over all the angles (defined as the total flux at $z = L$ divided by the incident flux) can be approximated as [27]:

$$T(L, \lambda) = \propto \frac{\ell_t}{L} \quad (1.1.3)$$

The total light transmission through a multiple scattering slab in the absence of absorption is directly proportional to the *transport mean free path*, $\ell_t$, and inversely proportional to the slab thickness, $T(\lambda) \sim \ell_t(\lambda)/L$. The optical conductance (transmission) is inversely proportional to the (optical) conductor thickness. Doubling the thickness of the conductor halves the transmission. Following this statement, with static measurements of the total light transmission through a given slab with known thickness it is possible to obtain the absolute value of the transport mean free path, $\ell_t$ [27, 28].

## 1.2 Speckle pattern and its statistics

In this section we present a qualitative explanation on the statistics of the speckle pattern. A speckle pattern is an intensity pattern produced by the mutual interference of a set of wavefronts. Portions of light from a single source makes different optical paths passing through a disordered scattering media so that the output is a sum of different wavefronts with random phases that interfere (see Figure 1.1). Let be the analytic signal representation of single polarization component of the electric field at observation point *(x,y,z)* and time instant *t* for a monochromatic wave [29]:

$$u(x, y, z; t) = A(x, y, z) \exp(i 2\pi \nu t) \quad (1.2.4)$$

where $\nu$ is the optical frequency, and $A$ represents the phasor amplitude of the field, which is a complex-valued function of space,

$$A(x, y, z) = |A(x, y, z)| \exp[i\theta(x, y, z)] \quad (1.2.5)$$

where intensity (irradiance) of the wave $I(x, y, z)$ is given by its square.

Whether the speckle pattern arises by free-space propagation, or by imaging, the amplitude of the electric field at a given observation *(x,y)* consists of a multitude of contributions from different scattering regions, a schematic of the process is depicted in Figure 1.1. Thus the field in space, $A(x, y, z)$, is represented as a sum of many elementary phasor contributions $a_k(x, y, z)$ with $k = 1, 2, \ldots, N$:

$$A(x, y, z) = \frac{1}{\sqrt{N}} \sum_{k=1}^{N} a_k e^{i\phi_k} \quad (1.2.6)$$

We can write the intensity at given point of the observation plane as:

$$I(r,t) \propto \left|\sum_{k=1}^{N} a_k \, e^{i\phi_k}\right|^2 \qquad (1.2.7)$$

We wish to know the statistics (*probability density functions*) of the complex field, the intensity, and the phase of speckle pattern at point *(x,y,z)*. The problem is identical to a classical random walk, indeed according to the central limit theorem the statistics of the *N* independent random variables is asymptotically Gaussian at $N \to \infty$.

We shall derive the relevant results here, by specifying the underlying assumptions and their physical meaning. Let the elementary phasor have the following statistical properties: the amplitude $a_k/\sqrt{N}$ and the phase $\phi_k$ of the *k-th* elementary phasor are statistically independent of the amplitudes and phases of all other elementary phasors (i.e., the elementary scattering areas are unrelated and the strength of a given scattered component bears no relation to its phase).

Under these conditions $\langle I^2 \rangle = 2\langle I \rangle^2$, therefore the standard deviation $\sigma_I^2$ of a polarized speckle pattern is given by:

$$\sigma_I^2 = \langle I^2 \rangle - \langle I \rangle^2 = \langle I \rangle^2 \qquad (1.2.8)$$

This magnitude is used to measure the speckle *contrast* as the ratio $C = \frac{\sigma_I}{\langle I \rangle}$; when $C = 1$, the speckle pattern is said *fully developed* [29].

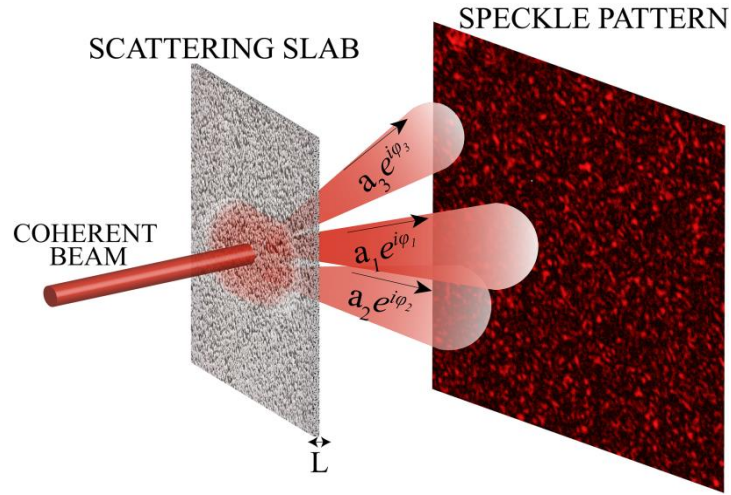

**Figure 1.1** *Schematic of light propagating through a scattering slab with thickness* $L > \ell_t$.

Moreover, it is possible to retrieve the *probability density function* of phase and intensity. We have that the phase obeys uniform statistics: the phases $\phi_k$ are uniformly distributed on the primary interval $(-\pi, \pi]$ (i.e., the surface is rough compared to the wavelength) [29]. On the other hand, the intensity follows a negative exponential statistics [29]:

$$p_I(I) = \frac{1}{\langle I \rangle} \exp\left(-\frac{I}{\langle I \rangle}\right). \tag{1.2.9}$$

## 1.3 Speckle grain size

In this Section we study the coarseness of the speckle pattern's spatial structure. To do so, we concentrate on what is called Second-Order statistics which will provide the typical speckle grain size at the back of scattering systems.

We describe the field reflected from (or transmitted through) a rough surface to an observation plane immediately adjacent to the same surface. The complex field $A(x, y)$ at the distance $z$ from the surface represents the field of interest. We are interested to calculate the autocorrelation function of the intensity distribution in the observation plane:

$$C(x_1, y_1; x_2, y_2) = \langle I(x_1, y_1) I(x_2, y_2) \rangle \tag{0.10}$$

In most of the cases is more convenient to deal with its normalized intensity called *correlation function*:

$$c_I(\Delta x, \Delta y) = \frac{C(x_1, y_1; x_2, y_2) - \langle I(x_1, y_1) \rangle \langle I(x_2, y_2) \rangle}{\langle I(x_1, y_1) \rangle \langle I(x_2, y_2) \rangle} \tag{0.11}$$

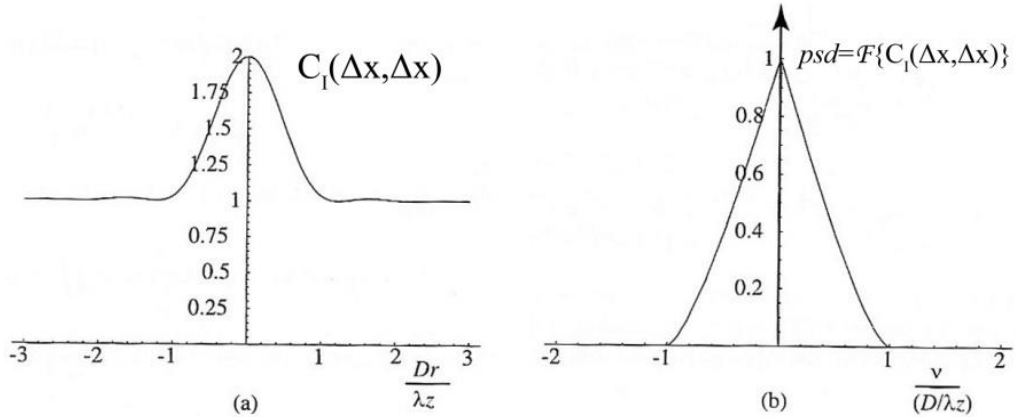

**Figure 1.2** *Cross section of speckle intensity. In (a) autocorrelation and in (b) power spectral density for free-space propagation from a circular scattering spot. Figure adapted from* **[29]**.

We can consider $c_I(\Delta x, \Delta y)$ as a measure of the average width of the speckle, the *typical speckle grain size*. Its shape strictly depends on the scattering system geometry, as an example, in Fig. 1.2 we report the correlation function of a speckle, generated from a circular scatterer with diameter $D$ as in Ref. [29]. In this case the speckle grain size calculated as the full width at half maximum of $c_I(\Delta x, \Delta y)$ is [29]:

$$w \sim 1.4 \lambda \frac{z}{D} \tag{0.12}$$

with typical longitudinal extension:

$$l_w = 6.7\lambda \left(\frac{z}{D}\right)^2 \tag{0.13}$$

On the contrary, when the light propagates through the scattering slab the grain size in the media is estimated as follows:

$$w_{IN} \sim \frac{\lambda}{2} \tag{0.14}$$

The other quantity of interest is the *power spectral density (psd)* of the speckle intensity distribution $I(x,y)$, which is given by the Fourier transform of $c_I(\Delta x, \Delta y)$. A representative *psd* curve of a speckle generated from a circular scatterer with diameter $D$ is depicted in Fig 1.2. In a more realistic case, the *psd* peaks at zero frequency ($\nu_x = 0, \nu_y = 0$) and has components extended over the frequency having the shape of a normalized autocorrelation function of the intensity distribution incident on the scattering spot. In practice, the *psd* describes the spatial frequency distribution of the scattered light, therefore the angular wave number $k$ associated to the fields that are interfering at the observation plane.

Our experiments and results are based on these concepts, in Section 4 we will show how to select those spatial frequencies in order to manipulate the speckle pattern statistics.

## *1.4 Memory effect*

When we face the problem of light scattering we, actually, face an unknown scattering matrix that connects the input beam to the speckle pattern at the output. In general, this matrix has a huge number of elements, making a complete characterization unattainable. Although the light propagation through strongly scattering media is considered a deterministic process that completely distorts the incoming wavefront, a little portion of information is always preserved. In particular, the incident and the scattered light has been demonstrated to be correlated.

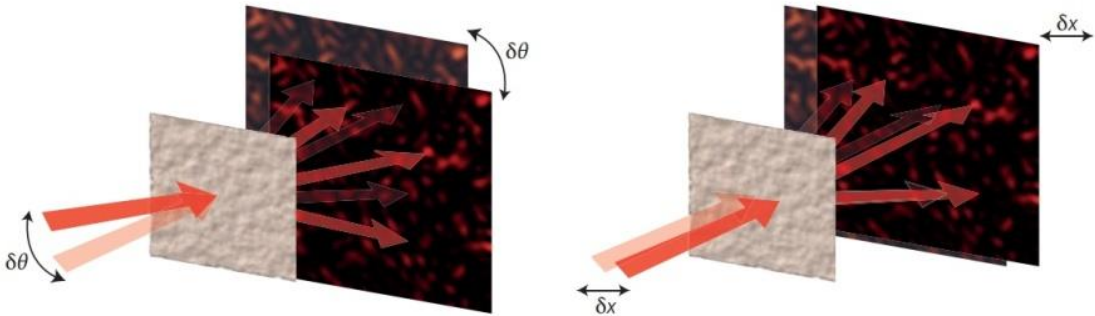

**Figure 1.3** *The optical memory effect principle. To a tilt angle **δθ** of the beam at the input corresponds a tilt **δθ** of the speckle pattern at the output. The same effect is observed when the beam is displaced at the distance **δx**. The figure is adapted from* **[30]**.

Recent studies have exploited the correlation known as the *optical memory effect* in order to estimate scattering transmission matrices [9, 31] and/or to retrieve object hidden behind obstacles [32]. The speckle memory effect states that if we slightly tilt (or shift) the incident beam, the speckle pattern produced on the other side of the obstacle tilts (or shifts) by the same angle (or displacement). A schematic of the principle is depicted in Fig. 1.3. By increasing the tilting angle (or shifting desplacement) the speckle continues shifting, but it gradually starts changing (decorrelating), and for an angle larger than the memory range the correlation with the original pattern is completely lost. The memory range depends on the sample thickness $L$ and the mean free path $\ell$. The larger the optical path along the scattering system, the smaller the memory range.

As described by Bertolotti in Ref. [30]: "*…the basic idea is that, due to the anisotropy in the scattering process, the light emerging from the turbid medium will not be dispersed isotropically as it is in the diffusion approximation, but will retain a significant forward component, that is, the scattered light will retain some memory of the original direction of the beam. Because the speckle produced by the scattered light in the far field is connected to the light emerging from the turbid medium by a Fourier transform, the angular memory due to the anisotropy is transformed in a correlation between the patterns produced by shifted beams…*"

Thus, due to the memory effect the elements of the scattering matrix are not fully independent, and it follows that the control of a limited number could be enough for light manipulation purposes in scattering environments. Indeed, the memory effect has been used to control the light propagation through scattering systems in adaptive optics. For instance, it allows the development of iterative genetic algorithms that aim to the control of light transmission through scattering sample. The algorithm converges to the desired solution even if the scattering matrix is not fully characterized [9]. This is the case of wavefront shaping, where a genetic algorithm shapes the wavefront of the beam impinging onto a scattering slab in order to compensate for scattering and to generate a focus at the back of the curtain [9]. This is the basic principle exploited in our experiments and its modus operandi will be thoroughly described in the next Chapter.

# 2 *Wavefront Shaping*

In this Chapter we introduce the concept of Wavefront shaping: the wavefront manipulation for controlling the light propagating through scattering media. We introduce the concept of focusing and we describe the optical geometry of opaque lenses. Then, their operation mode will be revised and benefits and limitations of the technique will be discussed. Finally, we theoretically analyze the focus shape exploiting a continuous field model.

## *2.1 Opaque Lenses (OL)*

In the previous Chapter we have seen that a laser beam trespassing a disordered scattering material produces a speckle pattern consisting of a set of maxima and minima of the intensity. In such a case, in fact, light travels following three dimensional random paths through the sample, and accumulating at the exit point random phases and directions. The result is that at any given point behind the sample the interference may be either constructive or destructive [33].

Light scattering may appear as a stochastic phenomenon, but it is rather deterministic. Due to the multiple scattering process there is an enormous number of degrees of freedom (DOF) to take into account when coherent light propagates trough strongly scattering media. For instance, the speckle pattern that arises from an illumination area *A* after propagation through a thick scattering medium is typically described by a number of parameters *N* (referred to the literature as number of modes) that scales as $2\pi A/\lambda^2$; in particular, for visible light *N* corresponds typically to 10 million modes per square millimeter [9].

On the other hand, during the last decade, adaptive optics modalities have been drastically advanced allowing unexpected control of light propagation in complex media [9]. Therefore, today it is possible to partially compensate for scattering and focus light through, or inside, scattering materials by spatially modifying the phase of the incident light wavefront, using a Spatial Light Modulator (SLM) to correct for the phase shift induced by disorder [9].

Due to the memory effect, controlling only a limited number of DOF ($N_{\text{DOF}}$) one can enhance the intensity in a determined region of the original speckle pattern [9]. The resulting intensity at targeted region over the average intensity of the speckle pattern $\langle I \rangle$ before the optimization gives the enhancement factor $\eta$. The $\eta$ *is indeed directly proportional* to the $N_{\text{DOF}}$ controlled and can be increased up to the value $\eta = 1000$ [8].

Wavefront shaping was first demonstrated by I. Vellekoop and Coworkers [8]; a schematic of their experiment is depicted in Figure 2.1. Their intuition was based on the following assumptions.

Light scattering is a deterministic linear process, therefore we can write the transmitted electric field $E_m$ at the CCD camera plane as the combination of the fields $E_l = A_l e^{i\phi_l}$ multiplied to the complex transmission matrix $t_{ml}$ of the scattering system:

$$E_m = \sum_{l=1}^{N_{DOF}} t_{ml} A_l \, e^{i\phi_l} \qquad (2.1.15)$$

where *the summation is over* the $N_{\text{DOF}}$ segments of the SLM and $A_l$ and $\phi_l$ are the amplitude and phase of the light reflected from the *l-th* segment.

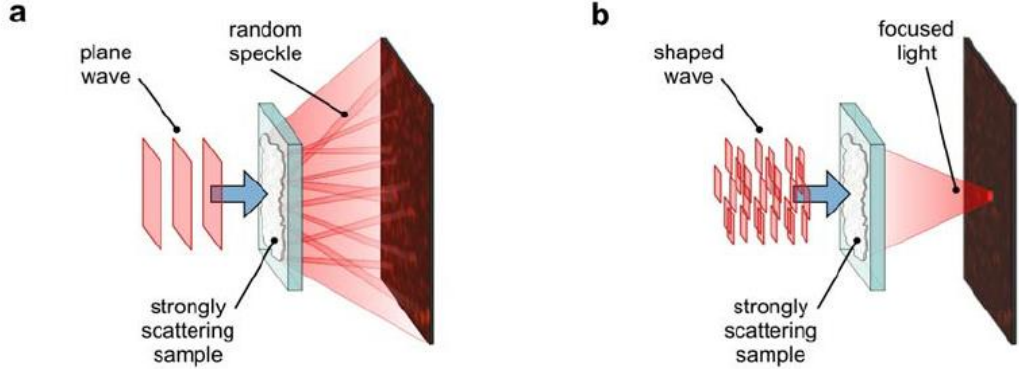

**Figure 2.1.** *Sketch of a standard wavefront shaping scheme as shown in very first experiment* **[8]**. *In (a) a plane wave suffers multiple scattering propagating through a strongly scattering sample, at the back of the sample a speckle pattern is generated and propagated in free-space to the observation plane. (b)The wavefront shaping of the incident beam allows focusing light through the scattering sample.*

The wavefront shaping technique requires two important constraints: the feedback and stationarity. The feedback (or guide-star) is necessary to measure transmitted output value at the target position. While the pioneering experiment was demonstrated using the camera counts [8, 15], many other feedbacks have been explored, e.g. fluorescence, acoustic signal, Raman scattering, ...etc. [34, 35].

In addition, wavefront shaping can be implemented if and only if the dynamics in the scattering system is slow enough to maintain $t_{ml}$ unruffled during the optimization process. Even if several approaches have been proposed, in order to improve the focusing speed and/or to adapt to different modulation schemes [36], the adaptive focusing process is still time consuming, a characteristic that drastically limits its direct application in biomedical imaging.

However, when those two constraints are respected the phase and/or amplitude (according to the type of control provided by the beam shaper in use) can be set to tailor the output field at the targeted position [9].

In our experiments the scattering samples are Titanium Dioxide (TiO$_2$) particles self-assembled in a disordered fashion. This material assures very low absorption while it is very easy to obtain a disordered matrix whit high structural homogeneity and little fluctuations in transmittance. We shape the beam wavefront by using a Phase Only SLM (model PLUTO from Haloeye, Germany). Taking advantage of the memory effect, we developed a genetic algorithm based on the pioneering work performed by I. Vellekoop and Coworkers [37].

In practice, each input field $E_l$ corresponding to the *l-th* mode and is sequentially diphased from 0 to 2π, *testing the corresponding intensity at the target position recorded by the CCD camera*. The phase values that maximize the intensity are chosen for each input mode. At the end of this process, since all the terms $t_{ml}A_l e^{i\phi_l^{optimized}}$ are in phase, all the input modes are set simultaneously to their optimal value, in order to get a strong constructive interference at the chosen target position [9], effectively generating a bright focus. In practice, the scattering medium serves as an *opaque lens* (OL).

In our experiments we adopted focusing algorithms based on the concepts just described. Starting the standard genetic algorithm for adaptive focusing [37] we applied minor modifications according to the specific needs encountered with each of the experiments developed; the operating principles of the final versions of the algorithms are fully illustrated in the Chapters 4 and 5.

## *2.2 OLs resolution*

From a theoretical point of view, the focus generated with a lens homogeneously illuminated with a collimated beam with circular aperture is an Airy disk. Its Half Width at Half Maximum is $w = 0.51\lambda/N.A.$, where *N.A.* is the numerical aperture of the lens and λ is the beam wavelength. Equivalently, for a given diameter *D* of the incident beam impinging on a lens with focal length $f$, the smallest achievable focus size (the best focal lateral resolution) is $\lambda f/D$. This threshold was first set by Abbe [38] and is called in literature "diffraction limit".

Regarding opaque lenses, when the modulator is programmed to generate a shaped wavefront, diffraction at the lens aperture will cause the beam to diverge. Moreover the light, while propagating through the sample, diffuses. It follows that at the scattering sample output the effective beam aperture is much larger than the one at the input as depicted in Fig. 2.2. For this reason, the classical thin lens approach is not directly applicable for opaque lenses.

I. Vellekop and Coworkers at the University of Twente [15] in 2009 have developed a model that studies the resolution of opaque lenses. In their experiment the wavefront illuminating the sample is shaped to increase the intensity on a single pixel of the camera. Therefore, to calculate the diffraction limit of the opaque lens is necessary to determine its numerical aperture.

First of all, they noticed that opaque lenses exhibit larger effective numerical aperture due to light diffusion, therefore they can confine the light in a smaller region compared to the ordinary lens.

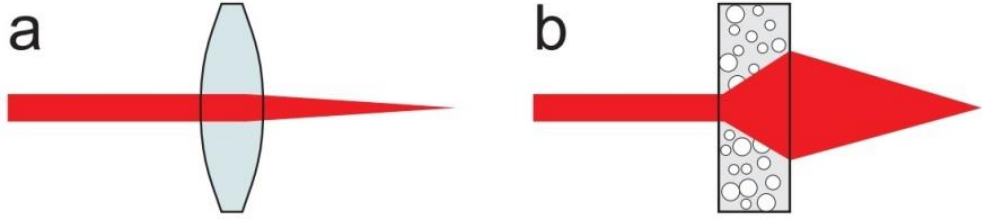

**Figure 2.2** *Comparison between conventional and opaque lens. In the first case (a) the lens' numerical aperture is given by the beam aperture, for the opaque lens (b) we have to take into account the diffusion encountered during the light propagation into the sample. The figure is adapted from* **[15]**.

However, it appears that the focus of an opaque lens always has the same size as a typical speckle grain, therefore the width of the focus is exactly the diffraction limit.

As we have seen in the previous Chapter, the speckle correlation function $C$ is a measure for the shape of a typical speckle [15] and it is defined as:

$$C(x_1, y_1; x_2, y_2) = \frac{\langle I(x_1, y_1)I(x_2, y_2)\rangle}{\langle I(x_1, y_1)\rangle\langle I(x_2, y_2)\rangle} - 1 \qquad (2.2.16)$$

where the brackets denote spatial averaging over all speckles. The profile of the focus overlaps exactly with the speckle correlation function, this correspondence is at the base of our experiments and in Chapter 4 we will study how the equality depends on the sample optical length and the spatial frequencies in use for focusing light.

## 2.3 *Adaptive focus and speckle correlation function: the identity*

I. Vellekop and Coworkers investigated (theoretically and experimentally) [15] the relation between the speckle correlation function and the generated focus. Their model is summarized in this Section and is the bedrock of our results showed in Chapter 4.

They developed a continuous field description of opaque lenses [15]. Considering a standard wavefront shaping scheme as depicted in Fig 2.3, we can define coordinates in the camera plane as $r_b$, on the back surface of the sample as $r_k$, and in the plane of the phase modulator as $r_a$. The optimization process (focusing) will enhance the photon counts of the camera pixel at the position $r_\beta$. Therefore, the field propagating through the scattering slab can be written as:

$$E(r_k) = \sum_{a}^{N} t_a(r_k) E_a, \qquad (2.3.17)$$

where *N* is the number of segments controlled with the phase modulator.

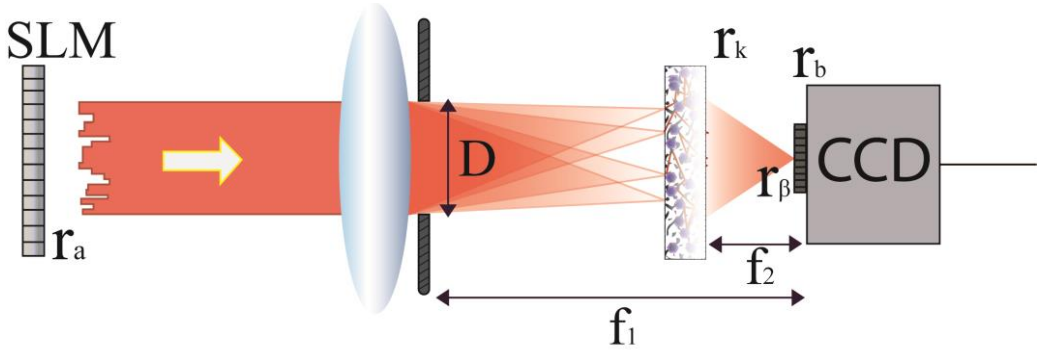

**Figure 2.3** *Schematic representation of the experiment developed in 2009 by I. Vellekop and Coworkers [15].The light reflected by the phase modulator SLM is imaged on the center plane of a lens. The aperture beam D is controlled by a pinhole. A CCD camera is positioned in the focal plane of the lens $f_1$. Between the camera and the lens a strongly scattering sample is placed.*

The light reflected by the SLM propagates perpendicular (in first approximation) to the device window, therefore, considering the Fresnel-Kirchhoff diffraction formula [39] the transmission coefficient $t_a(r_k)$ is:

$$t_a(r_k) = \frac{i}{\lambda} \iint_{S_a} d^2 r_a G(r_k, r_a). \qquad (2.3.18)$$

The integration is over the surface of a single segment $S_a$ of the SLM and $G(r_k, r_a)$ is the unknown Green's function for propagating from the modulator to the back of the sample. Light propagation from the back of the sample towards the focal plane is described by the free space Green's function $g(r_b - r_k)$

$$E(r_b) = \frac{i}{\lambda} \iint d^2 r_k \, g(r_b - r_k) E(r_k) \qquad (2.3.19)$$

It follows that the propagation from the SLM segment $a$ to the focal plane is given by the transmission coefficient

$$t_a(r_b) = \frac{i}{\lambda} \iint d^2 r_k \, g(r_b - r_k) t_a(r_k) \qquad (2.3.20)$$

The continuous field description (the term $t_a(r_b)$ is a continuous function of the spatial coordinate) has been chosen in order to define the shape of the focus of the opaque lens.
We consider the case where the modulator shapes the reflected field in phase and amplitude (even if the SLM in use in our experiment is a phase only modulator).
The field generated by the modulator at the end of the focusing process at the target position $r_\beta$ is

$$\tilde{E}_a = E_0 t_a^*(r_\beta) \qquad (2.3.21)$$

where the tilde sign indicates the optimized field $E_a$. The field $E_0$ can be considered as a constant that fixes the average intensity at the modulator to $I_{IN}$. In such a way, the intensity in the focal plane averaged over all possible configurations of the SLM that has an average intensity of $I_{IN}$ is defined as:

$$I_0(r_\beta) \equiv I_{IN} \sum_a^N |t_a(r_\beta)|^2 \tag{2.3.22}$$

Substituting Eq. (2.3.20) into Eq. (2.3.21) and then the result in Eq. (2.3.17), we obtain the field at the back of the scattering sample after optimization:

$$\tilde{E}(r_k) = \sum_a^N t_a(r_k) \left(\frac{i}{\lambda} E_0 \iint d^2 r'_k \, g(r_b - r'_k) t_a(r'_k)\right)^* \tag{2.3.23}$$

It follows that the average value of the optimized field at the back of the sample can be written as follows [15]

$$\langle \tilde{E}(r_k) \rangle = g^*(r_b - r_k) C_0 I_0(r_k) \tag{2.3.24}$$

where $C_0 = i\lambda E_0/(2\pi I_{IN})$ and $I_0(r_k) = I_{IN} \sum_a^N |t_a(r_k)|^2$.

Now, using the paraxial approximation [39], we can use the general result from Eq. (2.3.24) to calculate the intensity distribution in the focal plane after optimization [15]:

$$\langle \tilde{E}(r_b) \rangle = \frac{ie^{i\phi}}{\lambda f_2^2} C_0 \iint d^2 r_k I_0(r_k) \exp\left(\frac{ik_0}{f_2}[x_k(x_\beta - x_b) + y_k(y_\beta - y_b)]\right) \tag{2.3.25}$$

where $f_2$ is the focal distance of the opaque lens and $\phi$ is the phase factor.

Looking at the integral in Eq. (2.3.25) is possible to recognize the 2D Fourier Transform of the intensity. Indeed, we introduce the following term:

$$U_0(r_b) = \iint d^2 r_k I_0(r_k) \exp\left(\frac{ik_0}{f_2}[x_k(x_\beta - x_b) + y_k(y_\beta - y_b)]\right) \tag{2.3.26}$$

where

$$\mathcal{F}\{I_0(r_k)\} = U_0(r_b) \tag{2.3.27}$$

It follows that, expanding $C_0$, the intensity in the focal plane is given by:

$$\tilde{I}(r_\beta) = \frac{1}{(2\pi f_2^2)} \frac{N}{I_0(r_\beta)} |\mathcal{F}\{I_0(r_k)\}|^2 \tag{2.3.28}$$

which means that the opaque lens perfectly focuses a field that depends on the intensity distribution at the back of the sample.

In general, the intensity at the focus peak equals

$$\tilde{I}(r_\beta) = NI_0(r_\beta) \qquad (2.3.29)$$

which means that the final intensity in the target is proportion to the number of segments $N$ controlled with the SLM. For phase only modulation that is the modulation we use in our experiment, the focus enhancement reduces to $1 + (N-1)\pi/4$ [15]. In essence, the enhancement needs to be measured exactly at the focus peak and not, for instance, averaged over the whole speckle.

When the sample is thicker than a transport mean free path, the field at the back surface of the sample is uncorrelated on length scales larger than the wavelength and the transmitted light has no preferential direction. In this regime, in the paraxial limit the correlation function of Eq. (2.2.16) only depends on the coordinate difference $\Delta r_b \equiv r_b - r_{b'}$. According to the van Cittert-Zernike theorem [40], in this case, the correlation function is given by

$$C(\Delta x_b, \Delta y_b) = \frac{1}{(\iint d^2 r_k I_0(r_k))^2} |\mathcal{F}\{I(r_k)\}|^2 \qquad (2.3.30)$$

where $I(r_k)$ is the diffuse intensity distribution at the back surface of the scattering sample. When the SLM is designed to generate a random field, $\langle I(r_k) \rangle = \langle I_0(r_k) \rangle$. In this case, taking in account Eq. (2.3.25) and considering $\iint d^2 r_k I_0(r_k) = P_{TOT}$ as the average total power that is transmitted through the sample, we have that:

$$\tilde{I}(x, y, f) = C(x, y) N \frac{P_{TOT}}{2\pi f^2} I_0(r_\beta) \qquad (2.3.31)$$

where $f$ is the general opaque lens focal length.

This last equation predicts that the intensity profile of the focus is exactly equal to the speckle correlation function (up to a constant pre-factor). This identity seems to be counterintuitive because the intensity in the focus is the square of an amplitude, whereas the correlation function is a product of intensities. In any case, we will return on this equivalence in Chapter 3, where further discussion will be provided.

## 2.4 Discussion

The continuous field model only partially explains the results of this thesis.

In the previous Section we have seen that the size of the focus is predicted to be equal to the speckle size. The intensity distribution $I_0(r_k)$ at the back of the sample depends on the thickness and mean free path of the sample, the thicker the scattering medium, the smaller is the typical speckle grain, the sharper the final focus.

On the other hand, the extrapolation ratios that describe the boundary conditions, i.e. in the geometry in the model examined in the previous Section assumes that $I_0(r_k)$ is constant in a disk with a diameter *D* and zero outside of the disk [15]. The average field at the back of the sample is the complex conjugate of the free space propagator. In other words, the transmitted field is the same as if the sample time-reversed a wave coming from the target focus. Hence, the transmitted light focuses on the target as well as is physically possible for a system with aperture *D*.

In a more realistic and general geometry, for example when the incident light is focused sharply on a thick sample, the intensity distribution $I_0(r_k)$ will be a smooth function of the position. In such a geometry, an opaque lens differs from an ordinary lens. The disk aperture is not experimentally applicable as a boundary condition but strongly depends on the sample scattering mean free path and scatterer geometry. For an opaque lens, however, the amplitude of the controlled field at the back of the sample is proportional to the intensity distribution at the back of the sample.

This statement will become more clear at the end of the next Chapter, where the focus shape is estimated using a Fourier Optics approach.

# 3  Frequency Analysis of Optical Imaging Systems

In this chapter we threat the problem of the optical imaging system using the Fourier formalism. We describe the response of a focused optical system to a point source, thus its *point-spread function.* We demonstrate that as the numerical aperture of a thin lens determines the final focus resolution, the focus shape of a general optical system does depend on spatial frequencies in use for focusing. Since the optical response of a focusing wall (an opaque lens) is exactly the same of the one of a thin lens [22], the following treatise can be generalized to any kind of lenses. A fully extended version of the following notes can be found in [41].

## 3.1  Fourier Transforming property of a lens

In Chapter 1 we have seen that the speckle produced by the scattered light in the far-field is connected to the light emerging from the turbid medium by a Fourier transform. Moreover, its *power spectral density (psd)* is given by the Fourier transform of $c_I(\Delta x, \Delta y)$. This can be mathematically demonstrated turning to the Fourier optics formalism [41].

In more general terms, we can consider the field $U$ generated onto the observation plane *(x,y)* at the distance $z$ from the diffracting aperture situated on the plane $(\xi, \eta)$. $U$ can be described in the near-field approximation with the Fresnel diffraction formula as:

$$U(x,y) = \frac{e^{ikz}}{i\lambda z} e^{i\frac{k}{2z}(x^2+y^2)} \iint_{-\infty}^{\infty} \left( U(\xi,\eta) e^{i\frac{k}{2z}(\xi^2+\eta^2)} \right) e^{-i\frac{k}{2z}(x\xi+y\eta)} d\xi d\eta \quad (3.1.32)$$

or with the Fraunhofer diffraction pattern in the far-field (where $z \gg k(\xi^2 + \eta^2)_{max}/2$) as:

$$U(x,y) = \frac{e^{ikz}}{i\lambda z} e^{i\frac{k}{2z}(x^2+y^2)} \iint_{-\infty}^{\infty} U(\xi,\eta) e^{-i\frac{k}{2z}(x\xi+y\eta)} d\xi d\eta, \quad (3.1.33)$$

that is equivalent to the *Fourier transform* of the aperture distribution itself.

On the other hand, if we now consider a thin lens with focal distance *f*, we have that the field across the lens immediately behind it, $U'_l$, related to the incident complex field $U_l(x,y)$, can be written as:

$$U'_l = t_l(x,y) U_l(x,y) \quad (3.1.34)$$

where $t_l(x,y)$ represents the lens transformation. In the case of thin lenses we have:

$$t_l(x,y) = e^{-i\frac{k}{2f}(x^2+y^2)} \tag{3.1.35}$$

Moreover, due to the finite extent of the lens we introduce the pupil extent *P(x,y)* that takes value 1 within the lens aperture and 0 otherwise. In such a way we have:

$$U'_l = U_l P(x,y) e^{-i\frac{k}{2f}(x^2+y^2)} \tag{3.1.36}$$

Therefore, taking into account the Fresnel formula (3.1.32), onto the focal plane *(u,v)* at the distance *z=f*, we have:

$$U_f(u,v) = \frac{e^{i\frac{k}{2f}(u^2+v^2)}}{i\lambda f} \iint\limits_{-\infty}^{\infty} \left(U'_l(x,y) e^{i\frac{k}{2f}(x^2+y^2)}\right) e^{-i\frac{k}{2f}(xu+yv)} dxdy \tag{3.1.37}$$

Substituting (3.1.36) in (3.1.37) we obtain:

$$U_f(u,v) = \frac{e^{i\frac{k}{2f}(u^2+v^2)}}{i\lambda f} \iint\limits_{-\infty}^{\infty} U_l(x,y) P(x,y) e^{-i\frac{k}{2f}(xu+yv)} dxdy \tag{3.1.37}$$

Thus the field distribution $U_f$ is proportional to the 2-D Fourier transform of that portion of incident field subjected to the lens aperture. We notice that the complex amplitude distribution in the focal plane of the lens is the Faunhofer diffraction pattern of the field incident on the lens; in particular, the amplitude and phase of the light at coordinate *(u,v)* in the focal plane are determined by the amplitude and phase of the input Fourier component at frequencies ($f_x = u/\lambda f$, $f_y = v/\lambda f$ ) [41].

## 3.2 The Optical Transfer Function of a focusing system

These concepts can be applied to a general optical focusing system, for instance to our opaque lenses.

Let consider the black box imaging system illustrated in Figure 3.1. The output field is represented by a superposition integral:

$$U_i(u,v) = \iint\limits_{-\infty}^{\infty} h(u,v;\xi,\eta) U_g(\xi,\eta) d\xi d\eta \tag{3.2.38}$$

Where $h$ is the amplitude at the plane $(u,v)$ in response to the point source at $(\xi,\eta)$.

In this case, we can recognize in the argument of the integral the *amplitude impulse response* which corresponds to the *point-spread-function* of a focusing system, that according to eq. (3.1.38) can be described as [41]:

$$h(u,v) = \frac{A}{\lambda z} \iint\limits_{-\infty}^{\infty} P(x,y) e^{-i\frac{k}{2z}(xu+yv)} dxdy \tag{3.2.39}$$

where *A* is a constant amplitude and *z* is the distance from the exit pupil to the image plane.

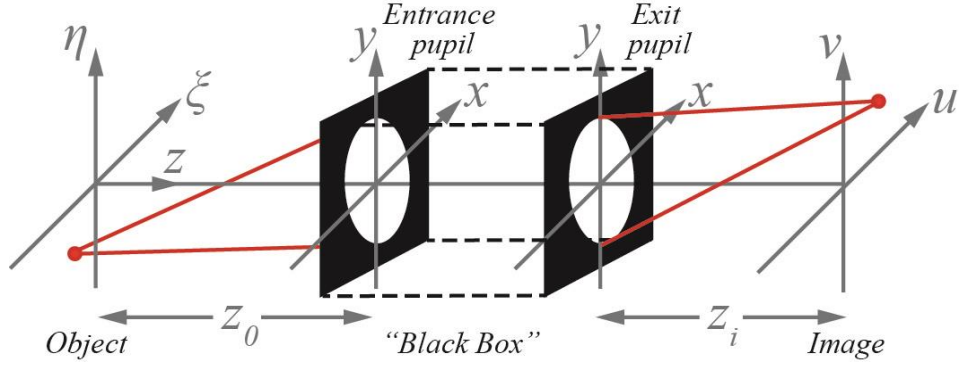

**Figure 3.1** *Generalized model of an imaging system. The figure is adapted from* **[41]**.

We can apply this concept to amplitude mapping of an imaging system. To do so we consider the following frequency spectra at the system input and output, respectively:

$$G_g(f_x, f_y) = \iint_{-\infty}^{\infty} U_g(u, v) e^{-i2\pi(f_x u + f_y v)} du dv$$

$$G_i(f_x, f_y) = \iint_{-\infty}^{\infty} U_i(u, v) e^{-i2\pi(f_x u + f_y v)} du dv$$

(3.2.40)

We define the *amplitude transfer function H* as the Fourier transform of the amplitude impulse response (the *point-spread function h*),

$$H(f_x, f_y) = \iint_{-\infty}^{\infty} h(u, v) e^{-i2\pi(f_x u + f_y v)} du dv$$

(3.2.41)

Therefore, the frequency spectra at the output is:

$$G_i(f_x, f_y) = H(f_x, f_y) G_g(f_x, f_y)$$

(3.2.42)

In eq. (3.2.40) we have seen that the *point-spread function h* is itself a Fraunhofer diffraction pattern and can been expressed as a scaled Fourier transform of the pupil function. It follows that the *Amplitude Transfer Function* can be written as:

$$H(f_x, f_y) = \mathcal{F}\left\{\frac{A}{\lambda z} \iint_{-\infty}^{\infty} P(x, y) e^{-i\frac{k}{2z}(xu + yv)} dx dy\right\}$$

$$= (A\lambda z) P(-\lambda z f_x, -\lambda z f_y)$$

(3.2.43)

For notational convenience, we can set $(A\lambda z)$ equal to unity and ignore the negative signs in the arguments of *P*. Thus

$$H(f_x, f_y) = P(\lambda z f_x, \lambda z f_y) \tag{3.2.44}$$

In practice, the scaled pupil function plays the role of the amplitude transfer function. Yet its Fourier transform provide the *point-spread function* of the optical system. The pupil limits the range of Fourier components passed by the systems, therefore it defines its ability to focus.

In most of the cases, however, we are interested at the intensity distribution in the focal plane. In this case we need to analyze the *normalized* frequency spectra of the intensities $I_g$ and $I_i$ at the system input and output respectively:

$$\mathcal{G}_g(f_x, f_y) = \frac{\iint_{-\infty}^{\infty} I_g(u,v) e^{-i2\pi(f_x u + f_y v)} du dv}{\iint_{-\infty}^{\infty} I_g(u,v) du dv}$$

$$\mathcal{G}_i(f_x, f_y) = \frac{\iint_{-\infty}^{\infty} I_i(u,v) e^{-i2\pi(f_x u + f_y v)} du dv}{\iint_{-\infty}^{\infty} I_i(u,v) du dv} \tag{3.2.45}$$

In a similar fashion, the normalized transfer function of the system can be defined by:

$$\mathcal{H}(f_x, f_y) = \frac{\iint_{-\infty}^{\infty} |h(u,v)|^2 e^{-i2\pi(f_x u + f_y v)} du dv}{\iint_{-\infty}^{\infty} |h(u,v)|^2 du dv} \tag{3.2.46}$$

That leads to the frequency-domain relation:

$$\mathcal{G}_i(f_x, f_y) = \mathcal{H}(f_x, f_y) \mathcal{G}_g(f_x, f_y) \tag{3.2.47}$$

The function $\mathcal{H}$ is known as the *optical transfer function* (OTF) and its modulus $|\mathcal{H}|$ as the *modulation transfer function* (MTF).

Since all the transfer functions are defined with *h*, we have to determine the exact relation between the *amplitude* and *optical transfer function*. From eq. (3.2.47) we recognize

$$\mathcal{H}(f_x, f_y) = \frac{\mathcal{F}\{|h|^2\}}{\iint_{-\infty}^{\infty} |h(u,v)|^2 du dv} \tag{3.2.48}$$

According to the Rayleigh's theorem, it follows that:

$$\mathcal{H}(f_x, f_y) = \frac{\iint_{-\infty}^{\infty} H(p', q') H^*(p' - f_x, q' - f_y) dp' dq'}{\iint_{-\infty}^{\infty} |h(p', q')|^2 dp' dq'} \tag{3.2.49}$$

And the change of variables:

$$p = p' - \frac{f_x}{2} \qquad q = q' - \frac{f_y}{2} \tag{3.2.50}$$

results to the symmetrical expression:

$$\mathcal{H}(f_x, f_y) = \frac{\iint_{-\infty}^{\infty} H\left(p' + \frac{f_x}{2}, q' + \frac{f_y}{2}\right) H^*\left(p' - \frac{f_x}{2}, q' - \frac{f_y}{2}\right) dp\, dq}{\iint_{-\infty}^{\infty} |h(p,q)|^2 dp\, dq} \qquad (3.2.51)$$

*Thus the OTF is the normalized autocorrelation function of the amplitude transfer function* [41].

## 3.3 Spatial filters

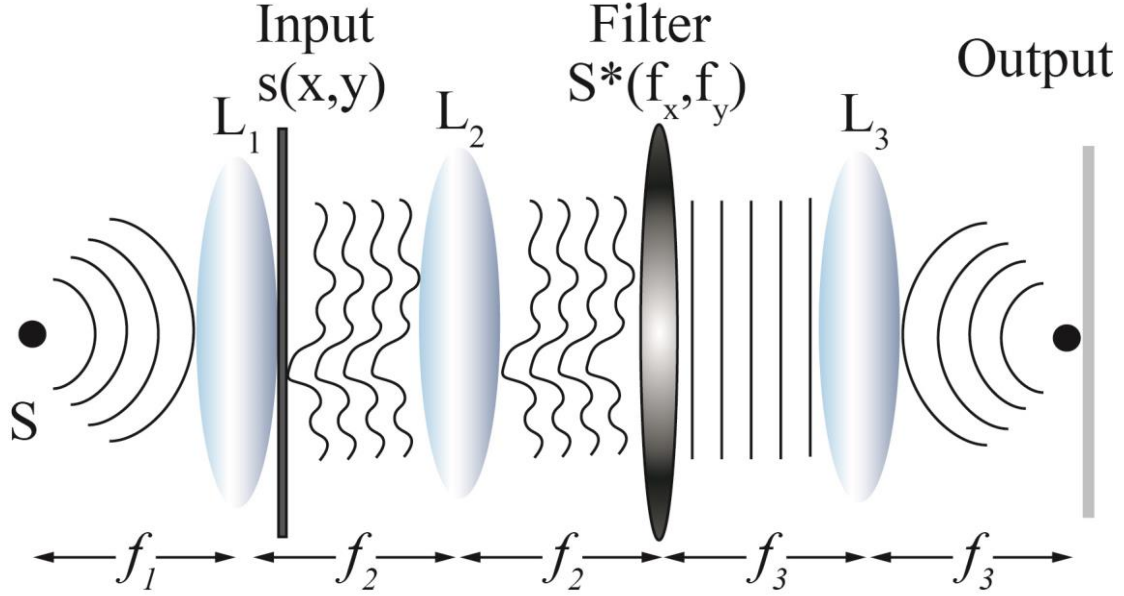

**Figure 3.2** *Schematic of the spatial frequencies filter operation. The Filter is placed at the Fourier plane of the 4-f system composed with lens L2 and L3. The figure is adapted from Ref.* **[41]**.

In the previous section we have seen that the amplitude transfer function is directly related to the pupil function as described in eq. (3.2.44). In many experimental applications the spatial frequencies of the optical system are selected using physical spatial filters. A schematic of the spatial filter operation is depicted in Figure 3.2.

By definition, a linear space invariant filter is said to be matched to a particular signal $s(x, y)$ if its impulse response $h(x, y)$ is given by:

$$h(x, y) = s^*(-x, -y) \qquad (3.3.52)$$

Then, if an input $g(x, y)$ is applied to a filter matched to $s(x, y)$, then the output $v(x, y)$ is found from eq. (3.2.39) as:

$$\begin{aligned} v(x,y) &= \iint_{-\infty}^{\infty} h(x-\xi, y-\eta) g(\xi, \eta) d\xi d\eta \\ &= \iint_{-\infty}^{\infty} g(\xi, \eta) s^*(\xi - x, \eta - y) d\xi d\eta \end{aligned} \qquad (3.3.53)$$

which is recognized to be the cross-correlation function of $g$ and $s$. Therefore, a filter matched to the input signal $s(x, y)$, is to be synthesized by means of a frequency-plane mask in a 4-f system. The Fourier transformation of the impulse response (3.3.45) shows that the required transfer function is [41]:

$$H(f_x, f_y) = S^*(f_x, f_y) \qquad (3.3.54)$$

where $H = \mathcal{F}\{h\}$ and $S = \mathcal{F}\{s\}$. Thus, the frequency plane filter should have an amplitude transmittance proportional to $S^*$, meaning that the frequencies allowed by the filter correspond to the optical system transfer function.

In this specific case we have assumed a single input signal $s(x, y)$ matched with the filter consisting in the frequency-plane mask. In this particular case, the field transmitted across the filter consists in a plane wave which is converted to a focus by the lens L3. In our experiments (described in the next Sections) this assumption is not valid. We deal, in fact, we linear combination of numerous fields, the focus at the output will be generated if and only if the fields trespassing the filter are in phase.

Anyway, from eq. (3.3.55) we understand that the point-spread function of a focusing system is the Fourier transform of the filtered spatial frequencies, therefore to different filter will correspond different focus shapes.

# 4  Spatial frequencies selection for structured focusing through scattering media

In this Chapter the first half of the work performed during this PhD is presented demonstrating that the focus generated via opaque lenses can be structured by controlling the scattering light. In particular, tailored spatial filtering is used for forming sub-correlation focusing and non-diffractive beams through scattering systems. In the first case, we studied the adaptive focusing resolution at given regime of scattering: lower the system transmittance, smaller the focus size. We prove our method to obtain super confined foci even in presence of mediocre scattering systems, the regime of semi-transparent media. In the second case, we take advantage of the opaque lenses configurability for demonstrating the formation of Bessel-like beams at user selected location at the back of scattering barriers. The Sections correspond to published articles that were produced by the experimental and theoretical studies of this project.

## 4.1 Enhanced adaptive focusing through semi-transparent media

Adaptive optics can focus light through opaque media by compensating the random phase delay acquired while crossing a scattering curtain. The technique is commonly exploited in many fields, including astrophysics, microscopy, biomedicine and biology. A turbid lens has the capability of producing foci with a resolution higher than conventional optics, however it has a fundamental limit: to obtain a sharp focus one has to introduce a strongly scattering medium in the optical path. Indeed a tight focusing needs strong scattering and, as a consequence, high resolution focusing is obtained only for weakly transmitting samples. In the In Vivo Imaging Lab (IVIL) at FORTH-IESL (Crete, Greece) we have demonstrated a novel method allowing to obtain highly concentrated optical spots even by introducing a minimum amount of scattering in the beam path with semi-transparent materials. By filtering the pseudo-ballistic components of the transmitted beam we were able to experimentally overcome the limits of the adaptive focus resolution, gathering light on a spot with a diameter which is one third of the original speckle correlation function [42]. The experiment and results are thoroughly described below.

### 4.1.1  Introduction

By properly adjusting the wavefronts it is possible to correct for the de-phasing acquired due to the random propagation in a disordered medium. The key enabling technology is the Spatial Light Modulator (SLM), a device which allows a point per point control of the wavefront of a coherent light beam. In practice by adjusting with the SLM the input beam shape it is possible to control the wavefront at the output of an optical system with an unknown scattering matrix. Various strategies have been developed to obtain light focused at a user defined location: time reversal phase conjugation [43], transmission matrix measurement [31] or phase scan based algorithms [37]. Vellekoop and Colleagues demonstrated that the minimum spot size achievable through adaptive focusing in strongly scattering materials is defined by the speckle correlation function [15]: in practice its limit is the speckle grain. The speckle pattern is a random distribution of bright and dark areas due to random interference of countless light paths transmitted through disorder and its grain size depends on the length $L$ and scattering mean free path $\ell$ [44, 45]: in transmission geometry, when the scattering strength increases, the typical grain of the speckle pattern becomes smaller. In order to obtain a tight focus through a scattering sample one has to exploit thick samples because high resolution is obtained in exchange of throughput [46], which is a critical obstacle for modern adaptive super-resolution techniques [16, 17], currently limited to low transmittance experiments only. Approaches based on dark field configuration improve visibility at the target [16], but do not allow selecting the speckle components in order to improve the effective resolution of the system. Optical Eigenmode (OEi) approaches were tested to achieve sub-diffraction optical features in free space for minimizing the size of a focused optical field [47]. The combination of particular photonic structures and wavefront correction by OEi methods have been exploited to produce subwavelength foci [48]. However, adaptive foci with a resolution under the limit of the speckle pattern correlation function were considered.

The current optical techniques adopted for bio-imaging are efficient enough up to the first millimeter in depth (1 transport-mean-free-path) [2] and wavefront modulation appears to be the best solution for correcting imaging quality and for suppressing the turbidity [49] at any diffusive regime. Therefore, an experimental technique which permits to achieve sharp adaptive focusing through weakly scattering samples was needed.

On the other hand, the speckle pattern resulting from semi-transparent media has larger speckle grains corresponding to higher intensities, and hence the focusing process which exploits intensity as a feedback automatically selects configurations with larger grains to increase intensity and thus producing larger foci. In our work, we demonstrated that using the high-pass spatial filter we can select those small grains, which are hidden due to dynamic range limitations, but nevertheless are present within the speckle pattern and not accessible otherwise. In other words, by appropriately selecting some of the components during the optimization we exploit hidden degrees-of-freedom that maximize the effective numerical aperture of the opaque lens. Indeed, we produce a focus smaller than the speckle pattern correlation function, effectively overcoming the theoretical limit previously proposed by Vellekop and coworkers [15] for strongly scattering media. The core idea consists of selecting only those light paths which experienced multiple scattering events by filtering the ones which have experienced only weak scattering. A spatial filter is employed for the selection of the appropriate light paths (modes) [41] and by exploiting a standard phase scan method [37] a stronger focusing with a smaller size speckle pattern is obtained. Furthermore the focus persists also if the filter is removed so that a sharp light spot is obtained within a speckle pattern with a much larger grain. By exploiting our protocol we are able to obtain a focus size approximately 68% smaller than the average speckle grain.

### *4.1.2 Description of the experimental system*

In Fig. 4.1 a schematic representation of the experimental setup is shown. A coherent laser source emitting at 594nm is used while a homemade telescope (lens L1 + lens L2) magnifies the laser beam by 10X. Modulation is performed by a phase only Spatial Light Modulator (SLM) (Holoeye, Pluto, Berlin-Adlershof, Germany) that shapes the wavefront of the beam; a 50:50 beam splitter (BS) guarantees the beam and the SLM are perpendicular to each other. Hence, 50% of the light reflected by the SLM is directed along a perpendicular axis where a second telescope (lens L3 + lens L4) reduces the beam by 15X to 0.24mm. A collimated beam impinges onto the scattering sample (S). During propagation through the turbid material, scattering decomposes an incident wave into multiple components which generate a speckle pattern by randomly interfering at the output of the sample. The speckle pattern is collected by a 10X infinity corrected microscope objective (OBJ) with 0.25 numerical aperture.

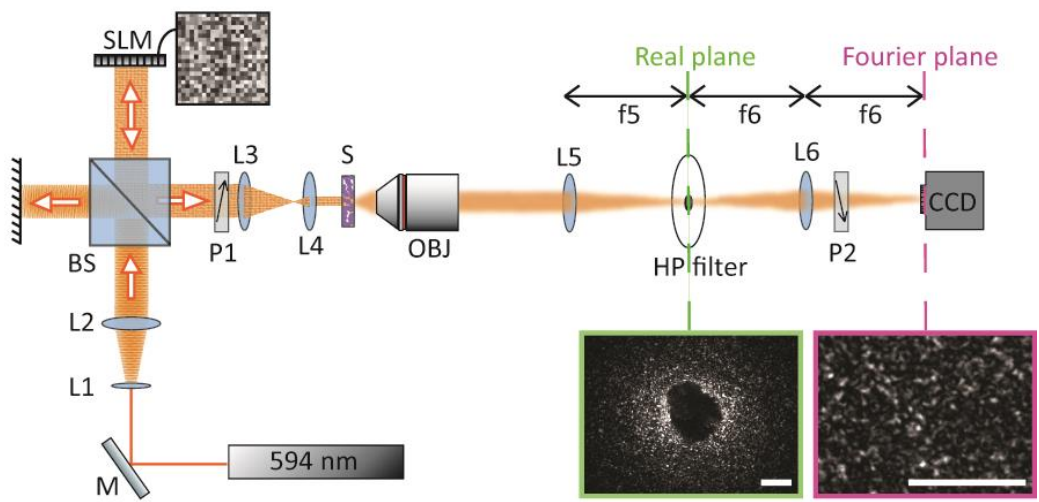

Figure 4.1 *A schematic representation of the experimental setup. The light transmitted through S is used to produce a real image of the sample by lens L5. The image is filtered from its central components by a Spatial High Pass filter (HP filter). The result of the filtering is Fourier transformed onto the camera plane by the lens L6. Polarizers P1 and P2 have perpendicular orientation in order to filter the ballistic contribution. Scale bars correspond to 0.5mm.*

At a distance of 150mm from the rear face of the OBJ a lens L5 is placed to reproduce the speckle pattern (and a real image of the sample) at its focal length (see Fig. 4.1). On this plane the High-Pass (HP) filter is aligned in order to block the central components of the speckle pattern: these components are related to modes which underwent a few scattering events (hence they are weakly scattered from the ballistic trajectory; we refer to them as *"pseudo-ballistic"*). Finally a 400mm lens L6 produces an image on the CCD camera plane which is the conjugate plane of the image from L5. The image on the camera is the result of the superposition between modes not blocked by the HP filter or by the Polarizers (P1 and P2 with perpendicular orientations serve to eliminate ballistic contribution). The combination of OBJ, L5 and L6 creates a magnification on the camera that corresponds to X15.

The scattering samples are $TiO_2$ thin layers fabricated by sedimentation of water suspensions and subsequent evaporation of the water (or ethanol). We exploited $TiO_2$ Anatase nano-powder with particle size <25nm (Sigma-Aldrich, St. Louis, MO,USA). Samples were let to sediment and to dry at $30^oC$ in an oven using the evaporation method [45]. In $TiO_2$ slabs absorption can be neglected and elastic scattering is the only loss mechanism. Employing a modified Beer-Lambert Law [44, 45, 28, 50] we have that after a thickness *L*, a ballistic beam attenuates as:

$$I_{OUT} = I_{IN} \cdot e^{-L/\ell} \tag{4.1.1}$$

where $I_{OUT}$ represents the intensity exiting the slab when intensity $I_{IN}$ impinges on it. Inverting the equation we can estimate the optical length of the sample ($L/\ell$), which provides a measure of the scattering properties.

In order to control the thickness and the area of the scatterers we used a specific protocol which allows the fabrication of reproducible thin layers, in particular we avoided the formation of cracks and evident surface defects. We directly deposit the colloidal solution on the glass slide and we register the number of drops poured. A single drop corresponds to a volume of $5\mu l$ of colloidal solution. This methodology produces homogeneous samples up to a certain thickness. The maximal thickness depends on the type of particles (volume, shape, dispersity and charge). If a large volume of solution is poured the surface of the sample results convex and cracks are more likely. For this reason we pour a maximum of 3 drops of solution, but we vary its concentration ($\rho(\%)$) in order to regulate the final samples thickness. The concentration $\rho$ is defined with the percentage of particles in the total volume of solution, in our case we varied $\rho$ from 0.5% to 2%. Following these precautions, we obtain that the inhomogeneities are much smaller than the size of the wavelength producing a much smaller scattering efficiency and less inhomogeneity. We fabricated samples ranging from semi-transparent (an example is shown in Fig. 4.2(2)) up to strongly scattering, a very large range of transmittance, and their shape and composition allow a good comparability. To demonstrate this we report in panel (1) of Figure 4.2 our sample thickness profile measured with the "Perthometer PRK" (Mahar-Perthen, Providence, USA) Surface Profilometer.

Table 4.1 presents the characteristics of the different samples we have produced. The first column of the table shows the density $\rho$ of the colloidal solution, which is described in terms of percentage of dispersed spheres with respect to the liquid. In the second column we show the amount of solution poured in terms of volume ($\mu l$). The third column gives the optical length of each sample.

**Table** 4**.1:** *Table describing the samples tested in our experiments and their relative optical thicknesses.*

| $\rho(\%)$ | Volume ($\mu l$) | $L/\ell$ |
|---|---|---|
| 0.5 | 5 | 0.70 |
| 0.7 | 5 | 0.81 |
| 1 | 5 | 1.67 |
| 1 | 10 | 2.92 |
| 2 | 10 | 5.74 |
| 2 | 15 | 6.33 |

The mode filtering relies on custom-made Spatial High Pass filters (HP filter) with diameters ranging from 0.35mm up to 1.3mm, an example of filter used in the experiment is shown in panel (3) Figure 4.2. The HP filters are fabricated by a mixture of 85% black dye used for solvent free resins (Pentasol No 3312, Prochima, Pesaro-Urbino, Italy), 14% Crystal Super Transparent resin (Prochima) and 1% of its catalyst. The clear resin that was used has minimum absorbance hence it is considered as perfectly transparent basic material. It must be noted that black pigment was used not only because it is perfectly dissolved in resin but also due to the fact that it has very smooth absorbance across the visible regime.

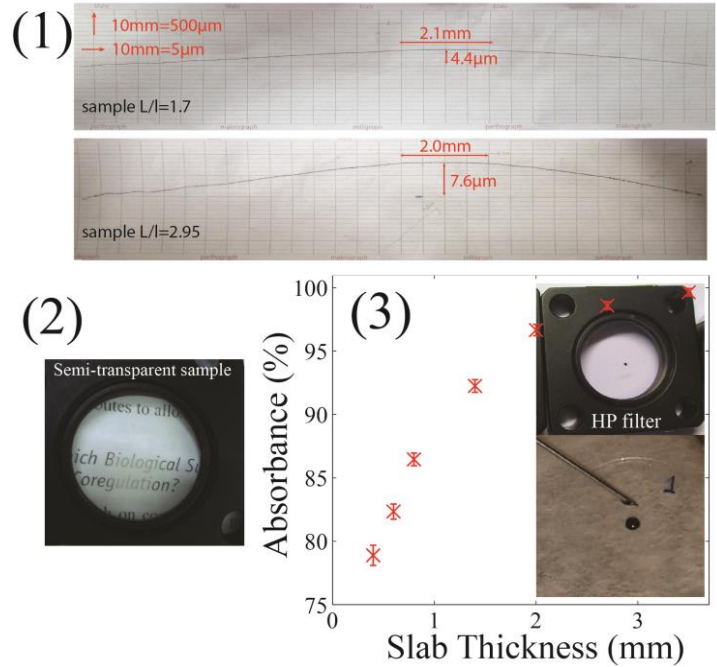

**Figure 4.2** *In panel (1) we report images of the samples' profiles with different scattering strengths. Using low concentration of $TiO_2$ in the colloidal solution we obtained semi-transparent curtains as showed in panel (2). In panel (3) we report the filter absorbance measured at different slab thicknesses. For fabricating the HP filter, we poured the mixture on the top of a cover slip as shown in the insets.*

We have always followed the fabrication method described below in order to achieve homogenous filters. The liquid mixture is placed in a vacuum chamber for 10 minutes in -30 inHg so that air bubbles that have entered the mixture while stirring are completely removed.

For producing the HP filters the bubble-free liquid mixture is deposited on thin microscope cover slip glasses with a micro-pipette (inset in panel (3) of Figure 4.2), a process that allowed controlling the diameter of the droplet: to larger volume poured corresponds larger diameter *D* of the filter. The mixture is then solidified in an oven at 30$^o$C for 5 hours. Once the droplet has reached the solid state it remains permanently stable and static (inset in panel (3) of Figure 4.2). To evaluate the optical properties of our filters, we fabricated thin films using the same protocol and we measured their total absorbance with a spectrophotometer (Lambda 950, PerkinElmer, Waltham-Massachusetts, USA). We tested films with different thicknesses as shown in panel (3) of Figure 4.2. The average thickness of the filters produced for our experiment range from 0.5mm up to 1 mm, with a nominal absorbance of (A=85%), reflection (R=14.5%) and total transmission (T=0.5%).

The optimization protocol for focusing starts from a random SLM mask, splitting the wavefront into segments with random de-phasing. The mask is composed on the SLM by 40x40 segments with a 255 grayscale each, corresponding to a fixed de-phasing of the light reflected within a range from 0 up to 5π. After setting the target, the algorithm tests a series of 50 random masks picking the one providing the best intensity value at the target position (preliminary optimization). Starting from this configuration, the algorithm starts a routine which tests a phase shift of π and -π for a single segment accepting it only if an intensity enhancement on the target is measured, otherwise the previous configuration is restored. This routine is repeated for each segment. The same process is used for testing a phase shift of π/2 and –π/2. This algorithm allows for an *enhancement* factor ($\eta$) [9] of 250, calculated as the ratio between the intensity at the target and the average intensity of the speckle pattern before optimization.

### 4.1.3 *Sub-correlation focusing with Spatial High-Pass filter*

Having established the methodology for sample preparation and the optical setup we studied the effects of the spatial filters on light transmission through our media. Different samples *S* were illuminated with the same beam waist (0.24mm) and the back surface of each sample has been aligned on the focal plane of the OBJ. We define the Full Width at Half Maximum, FWHM (*w*) as the width of the intensity peak around the target spot.

In Fig. 4.3 we compare the values *w* obtained with the standard focusing approach [15] (without filtering) for different samples characterized by the optical length $L/\ell$ (blue circles in Fig. 5.3(b)) with the values obtained when the HP filter is used (red squares in Fig. 4.3(b)). We study the focusing resolution, measured as the full width at half maximum *w* for different scattering samples with controlled optical transmittance (sample thickness is varied from a few micrometers to hundreds of micrometers) corresponding to optical lengths between one and ten scattering mean free paths [45], as reported in Table 4.1.

The results of Fig. 4.3 obtained for samples with different optical thicknesses $L/\ell$, and presented by the blue circles, demonstrate that opaque lenses reach the optimum efficiency for an optical length of the order of multiple of scattering mean free paths. The results of the same experiment with a HP filter applied as presented in Fig. 4.3(b) by the red squares, demonstrate that the FWHM of the foci obtained with the filter blocking the pseudo-ballistic modes (see sketch Fig. 4.3(a)) is always smaller than the non-filtered case.

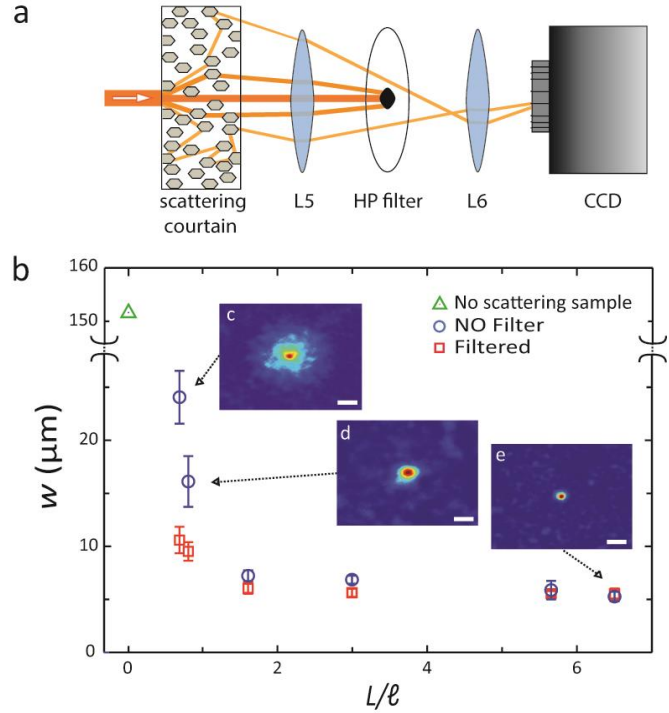

**Figure 4.3** *a) Graphical representation of the operating principle of the HP filter. Only light that experiences few scattering events is blocked by the filter, the rest contributes to the speckle detected. b) Full Width at Half Maximum (w) of foci obtained at the end of the optimization process plotted as a function of the optical length $L/\ell$ of the samples. The green triangle represents the beam waist impinging onto the sample. A comparison between the ordinary approach (no filter, blue circles) and the case with the D=0.8mm HP filter (red squares) is shown. The three insets, starting from the top, correspond to foci obtained at $L/\ell$=0.7, 0.81 and 6.3. White scale bars correspond to 30μm.*

We characterized the effects of the HP filters demonstrating strong adaptive focusing through semi-transparent materials. We measured the focus width, *w,* as a function of the HP filter diameter *D,* through a semi-transparent sample. Fig. 4.4 reports the results for *D* varying from 0.35 to 1.3 mm. The *w* decreases by increasing the filter diameter and it is possible to obtain a focus much smaller than the size of the speckle grains. Panel a) on Fig. 4.4 shows the speckle pattern at the back of the disordered sample (optical length $L/\ell$=0.81). Panels b) and c) show the same pattern filtered by a 0.7mm and a 1.3mm diameter beam stop, respectively. By eliminating the pseudo-ballistic components, our filter selects modes which have suffered stronger scattering and thus produce a larger effective numerical aperture yielding a smaller speckle and a smaller focus. On average (statistics over 10 measurements), we obtained a focus which is of 0.32±0.15 (one third) of that obtained when the filter is absent.

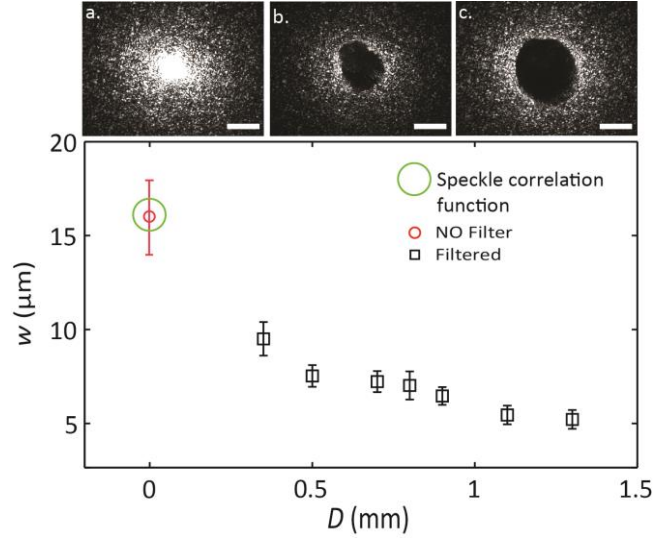

**Figure 4.4** *In the upper panel the speckle pattern in the plane generated by L5 under three different configurations. Panel a. is the free speckle pattern through a sample with optical length* **L/ℓ**=0.81*; b. and c. are filtered patterns with 0.7mm and 1.3mm diameter spatial filter, respectively. Scale bars are equivalent to 0.5mm. The graph shows the dependence of the focus width w (in black square) versus the beam stop filter diameter D. All data has been collected with sample length* **L/ℓ**=0.81 *and is compared to the FWHM of the focus obtained in the absence of a filter (red circle).*

Fig. 4.5a presents a snap-shot of the initial speckle pattern generated through a semi-transparent sample, before the optimization, when the HP filter is not inserted; the pattern composed of large grains with an average diameter of 25μm estimated from the speckle correlation function. In Fig. 4.5b we report an image of the focus achieved with a 0.5mm HP filter. A remarkable effect is observed after the removal of the HP filter; the adaptive focus is not affected and remains smaller than the speckle size as shown in Fig. 4.5c.

The filtering of the pseudo-ballistic modes is only needed during the optimization procedure, while the sharp focusing is retained after the filter removal. In Fig. 4.5d we compare the foci profiles. The solid curve corresponds to the measurement obtained with the HP spatial filter, while the blue dashed curve corresponds to the standard focusing approach [15] without filtering. In this case the effect of the filter is a reduction of the focus size by a factor of 0.52±0.11; moreover, removing the spatial filter does not alter the focus, which maintains the same shape, while being surrounded by larger speckle grains. The focus waist obtained is 12.3±0.8μm (blue profile in Fig. 4.5e) which is approximately the same as the size of the focus obtained with the filter (FWHM is of 12.1±0.8μm, represented by the blue dashed curve in Fig. 4.5d and 4.5e) and is approximately half of the speckle size grains (FWHM of the speckle correlation function is 23.5±2.5μm, represented by the dotted violet curve in Fig. 4.5e).

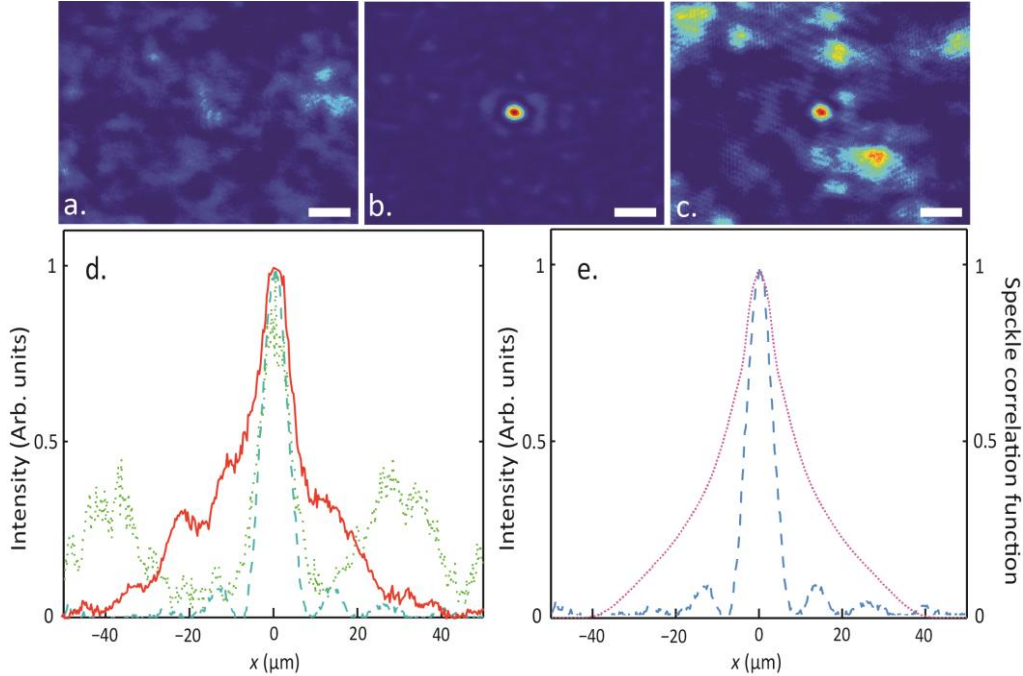

**Figure** 4.5 *a) speckle pattern obtained through a $L/\ell=0.7$ scattering sample. b) the optimized focus with a 0.5mm diameter filter, c) the same as in b but after the filter is removed. Note: the focus still present and smaller than the speckle grain. White scale bars correspond to 30μm. d) Focusing profiles at the target position without filter (solid red curve), with filter (dashed blue curve) and when the filter is removed (dotted green curve). e) The focus obtained with the filter (dashed blue curve) is compared to the speckle correlation function of the original speckle pattern in the absence of a filter (dotted violet curve).*

When the filter is absent the presence of the pseudo-ballistic modes increases the background signal with respect to the focus intensity decreasing the Peak-to-Background Ratio *($\eta_{PBR}$)* calculated as the maximum intensity at the target divided by the average intensity of the background. We measured $\eta_{PBR}$ equal to 196±35 with filter and equal to 16±5 without filter (statistics over 10 measurements). The pseudo ballistic modes (with small numerical aperture) hide the modes associated with strongly scattering light paths (which produce a less intense contribution but a larger effective numerical aperture) making impossible to obtain a sharp focus when both the contributions are interfering on the image plane. This is demonstrated by measurements shown in Fig. 4.6, where we report the radial distribution of the intensity in the plane of the filter: apparently, when the filter is absent (red curve) the intensity at *r*=0 (pseudo ballistic modes) is large. The filter displaces the maximum at larger values of *r* thus it selects highly scattering modes.

In order to measure the effective intensity modulated by the SLM during the focusing process when the spatial filter is inserted we grabbed a direct image of the plane of the filter, the total camera counts correspond to the modulated intensity $I_M$. We compared $I_M$ with the total camera counts from the same plane in absence of filter *I*. The ratio $I_M/I$ evaluates the factor of loss for the Peak-to-Background Ratio enhancement *($\eta_{PBR}$)* of the focus for the two case without and with the filter. The ratio of the modulated versus non modulated intensity ($I_M/I$) depends on the filter size *D* (Fig. 4.6b red open circles).

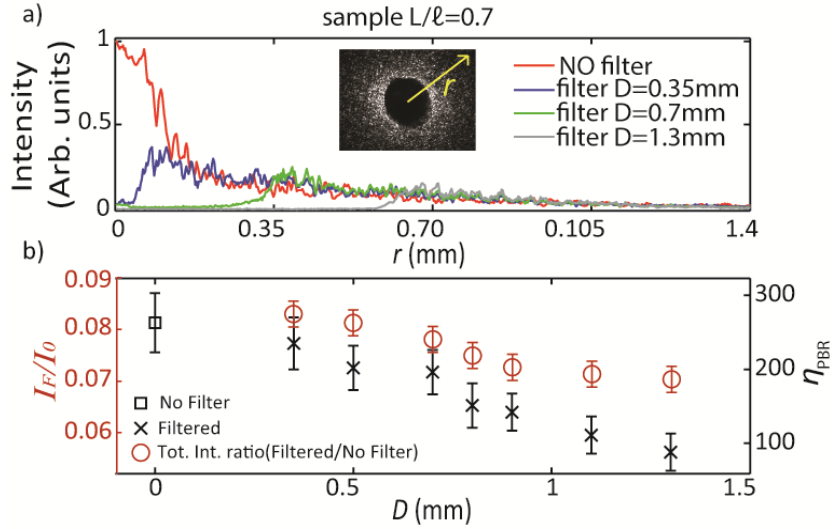

**Figure 4.6** *a) the intensity profile along the direction r of the speckle pattern in the Real Plane (indicated in Fig. 1) through a **L/ℓ**=0.7 scattering sample. Four different configurations are reported: the red line is the free speckle pattern, blue, green and gray are respectively the profiles with 0.35mm, 0.7mm and 1.3mm diameter spatial filter. b) red circles show the ratio $I_M/I$ as a function of the filter diameter D. The foci enhancements $\eta_{PBR}$ (black square and black crosses) decrease with the size of the filter since a part of the light controlled by the SLM has been stopped.*

The black crosses in the same panel show the enhancement $\eta$ *(with filter)* as a function of *D*. In the case studied in Fig. 4.5, the enhancement drops from $\eta_1$ =196±35 (with filter) to $\eta_2$ =16±5, a factor $\eta_2/\eta_1$ =0.082, in perfect agreement with the value of $I_M/I$= 0.080. In a nutshell, when the filter is removed the signal-to-noise ratio decreases by a factor equal to the amount of non-modulated intensity which is added to the previous speckle.

### *4.1.4 Numerical simulation*

The spatial High-Pass filter operating in our setup works by selecting those spatial frequencies which have experienced multiple scattering through the scattering layer, while the quasi-ballistic light is blocked and it does not contribute on the camera plane. It is possible to model the process numerically by using Fourier Optics theory and to relate the presence of the High-Pass spatial filter with the speckle size distribution on the camera plane. The Modulated Transfer Function (MTF) and the Point Spread Function (PSF) are connected by a Fourier transformation $\mathcal{F}$ of the input field; using this tool one can model the effect at the output due to the presence of an additional spatial filter at the input. Our numerical model starts from the acquisition of an unfiltered speckle pattern $S(x,y)$ where *x* and *y* are coordinates at the input plane at the back of the scattering sample. We assume *S* as the input of our model. Since we work with a 4-f system, we can assume that our optical system is a "black-box" (as in Fig. 3.7 from Section 3.2). Its PSF will depends on the intensity distribution in the Fourier plane defined as $S(k_x,k_y) = \mathcal{F}[S(x,y)]$, indeed the spectrum of the system. The spatial filter acts in $(k_x, k_y)$ space and it can be simply modeled with a circular mask centered in the Fourier plane:

$$M(k_x, k_y) = \begin{cases} 0 \text{ if } r \leq R \\ 1 \text{ if } r > R \end{cases} \quad (4.2)$$

where $r$ is the distance from the center of the mask and $R = D/2$ is the selected radius of the filter.

A schematic of the system is shown in Figure 4.7. Therefore, the effect of the filter can be represented with convolution products between $S(k_x, k_y) \cdot M(k_x, k_y)$, resulting in a cut of all the signal in a circular region around the center of the plane. By inverse Fourier transforming this quantity

$$\mathcal{F}^{-1}[S(k_x, k_y) \cdot M(k_x, k_y)] = S'_M(x, y) \quad (4.3)$$

we obtain the different speckle distribution $S'_M(x, y)$ on the camera plane where the ballistic light contribution has been filtered out, which represents the intensity detection at the camera plane in the setup.

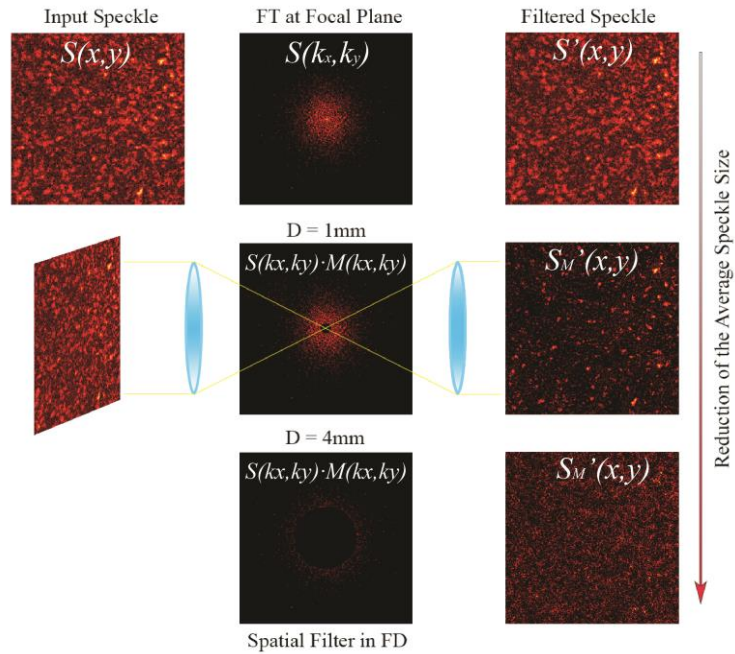

**Figure 4.7** *Effect of the high pass spatial filter in the average speckle size. Filtering the ballistic light and increasing the diameter of the filter reduces the dimension of the speckle grains, which allow us to obtain a sharper focus. The focus obtained in such way persists after removing the filter.*

The virtual speckle pattern obtained by the numerical model, corresponding to $|S'_M(x, y)|^2$, is then analyzed via correlation function, $c_I(\Delta x, \Delta y)$, to obtain the average size of the grains (*speckle grain*): an automatic routine performs the correlation function of the $|S'_M|^2$ at different filter diameters and calculates the correlation function's Full Width Half Maximum $w_{FWHM}$ as an indication of the speckle diameter (blue curve in Fig. 4.8(a)).

Since the speckle average size is not affected by the shape of the wavefront, the calculated function $f_{FWHM}(D)$ represents also the smallest focus achievable using a filter of diameter $D$. The model reasonably fits the corresponding focus width, *w*, obtained from the experiments. In such a way, considering the intensity distribution of the spatial frequencies of a filtered speckle (see panel (b) of Fig. 4.8), we can assume it as the MTF of our optical system as illustrated in panel (c) of Fig. 4.8.

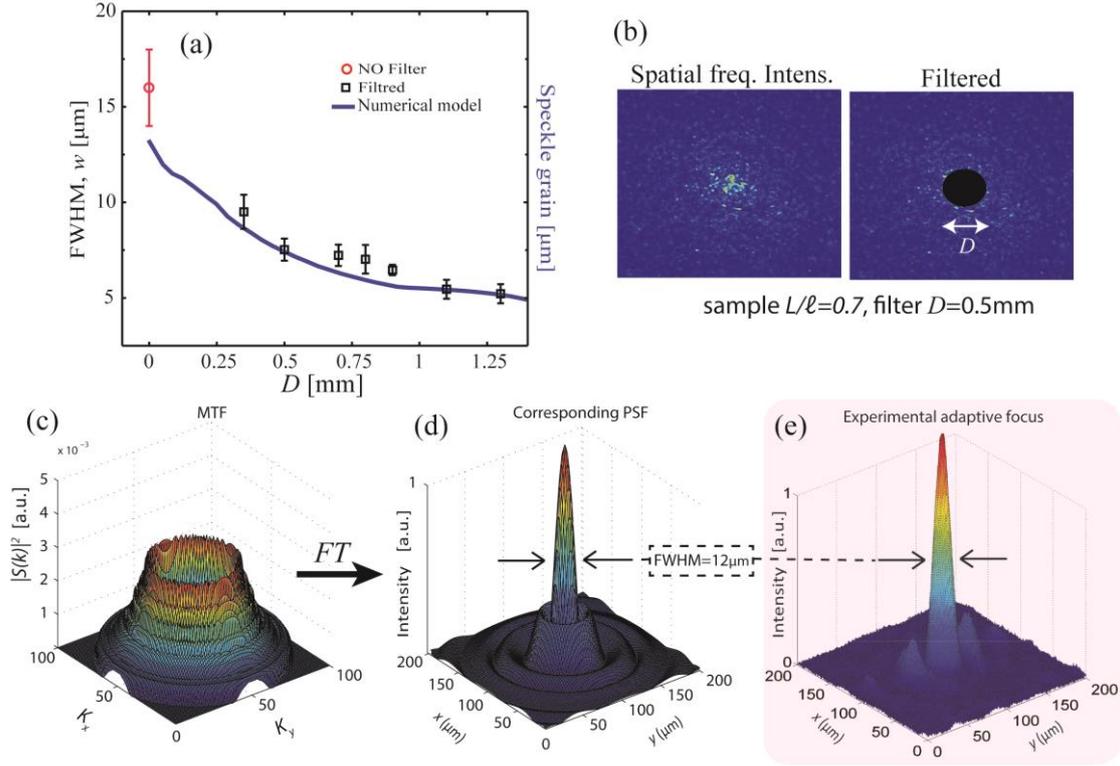

**Figure 4.8** *In (a) Blue line represents the speckle grain size, the values are obtained from the FWHM of correlation function of the synthetic speckle pattern obtained by Fourier transform (FT) of the filtered spatial frequencies, as shown in (b). The red circle is the focus size obtained without filter and the black squared are the FWHMs' of the foci obtained in presences of the filters. Intensity distribution of the filtered spatial frequencies is shown in (c). Its FT provide the corresponding PSF (d), which equals the final intensity shape of the adaptive focus (e).*

Its *FT* provides the corresponding PSF (Fig. 4.8(d)), which is equivalent to the $c_I$ of the speckle pattern (after filtering) and predicts the shape of focus obtained at the end of the wavefront shaping process (Fig. 4.8(e)).

### *4.1.5 Discussion*

We have demonstrated the enhanced adaptive focusing through weakly scattering media, by the introduction of a spatial filter in the image plane of the produced speckle pattern. The effect is a significant increase of the focusing resolution with a reduction of the spot size below the speckle size defined by the speckle correlation function. The method strongly improves the focusing resolution for turbid samples (turbid lenses) with optical lengths smaller than two scattering mean free path. The introduction of a spatial filter reduces the speckle size because pseudo-ballistic modes possess a reduced span of momentum components (thus producing a smaller effective numerical aperture) with respect to strongly scattered modes. This means that in the weak scattering regime, speckles result from a superposition of patterns possessing different grain size. Our filtering technique, selects a speckle with a smaller size during the optimisation protocol, and this allows the algorithm to exploit degrees of freedom which are "hidden" in the standard focusing optimisation. In fact, in the extremely weak scattering configuration, pseudo-ballistic modes are much more intense than the strongly scattering modes, thus producing a large grained speckle which hides the small grained speckles produced by the strongly scattering channels. With our approach we are thus able to select the components of a speckle pattern producing a focus with the maximum resolution. The presented results not only implement the state-of-art of the adaptive focusing process but may also open the way to a novel generation of high transmission, semi-transparent turbid lenses with high focusing resolution.

## *4.2 Bessel beams through scattering media*

Inspired by the results described in the previous Section, we explored a new type of filtering able to provide interesting light features. The focus shape depends on the particular filter in use and therefore, we were challenged by the possibility of producing Bessel beams with opaque lenses. Bessel beams are non-diffracting light structures, which maintain their spatial features after meters of propagation and are realized with simple optical elements such as axicon lenses, spatial filters and lasers. In this Section we demonstrate a method for generating non-diffracting Bessel-like beams through a heavily scattering system, exploiting wavefronts shaped by a spatial light modulator [51]. With the proposed method starting from amorphous speckle patterns, it is possible to produce at user defined positions configurable and non-diffracting light distributions which can improve the depth-of-field in speckled illumination microscopy.

### *4.2.1 Introduction*

Recent advancements have pushed optical imaging beyond the diffraction limit, increasing drastically the resolution of optical investigation in many fields. The plethora of new systems and methods has changed the paradigm of biological imaging from the traditional microscope to sophisticated techniques covering the wide range of imaging scales from molecules to cells to whole organisms. It is possible today ro reach nanoscale resolution [52] with commercial equipement, exploiting near-field methods [53] or Structured Illumination Microscopy (SIM) [54, 55]. An alternative approach is to exploit speckle patterns, random light structures generated by a coherent light beam inpinging on rough surfaces or trespassing strongly scattering layers and auto-regenerating behind an absorber [56]. It has been recently demontrated that speckle-light illumination microscopy methods [57, 58, 59] are capable of super-resolution with a wide Field of View (FOV) [60] exploiting wavefront shaping to compensate for light diffusion through highly disordered media [9]. This approach is becoming increasingly relevant to biomedical and biological investigations in both imaging and therapy [11] and numerous other fields [61, 20, 62, 63]. The scattered light can be engineered by the segments of a Spatial Light Modulator (SLM) in order to produce configurable optical elements [64], indeed, a scattering material with an adaptive optical system is an "opaque" lens [9] or polarizer [65] or wave plate [66] or spectral filter [67]. Regarding optical focussing, opaque lenses can be used for sub-diffractive resolution through a scattering layer [16, 17] and to generate light foci controllable in space and time [9, 15].

Wavefront correction has been exploited for enhancing and correcting the quality of Bessel beams [68, 69, 70, 71]. From such approaches fluoroscence imaging techniques can benefit due to the increased depth-of-field with a drastic improvement in volumetric imaging quality, increased acquisition frame rate and reduction of acquisition time [72]. Bessel beams have the property to be "self-reconstructing" meaning that the light-structure is immune to scattering encountered along the direction of propagation or beam trajectory [73], (thus enabling applications such as deep laser writing [74]), but not to extended scattering barriers [75]. We have demonstrated a method for manipulating speckle patterns in order to generate Bessel beams at the back of a scattering wall exploiting amorhous speckle patterns [51]: recently discovered light structures, which are generated by filtering the light transmitted at the back of a scattering sample with a ring-shape spatial filter [76, 77]. Amorphous speckles are disordered distributions of light intensity characterized by a short–range order which may be exploited to generate static scattering structures for example in photorefractive crystals [78]. In a way similar to Bessel beams, amorphous speckles exhibit an extraordinary Depth of Field (DOF). We demonstrate here the production of reconfigurable Bessel-like beams at user defined positions through scattering media by exploiting togheter adaptive focusing and amorphous speckles.

### *4.2.2 Amorphous speckle patterns*

The experimental setup used for the production of the amorphous speckles is schematically represented in Fig. 4.9. A He-Ne laser emitting at 594nm is used as a source, while the beam is magnified by a10x telescope and corrected only in phase by a SLM (Pluto, Holoeye Germany). A second 20x telescope de-magnifies the beam impinging on the scattering sample (S) with a waist of 0.80mm. The scattering sample consists of a dielectric slab of $TiO_2$ Anatase nano-powder with polydispersed particle size <25nm (Sigma-Aldrich, St. Louis, MO, USA), fabricated via sedimentation of aqueous suspensions and subsequent evaporation of the water. In our experiments we used a sample with optical length $L/\ell = 8.5$ calculated as the ratio between the thickness ($L$) and the scattering mean free path ($\ell$), $L/\ell$. Lens L1 collects light at the back of the sample and generates the Fourier Transform of the output (Fig. 4.9(a)) onto its focal plane (focal distance $f_1$=25.4mm). On this plane, a ring-shape spatial filter (RSF) is placed in order to select spatial frequencies in a $\delta r$ region as depicted in Fig. 4.9(b) and 1(c). Lens L2 (focal distance $f_2$=200mm) conjugates this "Fourier plane" onto a "real plane" corresponding to the position of the camera sensor. We will address the ensemble consisting of the SLM, disordered system and ring-filter, as an Axicon Opaque lens (AOL). The camera is mounted on a translation stage oriented along the direction of propagation of the light ($z$-axis). Two polarizers (P1 and P2) with reciprocally perpendicular orientation are used in order to eliminate any residual ballistic contribution through the sample.

The ring-shape filters are fabricated by placing a circular beam stop at the back of an iris. The beam stop consists of a circular film of dry mixture of resin with black dye on a transparent cover slip. Once the beam stopper is centered on the aperture of the iris, an annular aperture is formed with variable width $\delta r$ depending on the iris opening (Fig. 4.9(b)). We used three filters with radius $r$=1.45mm and $\delta r$= 0.25mm, 0.80mm and 1.50mm. As described above, lens L2 performs a Fourier transform of a ring-shaped momentum space, (which is populated by speckles) on the plane of the camera. In the absence of the filter (Figure 4.10.*I*A) a standard speckle pattern is generated.

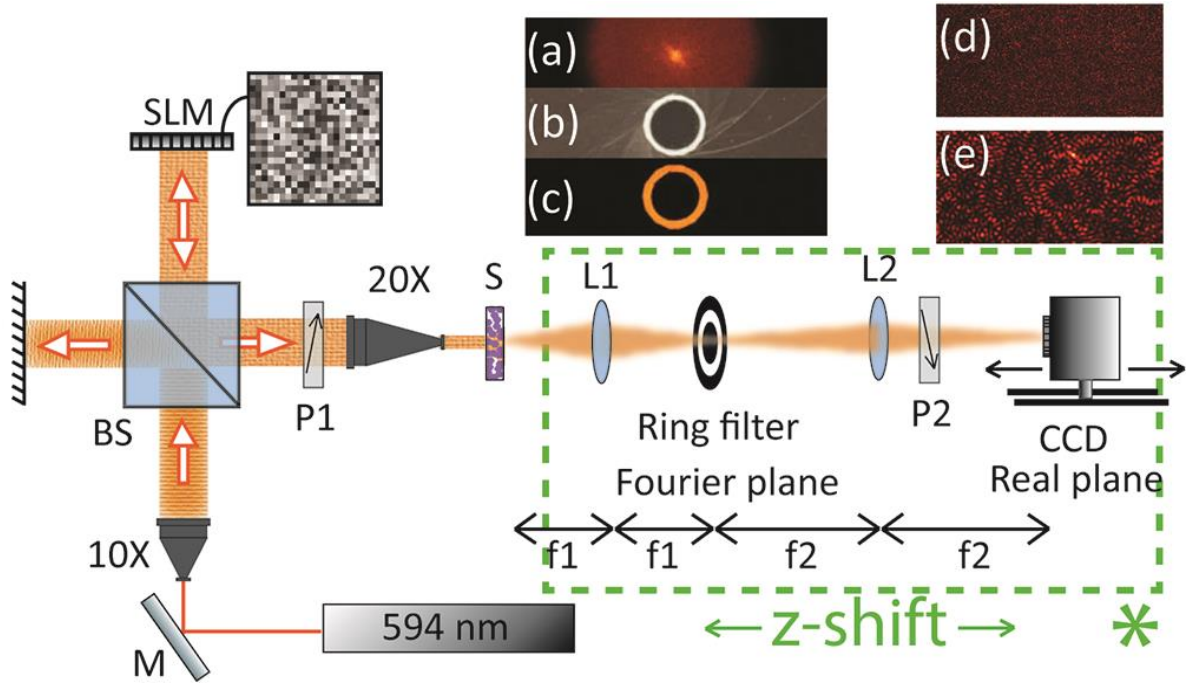

**Figure 4.9** *Experimental setup: a collimated beam is shaped by a SLM. Once the beam is de-magnified by a 20X beam reducer, it impinges on the scattering slab S. The output at the back of the curtain (inset d.) is imaged on the plane of the camera by a telescopic system composed by the lens L1 and the lens L2. The lens L1 generates the Fourier Transform (FT) of the back of the sample (inset a.) on a plane at its focal distance 25,4mm. When a ring filter (inset (b)) is inserted on this plane, spatial frequencies are selected (inset (c)). In inset (d) we report the pattern on the camera without filter and in (e) with filter. The green dashed rectangle tagged with the green ∗ identifies the detection system used to study the DOF of the Bessel beams generated just behind the scattering sample when the RF is removed.*

When the ring filter is applied instead, the pattern structure acquires correlation which becomes more evident if the filter's aperture $\delta r$ is decreased. Figures 4.10.*II*A, B, C & D present the interference patterns generated on the camera plane under different filtering conditions as shown in Figures 4.10.*I*A, B, C and D. We then studied the corresponding power spectrums obtained by Fourier transforming the speckles of Figures 4.10.*I*A, B, C & D, as shown in the series of Figures 4.10.*III*. In the absence of the filter, the complex distribution of speckles corresponds to a uniformly distributed power spectrum (Fig. 5.10.*III*.A). The introduction of the ring-shape spatial filter in the Fourier domain modifies the power spectrum profile as shown in Fig. 4.10.*III*.B. In practice, decreasing the width of the aperture, $\delta r$, the interference pattern becomes amorphous (see image Fig. 4.10.*III*.D) [76, 78, 79]. The ring shape in the power spectrum, visible in Fig. 4.10.*III*.C. & D, indicates light structures distributed with non-Gaussian disorder, in other words the amorphous patterns have local correlations between positions of the speckle grains [80].

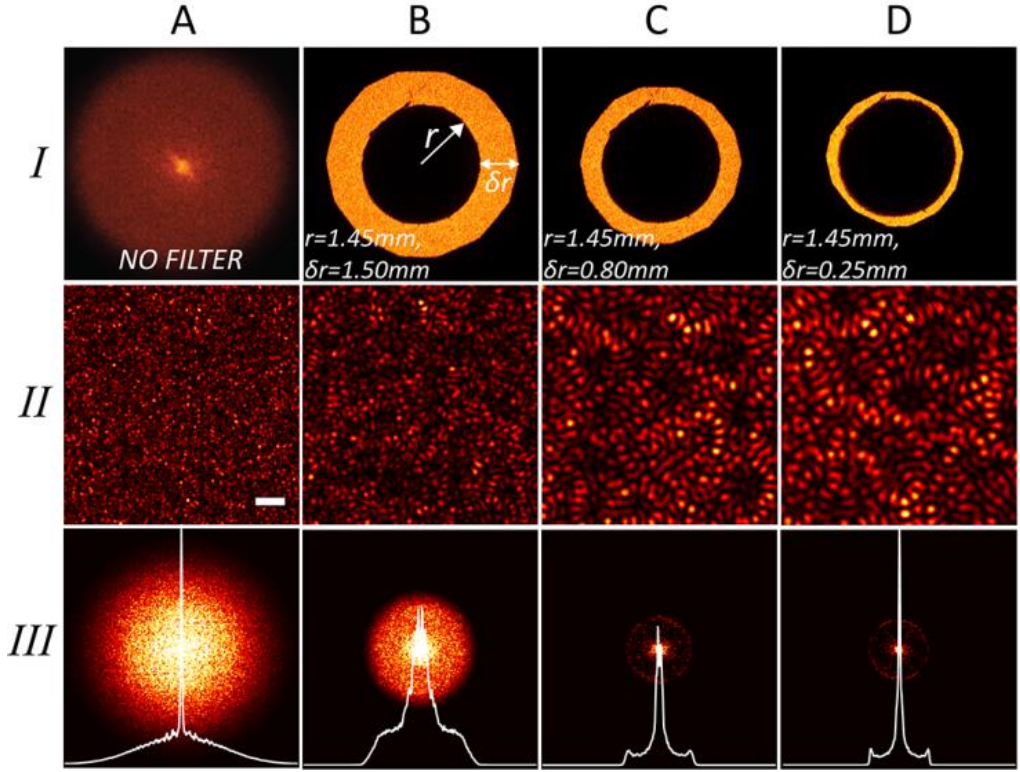

**Figure 4.10** *Images of the different filtering configurations: I.A. no filter; I.B, C & D filters with δr= 1.50mm; 0.80mm; & 0.25mm, respectively. The radius r is kept fixed at r=1.45mm in all cases. II.A, B, C & D the corresponding patterns generated at the camera plane. The white bar corresponds to 200μm. In III.A, B, C & D we calculate the power spectrum from the corresponding patterns. The superimposed white curves show the intensity profile calculated by radially averaging the power spectrum. Smaller δr results to a more defined annular pattern in the power spectrum.*

The following step is to focus light at an arbitrary (user defined) location with amorphous speckles. We exploited an iterative Monte-Carlo algorithm similar to those presented in previous works [37], although similar results can be obtained with the approaches based on transmission matrix measurement [31]. A mask composed of 40x40 segments is addressed to the SLM window, we assign to each segment a gray tone on 255 corresponding to a fixed de-phasing of the light reflected by the same segment within a range from 0 up to $3.1\pi$. A preliminary optimization tests a series of 50 random masks picking the one providing the best intensity value at the target position. The mask is taken as input for the second optimization routine, which tests a phase shift from $-\pi$ to $\pi$ (with $\pi/4$ as step size) for a single segment accepting the one providing the best intensity enhancement measured on the target. Each segment of the mask is tested. At the end of the focusing iteration with an annular filter (radius $r$=1.12mm and aperture $\delta r$=0.25mm), we achieve an *enhancement h=116* where $h$ is the ratio between average intensity before and the intensity at the peak after the optimization. As a comparison we performed the experiment also without filtering obtaining *h=55*7. As previously demonstrated [9] the great advantage of such approach consists in the ability to generate a spot at user defined positions without mechanically moving the optics. Once the focus is formed at the end of our optimization process we register the position of the camera plane as $z_0$. We collect the extension of the Bessel-like spot by translating the CCD camera along the propagation direction (see Fig.4.9); we then reconstruct the beam profile along the *z*-axis.

### 4.2.3  *Axicon opaque lenses*

In Figures 4.11(a), 4.11 (b) and 4.11 (c) we report images of the Fourier plane generated by L1, the speckle on the camera plane and the focus at the end of the optimization process with its intensity profile (inset white curve), respectively.

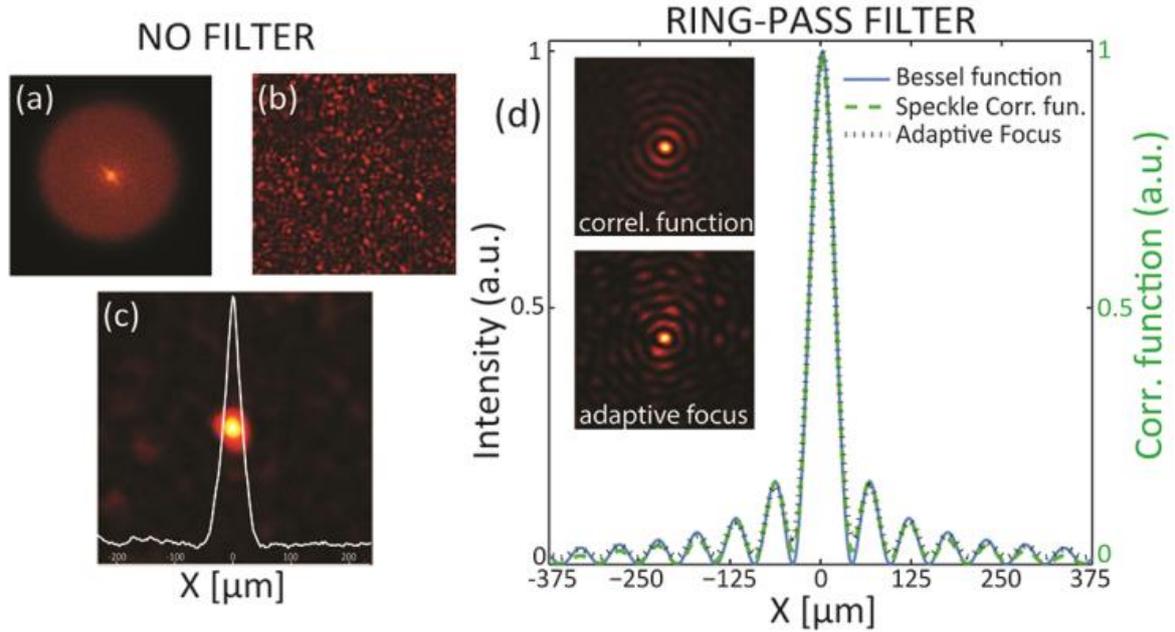

**Figure** 4.11 *(a) and (b) spatial frequencies and speckle pattern produced in the absence of the filter, (c) a zoom-in of the focus obtained at the end of the optimization process. (d) In the plot the squared of theoretical zero-order Bessel function in blue line, the focus intensity profile in dotted black line and the speckle correlation function before the optimization in green dashed line. In the inset the correlation function of the speckle pattern and the optimized focus in the presence of the ring filter.*

For this conventional configuration (with no filter adopted) the shape of the focus obtained via wavefront correction has the same shape of the correlation function of the speckle pattern from which it was initially generated [15]. In the case where a ring-shape filter is present, with $r$=1.68mm and $\delta r$=0.35mm, the final focus obtained at the end of the optimization process presents concentric rings appearing at the region around the central spot as shown in the inset "adaptive focus" of Fig. 4.11(d) and it strongly differs from the case in absence of filter presented in Fig. 4.11(c). We then compare the focus intensity to the correlation function of the speckle pattern profile (see the inset "correl. function" of Fig. 4.11(d)), the dotted black line and the dashed green line in Fig. 4.11(d) are their respective intensity profiles; again, also in this particular case, the focus profile is equivalent to the correlation function as predicted by Vellekop and co-workers [15]. Furthermore, we fit experimental data with the squared Bessel function profile (solid blue line) and observe that the three curves match consistently each other; the focus obtained from an amorphous speckle distribution is Bessel shaped.

Since Bessel beams are non-diffractive we expect a much larger Depth-of-Focus (*DOF*) with respect to Gaussian beams. In Fig. 4.12 we studied the *DOF* (measured as the distance along *z* for which the focus maintains the same lateral resolution $W_L$ while the Bessel beam intensity is higher than half of the intensity at the peak in the plane $z_0$) for several experimental configurations. The *DOF* is calculated from the *FWHM* of the focus profile along *z*, with the lateral resolution from the *FWHM* along x or y, averaged on 5 measurements.

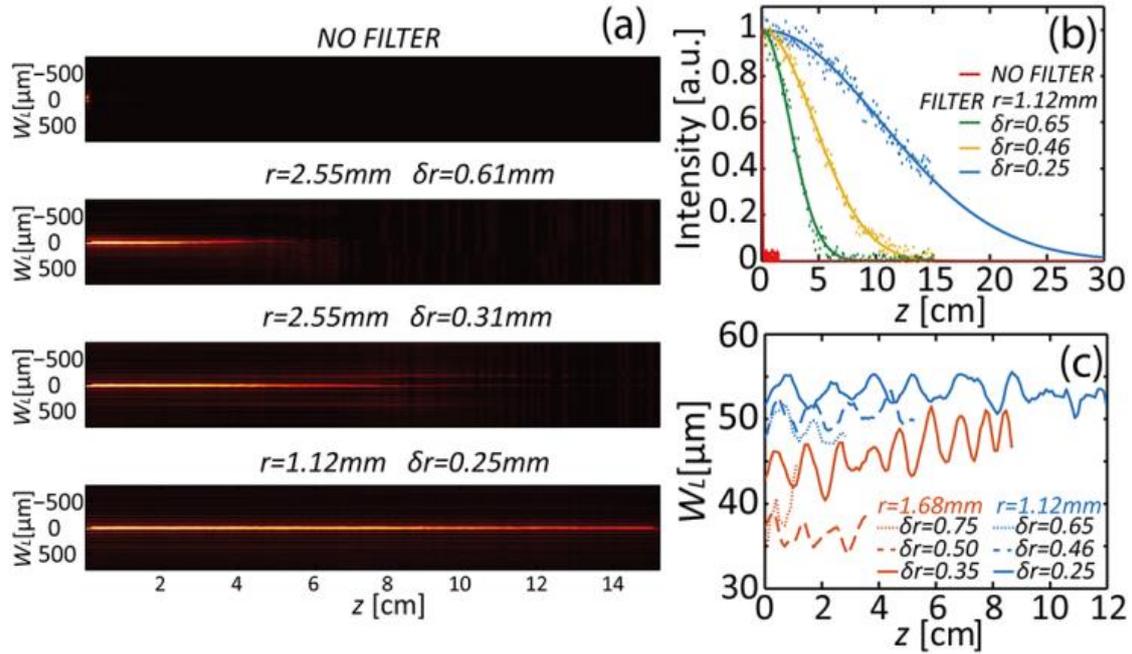

**Figure 4.12** *(a) x-z intensity profile (colormap: yellow = high intensity) of the beam transmitted after the filtering. To thinner apertures δr corresponds to larger depth of focus. (b) Intensity at the focus as a function of z position, the resolution is δz=100μm in the case without filter and δz=500μm with the filters inserted. Intensity is reported in absence of the filter (red dots), and in presence of filters (δr=0.65mm (green dots), δr=0.46mm (yellow dots) and δr=0.25mm(blue dots)). Continuous line is a fit with a Gaussian curve in order to evaluate the Full Width at Half Maximum (FWHM) of the intensity profile along z. In (c) the lateral waist ($W_L$) is monitored for different ring sizes when propagating along z.*

In the absence of the ring filter we measured a lateral resolution of $W_L$=23.25μm with a *DOF* of 0.8cm. On the other hand, in the presence of a filter with *δr*=0.25mm and *r*=1.12mm, the *DOF* is drastically increased; we obtained a lateral resolution of $W_L$=51.5μm and a *DOF*=29.4cm. In Fig. 4.12(b) each dot represents the intensity at the focus position taken at different planes *z* starting from $z_0$=0. The effective *DOF is* obtained from the *FWHM* of the Gaussian curves calculated by fitting the experimental data. We compared the values of the axial resolutions *DOF* obtained at different δ*r*: 0.65mm (in green), 0.46mm (in yellow) and 0.25mm (in blue) and with no filter (in red). The curves show that the *DOF* of the focus improves when the ring aperture decreases, the *DOF* ranges from 6.2cm at *δr*=0.65mm to 29.4cm at *δr*=0.25mm. As shown in Fig. 4.12(c) the beams with lateral waist ($W_L$) are monitored at different ring sizes when propagating along z; using very thin ring filters the beam propagates for tens of centimeter without diffracting (solid blue line, r=1.12mm) in comparison to a filter of r=1.68mm (in red); larger *r* correspond to higher lateral resolution $W_L$ and smaller *δr* correspond to longer beams propagating along z maintaining the same lateral resolution $W_L$. The curves in (c) show the beam waist oscillations along its propagation, typical finger print of truncated Bessel beams [70]. Similar results were achieved also targeting the focusing process onto different camera positions; we compared beams obtained at 5 different positions and we observed that all of them presented equivalent spatial dimensions.

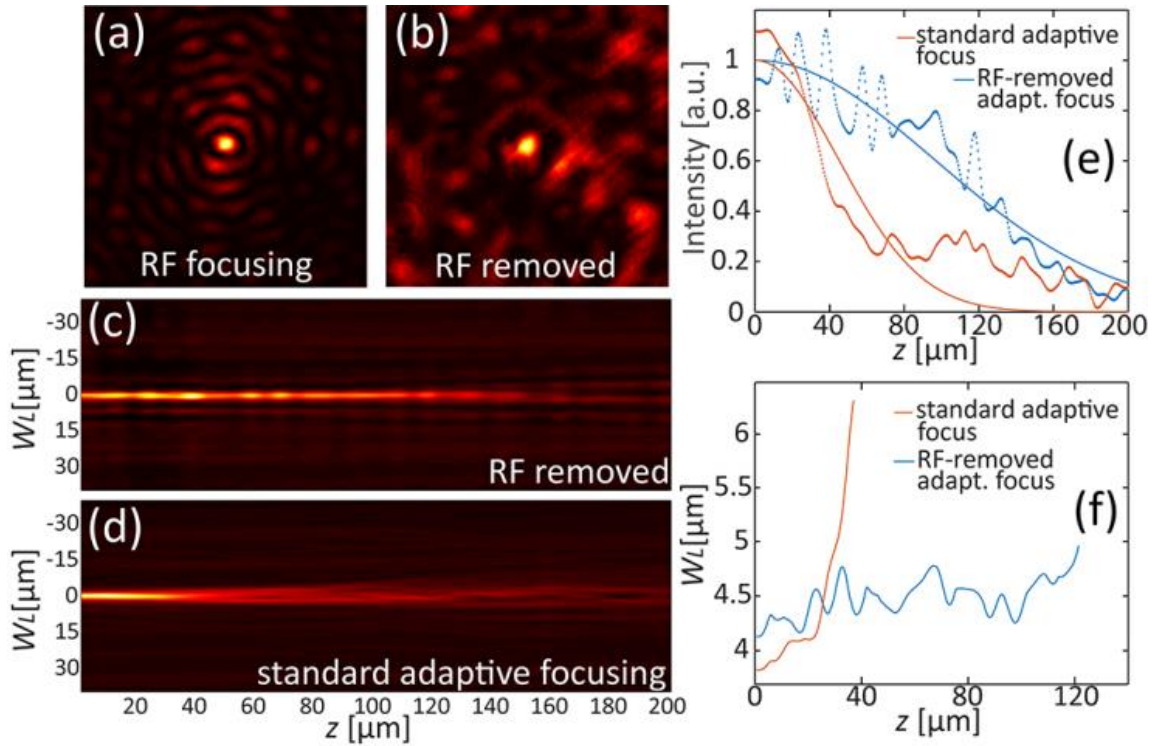

**Figure 4.13:** *The focus obtained in presence of the RF filter (a) and after removing the RF filter (b). We show x-z intensity profile (colormap: yellow = high intensity) of the beam which generates at 1.5 mm from the back of the scattering sample (the RF is used only during the optimization and then is removed) in panel (c) and the standard adaptive focus (without filter in the optimization procedure) in panel (d). In (e) their intensities along z are with red dots for the standard focusing, and in blue dots after filtering (with filter removed). Continuous line is a fit with a Gaussian curve. In (f) the lateral waist ($W_L$) is monitored for the two cases. Red curve correspond to the standard adaptive focusing and in blue the case with removed filter.*

As recently demonstrated for standard speckles [42] (see previous Section) the filter in the Fourier plane is strictly necessary only during the optimization procedure: if at the end of the iterative process the filter is removed, the L1+L2 lens system serves as a telescope and the scattering sample output is projected onto the camera plane, while the focus does persist on the camera plane (even if reduced in enhancement *h*). As shown in Fig. 4.13 the Peak-to-Background Ratio ($\eta_{PBR}$) decreases: we measure $\eta_{PBR}$=115.1 in the presence of filter as depicted in Fig. 4.13(a) (the filter with radius r=2.55mm and *δr*= 0.31mm has been used) and $\eta_{PBR}$=22.6 when the filter is removed (Fig. 4.13(b)). In Fig. 4(c), (d), (e), and (f) we compare the *DOF* of the focus after the filter removal (intensity along the direction of propagation *z* is shown in Fig. 4.13(c)) with the focus from the standard wavefront shaping (intensity along *z* is shown in Fig. 4.13(d)). In this case we have to take in account the effective magnification on the camera plane caused by the telescope (L1+L2). To scan the *DOF* we shifted the whole detection system composed of camera plus telescope device (green dashed rectangle in Fig. 4.9, translation stage with step size $\delta z$=5μm) which is a standard approach to characterize Bessel beams (24, 25).

The results are shown in Fig. 4.13(e) and Fig. 4.13(f); for the standard focusing (without filter) the lateral resolution is $W_L$=3.8μm and the *DOF* corresponding approximately to 20μm, while for the case of Fig. 4.13(b) we measured $W_L$=4.1μm and the *DOF*=121μm. This non-magnified focus is formed approximately 1.5mm after the scattering sample AOL and results in a shorter *DOF* when compared to the case in Fig. 4.12 where the focus is generated by the lens L2 which exhibits much longer focal distance.

### *4.2.4 Numerical simulation*

In this Section we propose again the numerical analysis from Section (4.1.4). In this case the filter is an annular aperture (the ring filter).

If a ring aperture is adopted as spatial filter, the wavefronts transmitted through the filter are selected in momentum around an annular distribution. In this case the mask at the center of the Fourier plane can be defined by modifying equation (4.3) as follow:

$$M(k_x, k_y) = \begin{cases} 0 \text{ if } r \leq R \\ 1 \text{ if } R < r < R + dr \\ 0 \text{ if } r > R + dr \end{cases} \quad (4.4)$$

where *dr* defines the thickness of the annular aperture in the ring spatial filter. In such a way we obtain the intensity distribution $|S'_M(x,y)|^2$ on the camera plane that is the result of superposition of the fields coming from the annular aperture only.

In a free space experiment when a focusing system is excited with a propagating annular beam, a Bessel beam is produced at its focal distance. Thus, to a modulated transfer function (MTF), consisting of a homogeneous ring, corresponds a Bessel-like point-spread function (PSF) as shown in Figure 4.14(a).

In presence of scattering, the fields trespassing the annular aperture of the ring filter are scrambled, therefore the resulting PSF is a speckle pattern (see Figure 4.14(b)). This speckle pattern is generated by a random superposition of Bessel beams, yet its correlation function, $c_I(\Delta x, \Delta y)$ reported in Figure 4.14(c), is still Bessel-like shaped; typical finger print of an amorphous speckle pattern. We also calculate its power spectrum, $\tilde{S}(k)$, as the Fourier transform of $c_I(\Delta x, \Delta y)$. As expected, the $\tilde{S}(k)$ is ring shaped with a peak in $\tilde{S}(k=0)$ as can be observed in Figure 4.14(d). In practice, the numerical model faithfully reflected the experimental results from Figures 4.10 and 4.11.

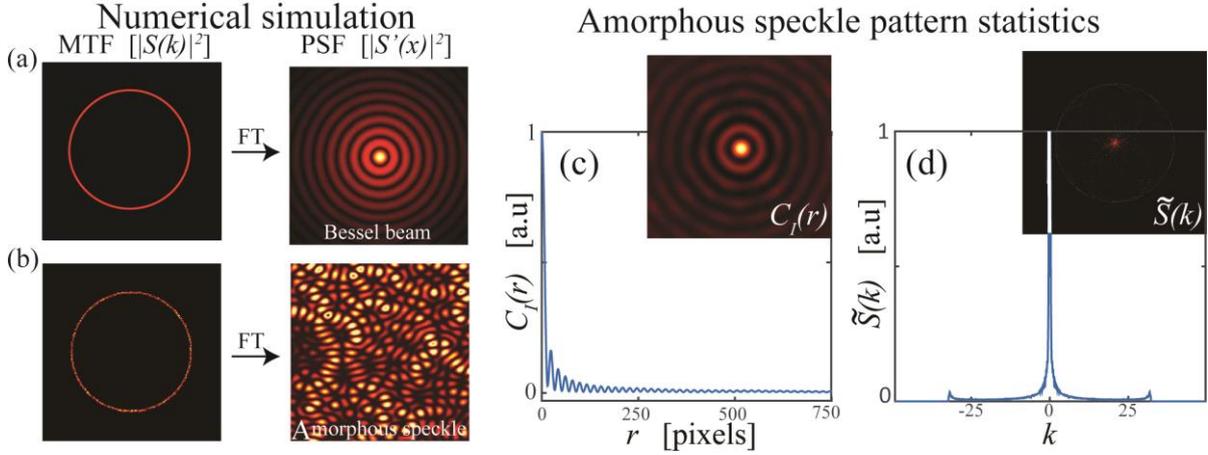

**Figure 4.14:** *Results from the numerical simulation. In (a) a homogeneous annular MTF generates a Bessel-like PSF. In (b), due to scattering, the intensity residing at the annular aperture is scrambled, therefore the corresponding PSF is an amorphous speckle pattern. In (c) and (d) we report respectively the correlation function and the power spectrum calculated from the amorphous speckle pattern in (b).*

Starting from this configuration, when a wavefront shaping algorithm is applied, the selected fields can be properly de-phased in order to obtain constructive interference at the target position. The produced focus in this case results Bessel structured due to the filtering imposed.

### 4.2.5  Discussion

In conclusion we demonstrated that amorphous speckle patterns can be manipulated by wavefront shaping, in order to enhance their intensity and produce foci at user defined positions. By exploiting adaptive wavefront shaping, from amorphous speckle patterns generated with a Ring Filter, we obtain a non-diffractive Bessel-like beam with a shape equal to that of the correlation function of the initial pattern. The mean distance between speckles is set by the radius of the ring in momentum space, whereas the symmetry along the propagation, indeed the property to be non-diffractive, is determined by the width of the ring filter aperture. In addition, we demonstrate a technique that allows for the generation of non-diffracting foci through a scattering wall. The focusing system described is an Axicon Opaque Lens (AOL), which enables the direct control of the pattern and the tuning of the non-diffractive beam's spatial position. The Bessel-like beam is obtained by introducing the RF during the optimization procedure, while it may be taken out at the end of it: in such a case the focus persists and is still non-diffractive. Our method for the generation of Bessel-like beams has potential applications for replacing mechanical scanning with a stable and fast phase control architecture, as well as for increasing the depth of field and penetration depth of techniques such as blind-SIM, SCORE and super opaque lenses.

# 5  Customized photonic glasses

We have seen that light scattering carries hidden degrees of freedom that can be explored for shaping the light in extraordinary structures. In a similar way to mechanical statistics saying that scattering allows numerous optical states, those states become accessible to the system when a field modulator is in use. We can pick the state of interest from the ensemble applying specific constraints to the system. Thus, by sorting spatial frequencies with filters and engineering the wavefronts, the scatterer converts to a stand-alone device able to replicate (eventually to improve) any conventional optical element. Those opaque systems are highly configurable, so they have immediately attracted the interest from many fields, finding them suitable for applications in integrated on-chip systems. The demand for compact and durable scattering systems has inspired the works presented in this Chapter. Here we describe the process to permanently print scattering structures and to explore complex geometries that give access to unconventional speckle patterns.

## 5.1 Robust authentication through stochastic femtosecond laser filament induced scattering surfaces

In this first Section we describe a reliable authentication method by femtosecond laser filament induced scattering surfaces. The stochastic nonlinear laser fabrication nature results in unique authentication robust properties [81]. The work presented here provides a simple and viable solution for practical applications in product authentication, while also opens the way for incorporating such elements in transparent media and coupling those in integrated optical circuits.

### 5.1.1  Introduction

Product authentication and anti-counterfeiting is an urgent task nowadays, due to the ever-increasing global trades. Secure authentications rely on secret keys that are unique and unclonable. Traditional encoding strategies utilize a deterministic and thus clonable coding process, such as holograms, digital bar-codes, and radio-frequency identification (RFID) codes, where the secret keys can be easily duplicated. Recently, a new robust encoding strategy called Physical Unclonable Function (PUF) has been proposed [82]. In PUFs, the secret keys cannot be precisely duplicated due to the uncontrollable fabrication errors. The PUF-based encoding strategy has been intensively investigated in the last decade, with many embodiments such as the silicon circuit wire-delay PUF, FPGA-butterfly PUF, and optical scattering PUF [83, 84, 85]. Optical PUFs are based on light diffusion from strongly scattering media. When a coherent light wave transmits through or reflects by a scattering medium, it is decomposed onto a large number of waves with various amplitudes and phases. These waves will interfere and form a complex speckle pattern. The speckle pattern is very sensitive even to minor changes in the scattering medium and can thus serve as a fingerprint for the authentication and communication purposes [86, 87]. Optical scattering is easily obtained from many materials like white papers, semiconductor powders, and opaque curtains, and thus optical PUFs based on such materials could provide an easy and low cost solution compared to the electronic PUFs, which are based on the complex integrated circuits (IC). Materials used for optical PUFs should be stable in time and the generated speckle patterns should be repeatable, which limits the candidate PUF materials to be mainly in the solid phase. Examples include opal diffusing glasses, white paints, and plastic cards, as demonstrated in the recent experiments [21, 88]. Furthermore, since the generated speckle pattern sensitively depends on the light illumination conditions, accurate positioning of the scattering medium is of paramount importance. Finally, until now, the demonstrated optical PUFs lack the effective controls over the topologies of the scattering media, which should be flexibly tailored according to specific engineering requirements. In this work, we demonstrate that by femtosecond laser filament micro-ablation optical scattering, PUFs can be fabricated on glass plate surfaces. The robustness of our PUF authentication is also enhanced through a simple optical positioning method. In addition, the versatility of femtosecond laser micromachining allows the integration of such PUFs in more complex optical circuits, while the technique can be extended in materials other than glasses allowing for a wide spectrum of applications. Previously reported optical scattering media were mainly fabricated by nano-particle deposition on glass plates, either through chemical vapor deposition (CVD) or through spray painting [89, 8].

### 5.1.2 *Printed photonic glasses and their reproducibility*

The optical scattering media used in this work are fabricated by direct laser writing on the surface of the glass plate. In our experiments, a series of linearly polarized laser pulses (~513 nm, ~190 fs, 60 kHz) provided by the second harmonic of an Yb:KGW regenerative laser system are used to micro-ablate the surfaces of glass plates (soda lime glasses with dimensions $75 \times 25 \times 1 mm^3$ ). The laser pulses are focused by a microscope objective lens ($NA = 0.40$, 20×) onto the glass surface, which is mounted on a high-resolution three dimensional air-bearing stage. The incident laser pulse energies are adjusted by a variable neutral density filter and measured before the objective lens. The femtosecond laser ablation of the soda-lime glass is first characterized at different laser energies and exposure times.

The laser exposure times are controlled by a mechanical shutter with a minimum opening time of about 20ms, corresponding to a least exposure of 1200 pulses (for the 60kHz repetition rate of our laser). Fig. 5.1(c) shows the microscope top-view images of the laser ablated holes on the glass surface; the corresponding laser energies are increased from 100nJ to 7μJ along the horizontal direction while the corresponding exposure is increased from 1200 pulses to 60000 pulses along the vertical direction.

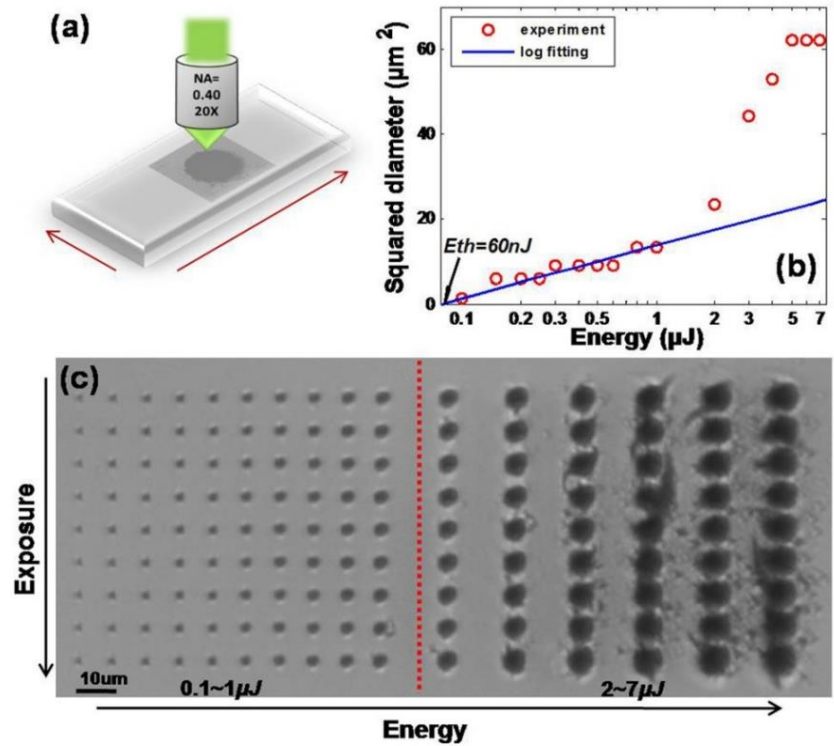

**Figure 5.1:** *(a) Scheme of the femtosecond laser ablation. (b) Squared diameter of the laser ablated holes as a function of the incident laser pulse energy. The red circles are the experimental measurements and the solid blue line is the logarithm fitting curve. (c) Microscope top-view image of the laser ablated holes on the soda-lime glass. The incident laser pulse energies are increasing along the horizontal direction and the corresponding exposures are increasing along the vertical direction.*

It can be clearly seen from Fig. 5.1(c) that the profiles of the laser ablated holes are very smooth at incident laser energies lower than 1μJ, while the hole-profiles become more irregular with the appearance of large cracks around the holes at higher laser energies. Femtosecond laser pulses have been shown to ablate glasses with very high precision (down to a few nanometers) at near-threshold energies [90], arising from the deterministic material breakdown ignited by nonlinear photoionization or photoinduced Zener tunneling ionization of valence band electrons. The smooth hole-profiles observed in the current experiment surely reflect the deterministic character of femtosecond laser ablation. However, as the laser power increases well above the critical value for self-focusing [91], nonlinear propagation effects come into play leading to filamentary propagation [92]. Filamentation itself will be significantly altered from one laser shot to another because the propagation medium is being changed between the successive laser shots. Beyond nonlinear propagation effects, at high laser powers, thermal diffusion processes become important and the material around the laser interaction zone is severely affected with the formation of vacancies and defect states [93]. These heat-affected-zones coupled with the complex nonlinear propagation dynamics will strongly impact the laser material ablation processes and result in stochastically distorted hole-profiles. Finally, femtosecond laser ablation is known to induce very complex surface textures with feature sizes spanning from a few nanometers to a few micrometers [94].

Actually, plasma plumes excited during the ablation contain a variety of nanoparticles [95], which are deposited as clusters in the surrounding area or directly adsorbed in the ablated holes. Such nanoparticle clusters create extra scattering centers, further complexing the femtosecond laser ablated structures and considerably increasing the degree of difficulty in reproducing them.

Returning to the analysis of Fig. 5.1, it has been demonstrated that the femtosecond laser ablation threshold can be approximately obtained by linear fitting the squared diameters of the ablated holes versus the logarithm of the corresponding laser energies [96]. Following this method, the laser ablation threshold energy for the soda lime glass is determined to be 60nJ and the corresponding average laser fluence is $0.7 J/cm^2$. Notice that only hole-diameters with laser energies lower than 1μJ are used for the fitting curve in order to be below the strongly nonlinear regime. One can easily see that above this limit, at higher energies, the diameter of the holes becomes much larger corroborating the huge contributions from thermal diffusion processes and nonlinear propagation effects as discussed above. One important merit of the PUFs is the manufacturer exclusion, i.e., even the manufacturer cannot reproduce it [82]. This indicates that the optical scattering PUF in our case should be fabricated at high input energies where the stochastic thermal and nonlinear processes cannot be reproduced. Meanwhile, the larger hole diameters at higher energies will also reduce the fabrication time. Based on the above considerations, the optical scattering PUFs are fabricated by ablating soda-lime glass surfaces with high energy laser pulses (7μJ) at randomly selected positions. The ablation positions are generated by random sequential addition of points within a round boundary of 100μm diameter (Fig. 5.2(a)), with restricted minimum distances between adjacent points larger than 4μm. The dimension of the designed ablated surface (the optical PUF) is chosen according to the parameters of the home-built optical positioning system, and it can be flexibly tuned based on specific engineering requirements. When the total number of ablated holes exceeds ∼120, the fabricated scattering surfaces present strong suppression of the ballistic-light transport, while the transmitted diffusive light forms a complex speckle pattern, as the one shown in Fig. 5.2(b). The remaining ballistic light intensity after the scattering surface is measured to be $0.012 \pm 0.004$ times the incident light intensity, and the thickness of the scattering surface is ∼20μm.

Thus, the scattering mean free path is estimated to be $4.5 \pm 0.4$μm, which means that the incident light will encounter 4–5 scattering events on average before leaving the scattering surface. Since the diameter of a single ablated hole, at 7μJ laser pulse energy, is about 7.5μm, for our chosen hole-density, the holes will randomly overlap, complexing further the structure. This can be clearly seen in Fig. 5.2(a), where a reflection microscope image of the ablated surface shows no clear perimeters of single ablated holes. Moreover, four cross-shaped structures are also fabricated around each ablated surface at near-threshold laser energies (∼100nJ). The width and the length of each line is 1μm and 30μm, respectively.

These four crosses are used to guarantee the accurate placement of the optical PUF at the same position after successive reintroduction of it in the imaging setup.

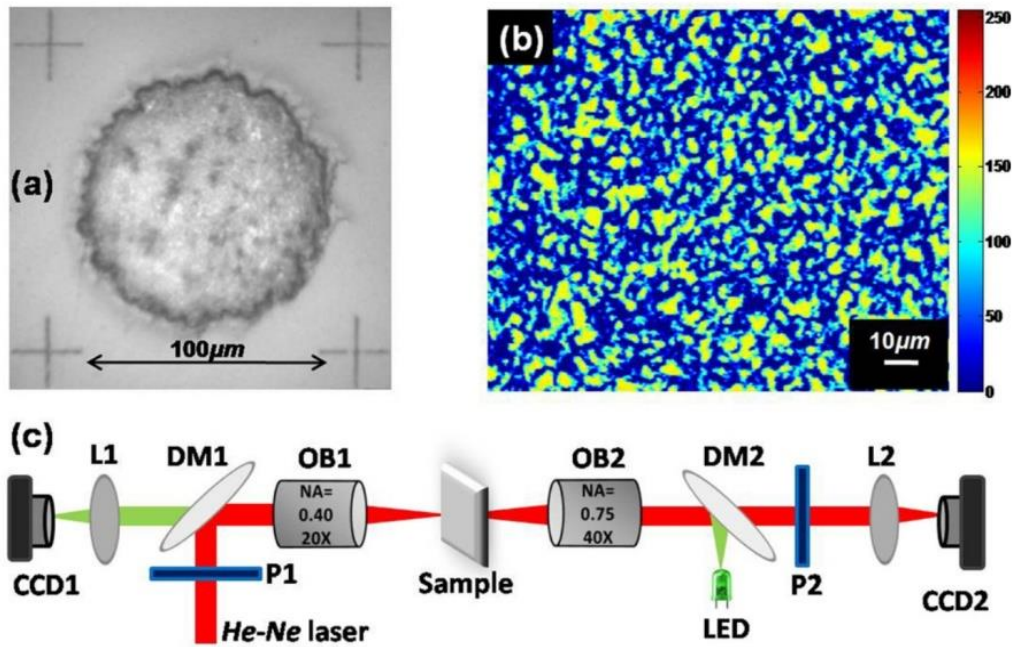

**Figure 5.2:** *(a) Microscope top-view image of the scattering PUF. (b) The measured speckle pattern presented on a color scale. (c) The experimental setup of the optical imaging and positioning system: fused silica lenses with 10 cm focal length (L1 and L2), dichroic mirrors (DM1 and DM2), Glan polarizers (P1 and P2), microscope objectives (OB1 and OB2), and CCD cameras (CCD1 and CCD2).*

A simple optical imaging and positioning system is built to characterize the optical scattering PUFs and also to authenticate them. The experimental scheme is depicted in Fig. 5.2(c). Objective1 is used to focus the coherent continuous-wave 632.8nm wavelength He-Ne laser beam on the optical PUF, while Objective2 collects and images the transmitted laser speckle patterns onto the camera CCD2. The speckle pattern is captured in a cross-polarization mode through the two polarizers P1 and P2 with orthogonal polarizations, in order to remove any remaining ballistic photons. Meanwhile, Objective1 also images the optical PUF, which is illuminated by a green LED light source, onto the camera CCD1. The optical images of the four crosses (Fig. 5.2(a)) set the reference points for aligning the PUFs, which are mounted on a high resolution three dimensional manual stage. It should be stressed that these reference points are very important, since the optical PUFs are unmounted and reinserted into the system many times during routine operations. The incident laser spot on the PUF has a diameter of 10μm, and the focal plane of Objective2 is ∼300μm away from the PUF. The positioning precision is mainly determined by the optical resolution of Objective1, which is ∼500nm in the lateral direction and ∼6μm in the axial direction. The speckle pattern sensitivity against the PUF displacement is investigated first. A series of speckle patterns are measured for different positions of the PUF away from the original position along the lateral direction. The speckle pattern cross-correlations, between the displaced speckle patterns and the original speckle pattern, are displayed in Fig. 5.3(a). It can be easily seen that the laser speckle pattern will first follow the movement of the PUF, until a displacement of 3μm, where the corresponding speckle pattern is completely uncorrelated with the original one. The peak values of the speckle cross-correlations for two speckle patterns can be regarded as the genuine correlation coefficients between them. Accordingly, the speckle correlation coefficients as a function of the PUF movements are drawn in Fig. 5.3(b). It can be deduced from the curve that the repeatability of the speckle patterns is retained only if the lateral positioning fluctuations of the optical PUF are less than 2μm, which is well within the resolution of the optical positioning system.

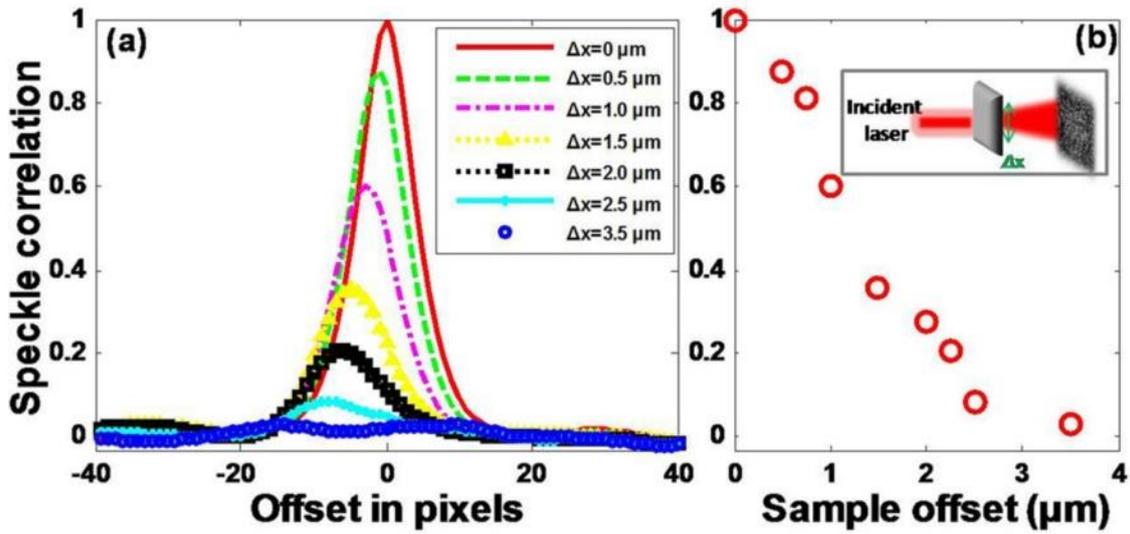

**Figure 5.3:** *The speckle cross-correlation traces between the original speckle pattern and the displaced speckle patterns. (b) The speckle correlations as a function of sample displacement. Inset shows the sample movement direction.*

However, it should be noted that our positioning system is incapable of resolving the slight tilts of the PUF, due to the low axial resolution. Fortunately, there is a memory effect (ME) in the optical scattering process [97], which dictates the underlying angular tolerances of the speckle patterns with respect to the PUF rotation. For a strongly multiple scattering medium, the memory effect range is inversely proportional to the medium's thickness: the thinner the scattering medium, the broader the memory effect range. Since the current PUF is an optically thin scattering layer as discussed previously, the corresponding effective optical thickness is even smaller than its physical thickness [98]. Thus, a broad memory effect range is expected, which can mitigate the angular positioning requirement, something that can also prove very useful for practical applications. In the following, we demonstrate the robustness of the technique for authentication purposes. Previous works on optical scattering PUF authentication were concentrated on extracting the characteristic bitwise codes from the speckle patterns and then comparing these codes instead of comparing directly the speckle patterns for the authentication purpose. This code-extraction process was applied to reduce the displacement noise within the speckle patterns, through complex algorithms such as Gabor-transforms and digital whitening [85, 87]. The current optical positioning system can largely reduce the positioning errors, and thus direct comparisons between speckle patterns are feasible, which avoids the need of time-consuming code-extraction processes and is much easier to implement. For our studies, 7 pairs of optical PUFs were fabricated with each pair being produced under exactly the same experimental conditions. Then all 14 PUFs were successively inserted into the positioning system, and the corresponding speckle patterns were registered.

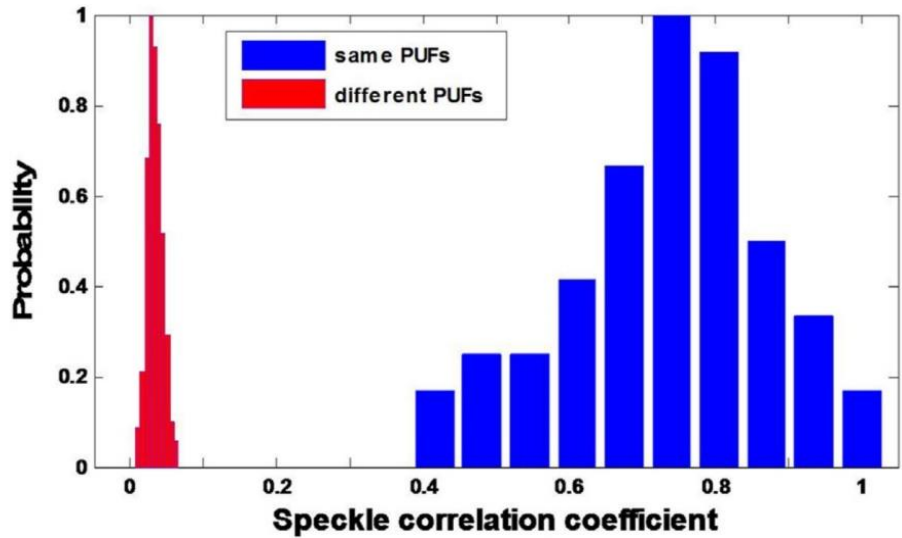

**Figure 5.4:** *The probability distributions of the speckle correlation coefficients derived from a set of 70 speckle patterns pertaining to 14 different optical PUFs. The left red columns represent the probability distributions for the speckle pattern pairs obtained from different PUFs, while the right blue columns represent the probability distributions for the speckle pattern pairs obtained from the same PUFs.*

Subsequently, each PUF was unmounted, reinserted, and adjusted at the same reference position four times, and each corresponding speckle pattern was recaptured for the repeatability tests. The speckle pattern correlations are calculated for every pair of the obtained speckle patterns ($14 \times 5 = 70$ frames in total), and the statistical probability distribution of the speckle pattern correlations are shown in Fig. 5.4. The blue columns represent the probability distribution derived from the speckle pattern pairs pertaining to the same origin (captured from the repeated insertions of the same PUF), while the red columns represent the probability distribution from the speckle pattern pairs of different origins (captured from the repeated insertions of different PUFs, including the PUFs fabricated at the same conditions). The remarkable discrepancy between the speckle correlations from the two sets and the well-defined separation between the two distributions clearly demonstrates the uniqueness of each optical PUF and also the good repeatability of the optical positioning system. These results are even more interesting as they imply that the PUFs produced under exactly the same conditions exhibit completely different optical scattering and thus uncorrelated speckle patterns; otherwise, the two distributions in Fig. 5.4 should have had even a small overlap. Therefore, the speckle pattern from each PUF serves as a unique fingerprint. We would like to note that recently tunable speckle patterns from a single PUF were demonstrated by simply shaping the wavefront of the illuminating laser beam using a spatial light modulator (SLM) [87, 21]. In this way, many challenge-response (phase mask-speckle pattern) pairs, instead of one single speckle pattern equivalent to the response of the plain mask demonstrated above, can be registered for one optical PUF. This method can also be applied with the current optical PUF to build a large database of challenge-response pairs, which will greatly increase the security levels.

### *5.1.3 Discussion*

In conclusion, the optical scattering media realized by femtosecond laser ablated glass surfaces in this work, compared to the popularly employed volumetric scattering media, are also very hard to be duplicated, although the light scattering is seemingly confined at the top interface between air and glass. We have demonstrated that the stochastic nature of the highly nonlinear ablation processes prohibits the reproduction of the same scattering medium even if the exact same experimental conditions are used. Last but not the least, due to the versatile femtosecond laser material interactions, the current method can be easily applied to other materials, such as semiconductors, polymers, and plastics. Besides, the monolithic integration of diverse optical components enabled by the femtosecond laser direct writing technique, including waveguides [99], optofluidic channels [100], and Fresnel lenses [101], can be combined with the current method to provide much more compact and robust authentication systems that can promote future products of anti-counterfeiting technologies.

## *5.2 Opaque cylindrical lenses*

Strong scattering in the optical paths can be proactively exploited for determining light propagation and focusing through turbid media and ultimately improve optical imaging and light manipulation capabilities. The use of light shapers together with strongly scattering structures enable the production of foci confined in the nanoscale or on the other side may provide significantly enhanced fields of view. Exploiting this concept, we have introduced *ad-hoc* engineered scattering structures, by direct femtosecond laser writing in the bulk of glasses, presenting isotropy in one dimension and together with precisely phase-shaped illuminating light we produce sub-micron thin light-sheets at user defined positions [102]. Our approach permits to focus light of different wavelengths onto the same defined position without moving any optical element and correcting for chromatic aberrations. Furthermore, we demonstrate that natural biological tissues presenting isotropy in one dimension can also be used for generating light sheets directly in the biological sample.

### *5.2.1 Introduction*

It is with immense interest that the scientific community follows the technological advancements of optical and photonic imaging and the creation of new and disruptive knowledge. It is in fact very recent that a Nobel Prize was awarded for overcoming the diffraction limit set by Abbe [38] and the creation of a new field [52]. This has been achieved by the advent of innovative approaches exploiting prior knowledge from other fields of applied optics to overcome the fundamental limits set by the very nature of light-matter interactions [25]. The main drawback of optical modalities originates from the multiple scattering that light suffers when propagating through complex media such as biological tissues. This strong scattering leads to impaired resolution, blurring our vision [2]. Novel concepts of adaptive optics can be exploited to reverse image distortion or retrieve otherwise hidden features [7]. In this context, wavefront shaping [9] is one of the most efficient methodologies in compensating for scattering. By controlling the phase of the wavefront impinging on a scattering medium one can control the energy density of light at the output. However, the breakthrough that paved the way for wavefront shaping concerned optical focusing using a combination of a Spatial Light Modulator (SLM) and a scattering system, which is referred in the literature as opaque lens (OL) [15, 22]. OLs can generate single and multiple foci at user defined positions by exploiting the speckle correlations, rendering in practice the focal position reconfigurable [15] if a feedback (guide-star) is provided [34]. In addition, OLs exhibit a larger effective numerical aperture than conventional lenses, achieving, thus, resolutions comparable to those of super-resolution techniques [16] and extended Fields of View (FOV) [103]. In practice, the focus has the same size as the speckle pattern correlation function at the beginning of the focusing process [15, 42]. However, wavefront shaping covers a significantly wider range of applications or fields apart from biomedical imaging [11]; manipulating the light transmitted through a complex medium has been demonstrated to allow communication control [62] or to improve cryptography [21]. In this work, we have exploited wavefront shaping to produce and manipulate improved illumination patterns in use for Light Sheet Microcopy (LSM) [104] or Selective Plane Illumination Microscopy (SPIM) [105] or Light Sheet Tomography (LST) [106]. These techniques have already found their way into biology labs in increasing numbers due to their advantages compared to traditional methods such as confocal microscopy [107]. The light sheets are used to excite fluorescence (or generate elastic scattering in LST) from a section of a large biological sample hence performing real time optical sectioning and 3D imaging with low phototoxicity [105]. The ideal scenario in LSM, as in optical microscopy in general, consists in the ability to generate a sub-diffractive focus at user controlled position with long axial resolution at different wavelengths. This key challenge has attracted the interest from many fields and different approaches have been recently demonstrated [108, 109, 110, 111]. In this sense, implementing metasurfaces [112], i.e. engineered phase-change materials one can correct for chromatic aberration [113] and generate reconfigurable lenses [114]. Stimulated by the wide range of applications and advantages derived from flat-lenses [114] we demonstrate a method for generating light-sheet illumination through *ad-hoc* laser engineered scattering systems [81]. Our intent is to exploit the enhanced effective numerical aperture of OLs [15] for generating super thin light sheets. Our system consisting of a SLM, and the scattering structure will henceforth be referred to as an Opaque Cylindrical Lens (OCL). We present a platform that enables the production of thin multi-color light sheets at user controlled positions without the need for mechanical scanning, exhibiting axial resolution comparable to the state of the art and with the potential to go even beyond that.

### *5.2.2 Description of the system*

When a coherent light beam travels through opaque structures experiences multiple scattering and at the output of the structure the light is fully scrambled with components presenting arbitrary k-vectors and phases. The superposition of these components generates a speckle-like interference pattern which consists of a set of intensity maxima and minima distributed in a complex pattern. A speckle pattern has been demonstrated to be complex rather than random, which means that it conserves memory of the scattering structure [115] from which it was generated.

In our case the scattering media are composed of randomly distributed rods piled along one direction in such a way to result parallel to each other as represented in Figure 5.5(a), we refer to such an optical element as *Anisotropic Photonic Glass* (APG). This structure presents symmetry in one dimension that can be reflected on the generated speckle pattern. The samples are fabricated by direct laser writing, a schematic of the fabrication process is illustrated in Fig. 5.5(a), a top-view optical transmission image of the APG is shown in Fig. 5.5(b) and the final structure imaged with a 15x microscope objective is depicted in Fig. 5.5(c).

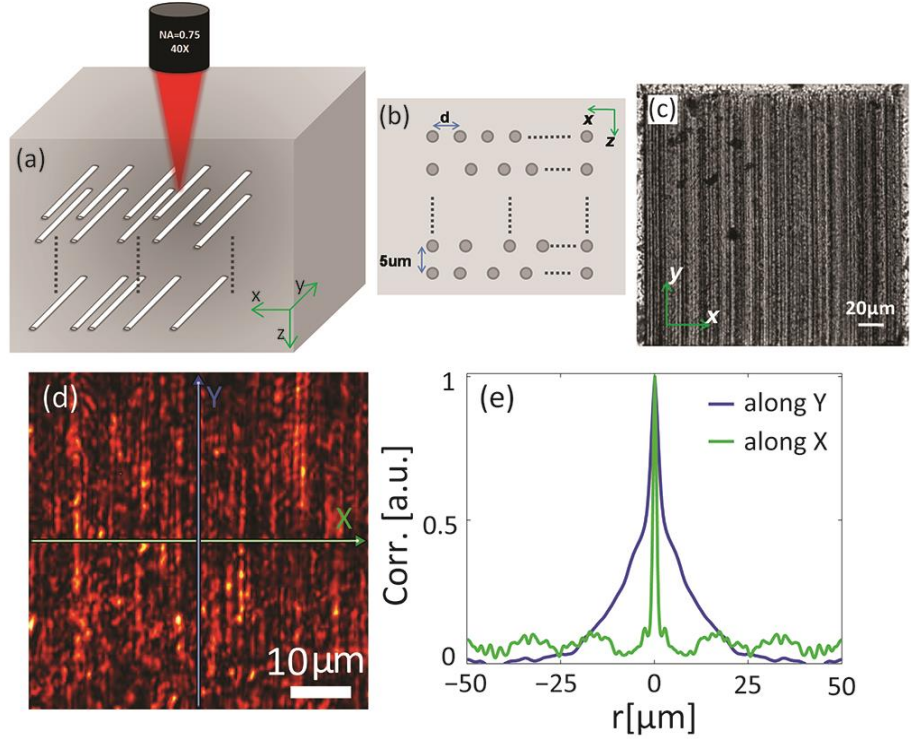

**Figure 5.5:** *The APG is fabricated in the bulk of the glass by direct laser writing as illustrated in panel (a). The distribution of the lattice structure on the x-z plane is depicted in (b). An optical transmission micrograph of a fabricated APG is shown in panel (c). In (d) the speckle pattern at the back of the photonic lattice: the isotropy along the y axis in the APG results in an elongated speckle pattern. In (e) the correlation function Corr(r) of the pattern along the direction x (green solid line) and the direction y (blue solid line).*

The interference pattern generated onto the camera plane placed at the back of the structures (Fig. 5.5(d)) is composed of elongated speckle grains oriented with the same direction as the rods in the structures. This observation was confirmed by estimating the typical grain shape/size of the speckle pattern that is given by the correlation function *Corr(r)* of the intensity distribution at the camera plane, this was calculated as the inverse Fourier transform of the speckle pattern (*I*) energy-spectrum:

$$\text{Corr}(r) = FT^{-1}\{|FT\{I - \mu_c\}|^2\} \quad (5.2.55)$$

where $\mu_c$ is the intensity mean value on the camera plane.

Fig 5.5(e) shows a comparison between the "typical" speckle grain dimensions along the *x* and *y* directions of the camera retrieved from the normalized *Corr(r)* profile along the 2 directions. The ratio between the Full Width at Half Maximum (FWHM) of the profile along *y* to those along *x* gives the *factor of speckle elongation* (FOSE), which in our case is: FOSE ≈ 3 (0.9μm along the *x* axis and 2.85μm along the *y* axis). This confirms that the typical speckle grain is indeed elongated along the *y* axis.

The scattering photonic lattice sample is fabricated by direct femtosecond laser writing in the bulk of a soda-lime glass plate [81] (a microscope slide with dimensions 76x26x1$mm^3$). The high power femtosecond laser pulses provided by a Ti:Sapphire laser system, with a central wavelength of 800nm, a Fourier-transform-limited duration of 35fs, and a repetition rate of 1kHz, are tightly focused in the bulk of the glass through a 40x microscope objective (0.75 numerical aperture). The incident pulse energy is set to be 8$\mu J$ (measured before the objective), which is high enough to induce micro-explosion and void-formation deep inside the soda-lime glass. Then the glass is scanned across the laser focus by motorized stages, in order to create the 3D structure. The lattice structure starts from 100$\mu m$ and extends to 600$\mu m$ underneath the glass surface, consisting of 100 layers of void-lines and each layer containing 70 void-lines located in random x-positions (see Fig. 5.5(a)).

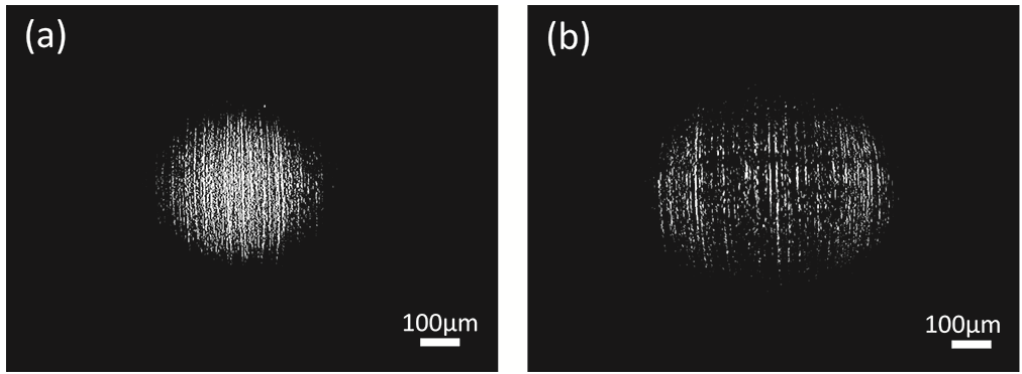

**Figure** 5.6: *Scattered light recorded 0.5mm (a) and 1mm (b) from the back surface of the APG. The speckle patterns are collected with a 10x imaging system. The symmetry in the APG structure is conserved in the speckle patterns which present elongated speckle grains.*

A typical void line has a width of 2$\mu m$ and a length of 200$\mu m$, which is inscribed by moving the glass straight along the y-direction at a scan speed of 700$\mu m/s$. The adjacent layers are 5$\mu m$ apart, and the minimum distance between adjacent void-lines in the same layer is 2$\mu m$ (see Fig. 5.5(b)). In essence, the structure consists of an ensemble of parallel rods randomly distributed in the volume of the glass, which forms a scattering medium with a refractive index mismatch of $\Delta n$=0.45 between the glass matrix and the void rods. We measure that the intensity of the light trespassing the APG ($I_{OUT}$) is 9% of the amount at the input ($I_{IN}$). In Fig. 5.6 we show the speckle pattern recorded 0.5mm (Fig. 5.6.(a)) and 1mm (Fig. 5.6.(b)) from the back surface of the APG, the images are taken with a 10x imaging system.

In such a configuration a phase-only SLM can shape the wavefront and modulate light traveling through the highly scattering APG. Indeed, using a scattering optical element one can improve the effective numerical aperture of the focusing system with respect to free space propagation [15] and herein we exploit this concept.

In Fig. 5.7 we show a detailed scheme of our setup. Coherent laser sources emitting at 594nm, 532nm and 488nm are used while a homemade telescope (lens L1 with 25.4mm focal length + lens L2 with 250mm focal length) magnifies the laser beams by 10X.

We use a quadratic slit filter (F) that produces a square flat-top beam, in such a way we set the beam's waist at 6mm. Modulation is performed by a phase only Spatial Light Modulator (SLM) (Holoeye, Pluto) that shapes the wavefront of the beams; a 50:50 beam splitter (BS) guarantees that the beams and the SLM are perpendicular to each other. Hence, 50% of the light reflected by the SLM is directed along a perpendicular axis where a cylindrical or spherical lens with 150mm focal length (L3) focuses the beams at its focal distance.

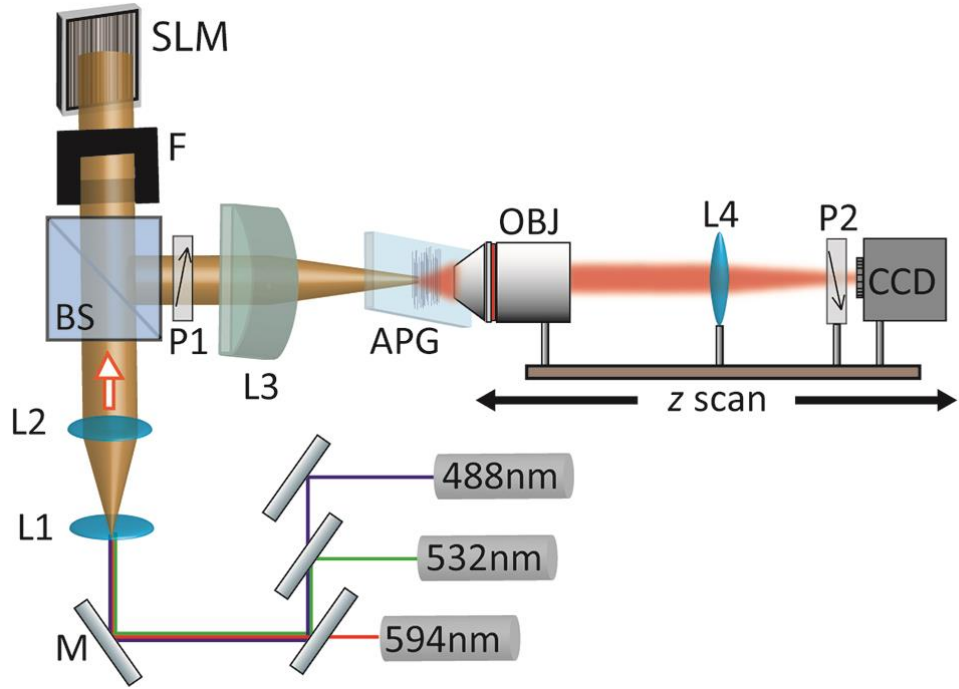

**Figure 5.7:** *Schematic of the experimental setup: three laser sources at 488nm, 532nm and 594nm are aligned and their wavefronts are shaped by a phase only spatial light modulator. All the beams have the same waist (6mm) and they are focused onto the APG. The transmitted light is collected with a 40x imaging system composed of an objective lens (OBJ) and a tube lens (L4). Polarizers P1 and P2 are perpendicularly oriented to each other in order to suppress any residual ballistic light passing through the APG.*

The beam impinges onto the scattering sample (S) and the output is collected by a 40x infinity corrected microscope objective (OBJ) with Numerical Aperture 0.65 and the tube lens with 200mm focal length (L4) which projects the sample output onto the camera plane. We used two polarizers with perpendicular orientation (the first in front of the lens L2 and the second in front of the CCD camera) to suppress any residual ballistic light passing through the APG. The OBJ, L4, P2 and the CCD are mounted on a stage that can move along *z* (the direction of propagation of the light), we use that to follow the depth of field of the focus generated behind the APG.

### 5.2.3 *Focusing through tailored photonic glasses*
A more schematic representation of our focusing system is depicted in Fig. 5.8, where a cylindrical lens ($f_1$=150mm) generates a light sheet at its focus, while a long working-distance microscope objective and a tube-lens produce a magnified (40x) image on the CCD camera plane. We consider the FWHM of the focus intensity profile as the focus width *w* which determines the nominal focusing resolution. In free space we obtain a light-sheet with a lateral resolution of $w_{0X}$=6.7μm (see Fig. 5.8(a)). For the configuration with the scattering medium, see Fig. 5.8(b), we simply add the APG at 148.9mm ($f_{2A}$) from the cylindrical lens. Using an iterative algorithm, we are able to find the phase mask that allows generating a light sheet at 0.1mm ($f_{2B}$) form the back of the structure. In this case the light sheet has a lateral resolution of $w_X$=0.9μm, a value which is the result of an average over 10 foci obtained at different positions on the camera plane (corresponding to different target positions). Thus, we have achieved an experimental effective diffraction limit for the light sheet with a consequent enhancing factor [15] of $w_X/w_{0X} = 0.13$.

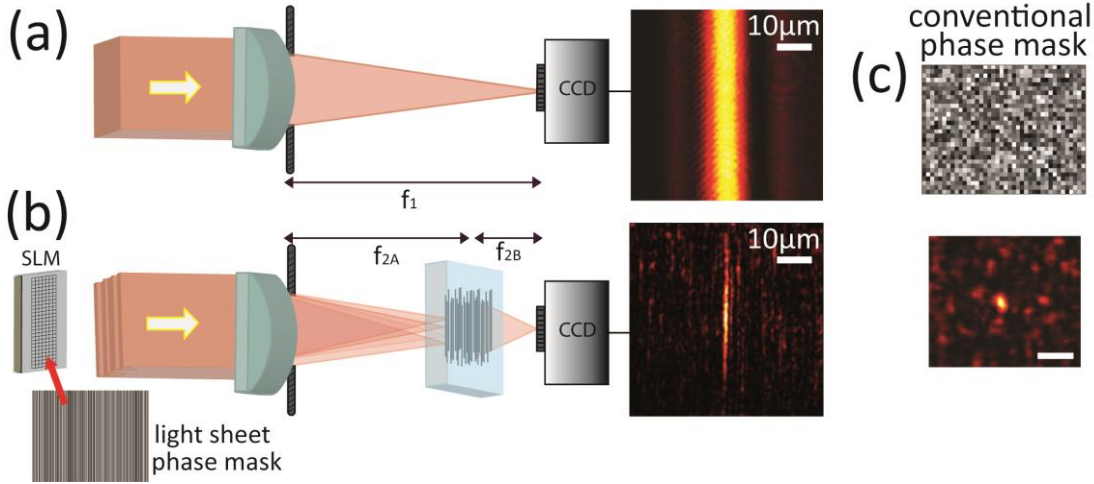

**Figure 5.8:** *A cylindrical lens generates a light sheet at its focus $f_1$=150mm. On panel (a) we show the free space configuration. On (b) we introduce the APG at 148.9 mm from the lens. We shape the beam by addressing to the SLM phase masks composed of parallel stripes to conserve the system isotropy. If segmented masks as in panel (c) are used for the optimization process the isotropy is lost and a conventional round focus (lower side of panel (c)) is generated.*

The focusing process is achieved by exploiting an iterative Monte-Carlo algorithm similar to those presented in previous works [37]. In general, a mask composed of 200x200 segments is addressed to the SLM window and we randomly assign to each segment of the first line a gray tone on a scale of 255 levels corresponding to a fixed de-phasing of the light reflected by the same segment within a range from 0 up to 4.5π. The same sequence is repeated for all the lines underlying the first.

In this particular case we address to the SLM a mask composed of parallel stripes/columns, and at each band corresponds a de-phasing that depends on the gray tone set on it. It follows that the beam impinging on the cylindrical lens is modulated along one dimension only. The result is a mask composed of $N$=200 columns with different gray tones as shown in Fig. 5.8(b). A preliminary optimization step tests a series of 50 random masks picking the one providing the best intensity value at the target position. The mask is used as an input for the second optimization routine, which tests a phase shift from 0 to 2π for a single column accepting the one providing the highest enhancement measured at the target. Subsequently, each column of the mask is tested. The focusing system for a single wavelength allows a peak-to-background ratio ($\eta$) corresponding to the factor $\eta$=100, calculated as the maximum intensity at the target (intensity of the light sheet at the peak) divided by the average intensity of the background (side lobes average intensity). If the optimal phase mask for one of the wavelengths is used for another, the focus is completely lost, meaning that due to the multiple scattering the three sources experience completely different optical paths through the APG and their respective optimal mask are completely independent. It follows that when one uses for all three wavelengths the same optimal phase mask that was optimized for one of the wavelengths then the peak-to-background ratio drops significantly $\eta=33.4$.

When the stripes of the phase mask are parallel to the rods of the APG the speckle pattern at the output persists in having elongated grains as shown in Fig. 5.6 while the light sheet foci are shown in Figure 5.8(b). On the contrary, when segmented masks are used, the geometry of the system is broken and the foci achieved at the end of the optimization process are confined in the typical circular region as shown in Figure 5.8(c).

We remark that the optimized parallel phase mask takes advantage of the directionality of the system reducing the complexity of the focusing algorithm from $N^2$ segments (the case of the conventional phase mask shown in Fig. 5.8(c) to $N$ stripes (the case for light sheet), resulting in a faster focusing process.

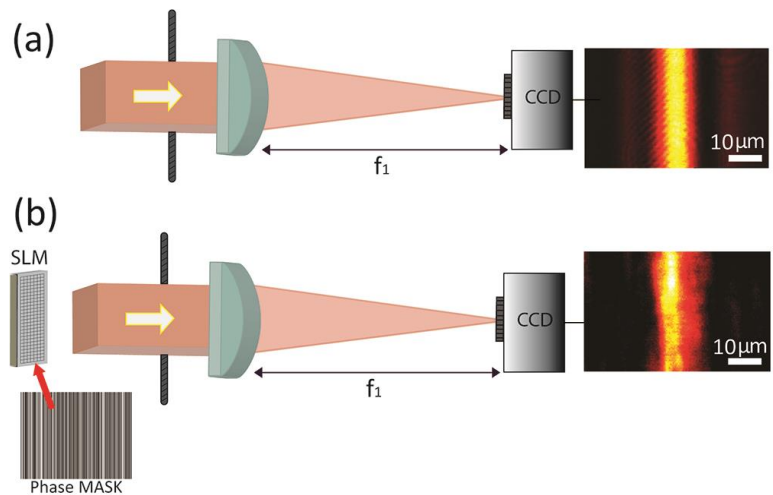

**Figure 5.9:** *A cylindrical lens generates a light sheet at its focus $f_1=150mm$, the quadratic slit filter (F) control the aperure D of the lens. Both in panel a. and panel b. the light sheet is generated in the free space configuration. On b. we shape the beam addressing to the SLM phase masks composed of parallel stripes. In b. we show the final light sheet obtained using the focusing algorithm starting from a random mask. The lateral resolutions of the 2 light sheets are similar.*

If we consider free space modulation, although a phase-mask in absence of the scattering structure can shape the focus generated by the cylindrical lens [111, 116], a narrower focus (onto the same plane) is not possible. This is demonstrated in Fig. 5.9. In Fig. 5.9(a) a plane wave is directly focused by a cylindrical lens on the camera plane. In Fig. 5.9(b) we maintain the free space configuration (absence of scattering media) and we impose a random mask composed of parallel strips on the SLM generating an optimized in size light sheet by running the same focusing algorithm as the one in use for the cases of Fig. 5.8(b). As expected both light sheets have very similar lateral resolution (despite the small anisotropy in the image obtained with the SLM, which is due to the pixelated nature of the applied phase mask), $w_{0X}$=6.7μm in a. and $w^{SLM}_{0X}$=6.81μm in b. (the case with the SLM). This supports further the argument that a scattering structure is necessary to eliminate the ballistic contribution and to introduce higher spatial frequencies (higher effective numerical aperture) which are responsible for the reduced focus size [42].

### 5.2.4 Fast multi-wavelength light sheet scanning with sub-micron resolution

Light Sheet Microscopy requires fast scanning (ideally in real time) of the samples at different wavelengths. In order to correct for the chromatic aberration one has to slightly change the position of the lenses at the illumination to maintain the sample on focus for the different wavelengths.

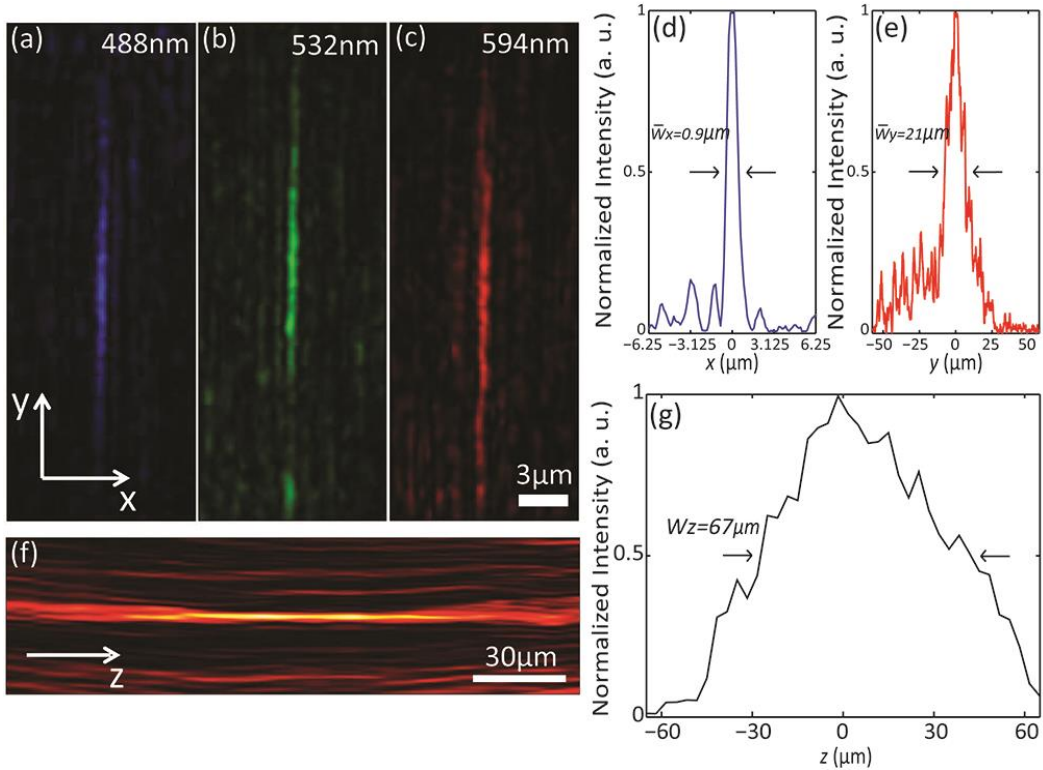

**Figure 5.10:** *Images (a), (b) and (c) show respectively light-sheets generated at 488nm, 532nm and 594nm laser source wavelengths. The intensity profile along the x axis (blue curve in panel (d)) and along the y axis (red curve in panel (e)) provide the light-sheet width ($\bar{w}_X = 0.9\mu m$) and length ($\bar{w}_y = 21\mu m$). In (f) we show the light-sheet along the z axis and we measured the FWHM of its intensity profile ($w_z = 67\mu m$) as shown in plot (g).*

This introduces delays and uncertainties, which are not suitable for fast *in vivo* scanning of the subjects. In our approach, we focus the light 100μm behind the APG and we use three laser sources emitting at 488nm, 532nm and 594nm. Running the focusing process we find the three phase masks which allow the light sheets to co-localize at the same position for the three different colors. In such a way, we can change the illumination wavelength by switching from one mask to the other without the need of moving the position of lenses or of inserting shutters or filters along the light path. In Figures 5.10(a), 5.10(b) and 5.10(c) we show the three light sheets obtained onto the same target position. For this we have maintained the same geometry shown in Figure 5.8(b) only replacing the cylindrical with a spherical lens with the same focal length. We then consider the intensity profiles of the final foci to obtain the average lateral resolution of $\bar{w}_X=0.9\mu m$ along the *x* axis (see Fig. 5.10(d)) and $\bar{w}_y=21\mu m$ along *y* axis (see Figure 5.10(e)). For the calculation we create 10 light sheets at different times onto 10 different target positions and we average to extract the final resolutions. We repeat the same process for each wavelength and evaluate a change of $\Delta w_X = 0.03 \mu m$ in the light sheet width passing from 488nm to 594nm wavelength. For applications with resolution on the micro-scale such small values of $\Delta w_X$ are insignificant. We also measure the axial resolution $w_z$ of our light sheet scanning along *z* the region around the plane where the focus was initially generated $z_0 = 100 \mu m$ in front of the APG. We move the system composed of the objective lens, tube lens and camera $100 \mu m$ in front and at the back of $z_0$ and we collect a frame each *2μm*.

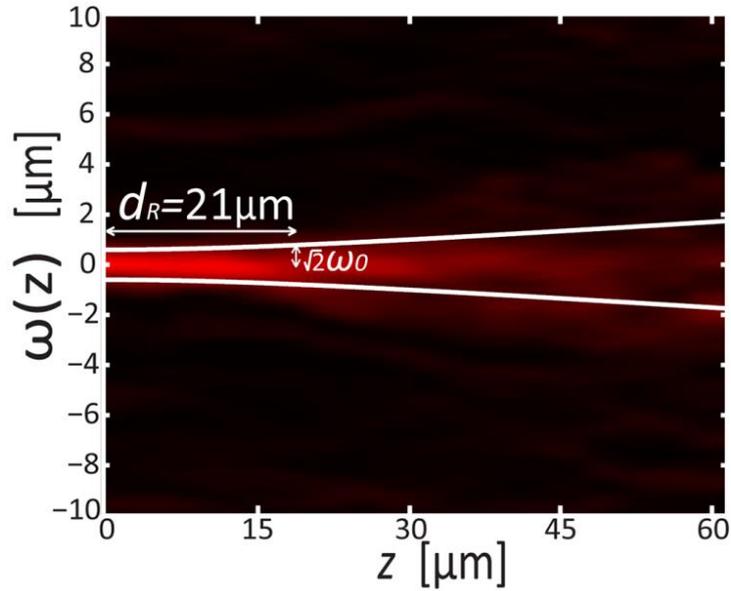

**Figure 5.11:** *Light sheet intensity along z. The beam width is fitted using Rayleigh formula in eq. (5.2.2) providing the two white curves. From the curves we calculate the distance $d_R$ where the beam width is $\sqrt{2}\omega_0$.*

By registering all the frames collected we can reconstruct the light sheet propagating along *z* as shown in Fig. 5.10(f). We measure the axial resolution $w_z = 67\mu m$ calculating the FWHM of the intensity profile along the direction *z* as shown in Fig. 5.10(g). One can easily find the optimal masks for focusing at defined positions for each wavelength, and thus is able to switch from one color to the other simply by addressing a different mask to the SLM. Considering that the modern SLMs are able to achieve refresh rates as high as 500Hz, our approach can equip the user with a very fast tool for scanning the specimens at different wavelengths in a small fraction of a second, i.e. in real time.

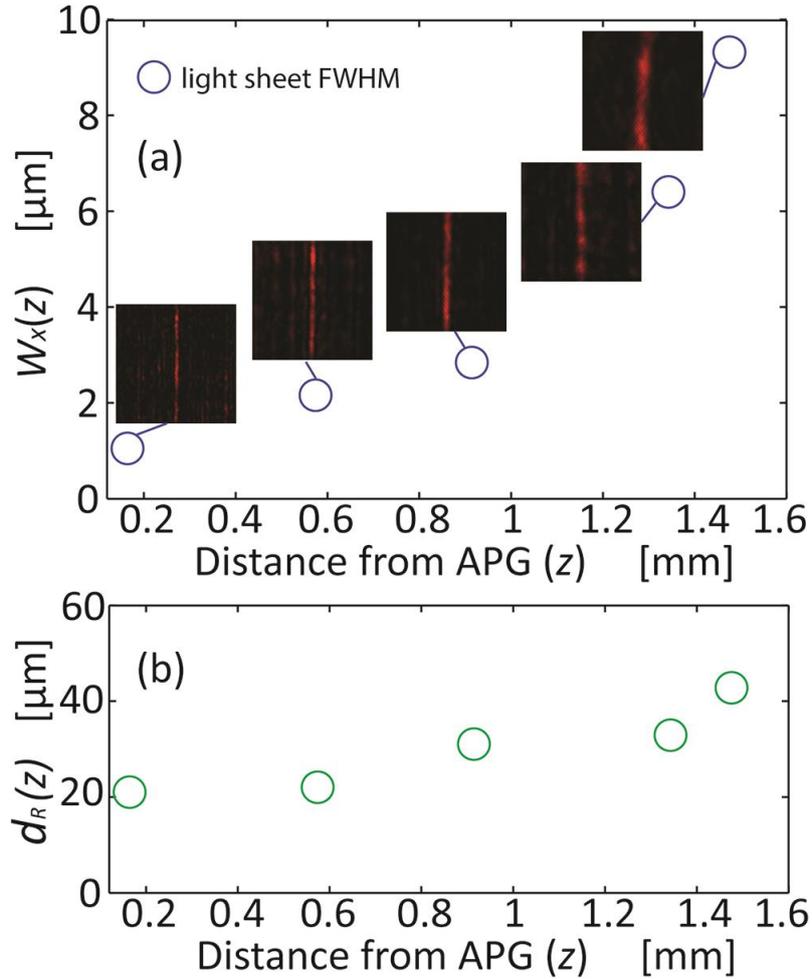

**Figure 5.12:** *In panel (a) the full-width half maximum of the light sheet (cross section of foci along x (Wx)) is measured as function of the distance from the APG. In panel (b) the respective Rayleigh distances ($d_R$) are measured.*

Good quality Light Sheet Microscopy (LSM) requires homogeneous light sheet illuminating a single section of the sample. For this reason the Field-of-View (FOV) in LSM is defined by the Rayleigh distance ($d_R$) of the light sheet used in the experiment. The $d_R$ is defined as the distance along the propagation direction of a beam from its waist to the place where the beam radius has increased by a factor of $\sqrt{2}$, where in our case the beam waist $\omega_0$ is defined as $\omega_0 = \bar{w}_x/\sqrt{2\ln 2}$. The beam width along $z$ of our light sheet is studied in Fig. 5.11. The propagating beam intensity from our experiment is fitted with the relation [117]:

$$\omega(z) = \omega_0 \sqrt{1 + \left(\frac{z}{d_R}\right)^2} \tag{5.2}$$

in such a way we obtain the two white curves in Fig. 5.11 that allow the estimation of $d_R$. We calculate $d_R = 21\mu m$, i.e. our light sheet exhibits an effective field-of-view $FOV \approx 40\mu m$. According to Rayleigh's formula [117], considering our experimental $\omega_0$, the $FOV$ measured results larger than the expected ($\sim 10\mu m$). On the other hand, extended $FOV$ in similar geometry and the extraordinary point spread function of opaque lenses [116, 103] have been recently demonstrated and are currently subject of study.

Furthermore, light sheet imaging requires fast volumetric scan across *x* and, for larger sample, at different depth *z*. In our system the light sheet is formed at user defined positions allowing for rapid volumetric scan: moving along *z* the detection system one can focus onto different planes along *z*. As a demonstration of the full control of the system we report in Fig. 5.12 light sheets focused at different distances *z* from the back surface of the APG. In panel (a) their $W_z$ (FWHM along *x*) is measured as a function of the distance *z*. In addition, we calculate their respective Rayleigh distances ($d_R$) and the results are shown in Fig. 5.12 (b).

### 5.2.5 *Focusing through Dentinal Tubules*

Particular biological structures have already been demonstrated to produce interesting sub-diffractive light features [48]. Structures similar to APGs can be found in nature, e.g. muscle or collagen fibers, dental enamel and others. Probing these structures with polarized light can disclose information on the molecular organization of the investigated specimen and for this reason, they are lately subject of studies in both the linear and non-linear regimes [118]. In this case, to test the approach described above on a biological sample, we chose histological Dentinal Tubule (DT) slices [119] as the scattering medium. DTs are microscopic sigmoid ('S') shaped curved channels which contain the long cytoplasmic processes of odontoblasts and extend radially from the dentinoenamel junction (DEJ) in the crown area, or dentinocemental junction (DCJ) in the root area, to the outer wall of the pulp, forming a network for the diffusion of nutrients throughout dentin [120]. Tubules run horizontally from the inside of the tooth to the outside located within the dentin and near the root tip, incisal edges and cusps, the dentinal tubules are almost straight. Gradually narrowing from the inner to the outermost surface of the teeth, the tubules have a diameter of 2.5μm near the pulp, 1.2μm in the middle of the dentin, and 0.9μm at the dentin-enamel junction. Their density is 59,000 to 76,000 per square millimeter near the pulp, whereas the density is only half as much near the enamel [120, 121]. In that respect, dentinal tubule slices resemble and can be used as natural 2D scattering structures. In our study we used a 0.7mm thick dentinal tubule ground section (DENT-EQ, Basavanagudi, Bangalore, India), enough to scramble completely the impinging light and generate a complex speckle pattern with negligible ballistic contribution.

The tubules run parallel to each other, a characteristic that makes them a natural APG. We used dry DT ground sections, such as those presented in the micrographs in Figures 5.13(a) and 5.13(b). Using thicker slices a highly scattering medium is formed. When we impinge onto it with a coherent beam an interference pattern composed of elongated grains is generated at the back of the slice, as shown in Fig. 5.13(c). Calculating the *Corr(r)* from the image in Figure 5.13(c) we can estimate the average orientation of the speckle grains, in the direction along which they present the elongation. The *Corr(r)* results with a central pick being elongated toward a direction which exhibits an angle $\vartheta$ with respect to the vertical axis of the camera. Therefore, we also rotate the SLM mask at the same angle $\vartheta$ in order to correctly align the phase bands parallel to the tubules direction (see Fig. 5.13(d)). Once we select a target on the camera plane and run the focusing process we are able to produce a light sheet at the back of the DTs, as shown in Figure 5.13(e), with a width of $w_x = 1.8\,\mu m$ and a length of $w_y = 17.5\,\mu m$. A misalignment of the parallel phase mask with respect to the directionality of the tubules produces a break to the intrinsic dimensionality of the system and leads to a conventional focal spot (not elongated).

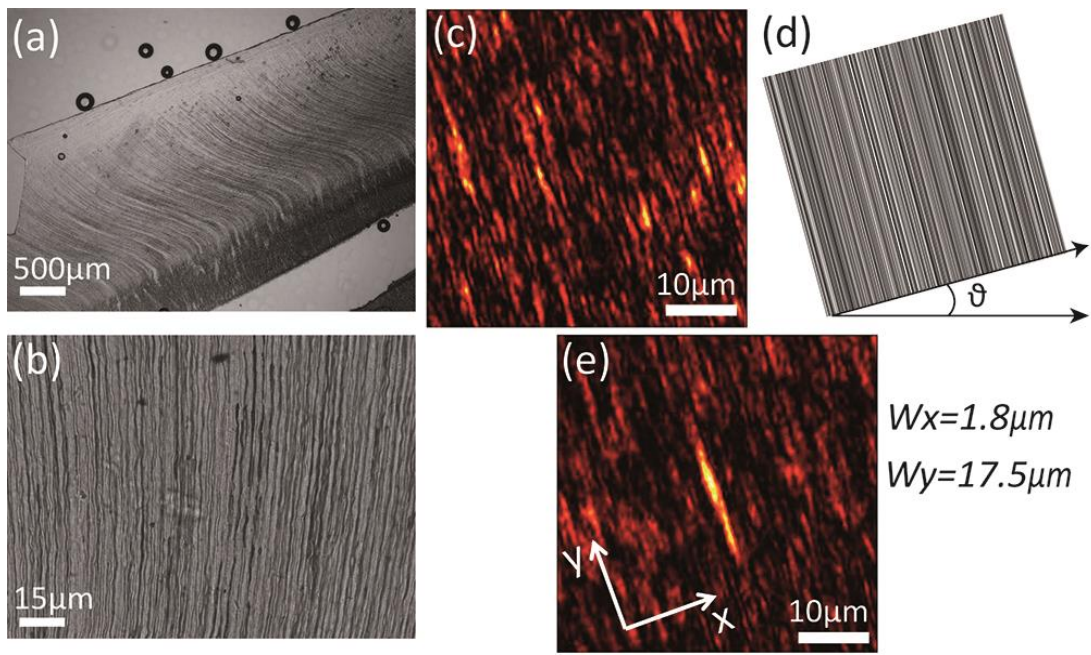

**Figure 5.13:** *Adaptive focusing through a DT. In (a) a thin DT shows parallel tubules running along the dentinal section. In (b) a thick DT forms a perfect natural APG. The light trespassing the DT forms elongated speckles isotropically oriented along one direction given by the tubules orientation as shown in (c). $\vartheta$ is the angular distance between the orientation of the grains and the vertical camera axis. To maintain the isotropy the SLM is rotated at an angle $\vartheta$ as illustrated in the sketch (d). Sketch (e) is the light-sheet at the end of the process*

## 5.2.6 Discussion

In our experiments we generate light sheets that improve the resolution of conventional cylindrical lenses at a given focal length by a factor 8. In addition, with our technique we can easily correct for chromatic aberrations as we demonstrated above by generating light sheets at different wavelengths on the same spatial position without moving any of the system components. In practice, for a given focal length the light sheets produced at different wavelengths will slightly differ in shape due to the difference in refractive index experienced through the structure, but they will be generated on the same plane giving the possibility to switch from one color to the other by only changing the phase mask addressed to the SLM. Moreover, OCLs can be considered for a new generation of tailored microscope cover slips, a platform able to provide direct structure illumination on the specimens. Furthermore, our technique can be effectively scaled: for achieving, for instance, higher resolutions one can reduce the diameter and the reciprocal distances of the photonic lattice rods. Lithographic techniques could push this limit quite low and significantly outperform present techniques.

# 6 Conclusions and Future Outlook

The image on the cover of this Thesis is a picture of our optical bench (captured by my colleague Daniele Ancora with camera "D3" and lens "50mm, f/1.8, AF-D", from Nikon, Japan). Although the image is taken out-of-focus, it revels the interferences between higher diffraction orders that are usually discarded in the experiments. Again, such unconventional prospective may disclose elements that contain a lot of information and that would be otherwise inaccessible. In similar way, today disorder is counter-intuitively tested to enhance the optical based communications [122, 123, 124].

During the development of this Thesis we have delved into "the scattered light" in order to investigate the potential of opaque optical elements. While this practice has resulted very effective to master the problem, it leaded to novel light features with interesting properties.

Appropriately filtering the scattering light, we disclosed "hidden" channels responsible for focusing beyond the correlation limit and Bessel beams formation behind walls. The amorphous speckle pattern, presented herewith, has been only recently explored [125], it is currently subject of studies [56, 126, 77] and used for increasing the depth-of-focus in optical imaging [127] as predicted in our work [51]. Besides, its non-disordered speckle grains distribution is very intriguing because presents important similarities with exotic states of matter [80], indeed, the fact that those states can be simulated with electromagnetic fields distributions is very intriguing. Therefore, we expect that amorphous speckles can inspire for infinite other classes of speckle structures. The key consists in finding the spatial frequencies filter that would generate innovative light lattices. It follows that to each of those speckles structures would correspond an exclusive focus shape (equal to their spatial correlation), when wavefront shaping is applied. We intent to investigate further in this sense to contribute at the creation of organized light patterns [128] with striking properties.

Nevertheless, the main target will remain the direct application of wavefront manipulation in the current optical imaging techniques at the presence of thick samples; the aim is to gain depth without turning down the optimal resolution. The current obstacles are mainly two: the speed of the optimization that, ideally, has to be faster than the intrinsic dynamic of the biological sample and the access to a guide-star to use as a feedback during the shaping process. Even if in the recent times we have seen incredible advances in this sense [34, 11, 129], the problem requires further hardware advancements, in particular faster spatial modulators are needed.

On the other side, we have shown that engineered light patterns can be also achieved designing particular scattering geometries. The light flowing through those photonic matrices is guided into complex light paths that induce spatially tailored speckles. In this scenario, we have demonstrated that scattering systems are widely configurable, therefore they can be now thought as new generation of optical element [130]. For this reason we have imagined them as advantageous photonic devices integrated on-chip. For the first time in the field of wavefront shaping, a complex scattering medium is fabricated *ad-hoc* to produce structured foci, with sub-micron resolution; a proof of principle that can be of interest in a variety of fields, ranging from microscopy, to secure communication and to integrated circuit manufacturing based on planar technology. It is, in fact, very recent that designed disorder has been implemented for overcoming the present limitations in photonic micro-platforms [131, 132].

A major advantage of our printed photonic glasses resides in the fact that they are inscribed in the bulk of a glass, a characteristic that makes them robust against deterioration especially if compared with the standard dielectric scattering slabs based on powder conglomerates (such as $TiO_2$ or PMMA), a significant advantage for commercial purposes.

In the specific case of our opaque cylindrical lenses, the production of a thin light sheet through scattering media aims at improving microscopic imaging of turbid biological samples and organisms. The next step will target into the incorporation of our setup to a conventional light sheet microscope and optimizing the hybrid adaptive microscopy workstation. The goal will be to employ this system in imaging bioengineered multicellular 3D large samples such as cancer cell spheroids [133] and other model organisms.

Because of those reasons we wish that the works presented with this Thesis can be of interest to a large scientific community and can partially contribute to the progress of new technology in different fields of Optics and Photonics.

# 7  *Bibliography*

[1] B. Abbott, *et al.*, "Observation of gravitational waves from a binary black hole merger," *Physical review letters,* vol. 116, no. 6, p. 061102, 2016.

[2] V. Ntziachristos, "Going deeper than microscopy: the optical imaging frontier in biology," *Nature methods,* vol. 7, no. 8, pp. 603-614, 2010.

[3] F. Pampaloni, Reynaud, E. G. and E. Stelzer, "The third dimension bridges the gap between cell culture and live tissue," vol. 8, no. 10, pp. 839-845, 2007.

[4] K. Howe, *et al.*, "The zebrafish reference genome sequence and its relationship to the human genome," vol. 496, no. 7446, pp. 498-503, 2013.

[5] R. Zhang, *et al.*, "In vivo cardiac reprogramming contributes to zebrafish heart regeneration," vol. 498, no. 7455, pp. 497-501, 2013.

[6] D. Zhu, K. Larin, Q. Luo and V. Tuchin, "Recent progress in tissue optical clearing," vol. 7, no. 5, pp. 732-757, 2013.

[7] R. K. Tyson, Principles of adaptive optics, Boc Raton, FL, USA: CRC press, 2015.

[8] I. Vellekoop and A. Mosk, "Focusing coherent light through opaque strongly scattering media," *Optics letters,* vol. 32, no. 16, p. 2309–2311, 2007.

[9] A. Mosk, A. Lagendijk, G. Lerosey and M. Fink, "Controlling waves in space and time for imaging and focusing in complex media," *Nature photonics,* vol. 6, no. 5, pp. 283-292, 2012.

[10] C. Maurer, A. Jesacher, S. Bernet and M. Ritsch-Marte, "What spatial light modulators can do for optical microscopy," *Laser & Photonics Reviews,* vol. 5, no. 1, pp. 81-101, 2010.

[11] H. Yu, J. Park, K. Lee, J. Yoon, K. Kim, S. Lee and Y. Park, "Recent advances in wavefront shaping techniques for biomedical applications," *Current Applied Physics,* vol. 15, no. 5, pp. 632-641, 2015.

[12] J. Carpenter, B. J. Eggleton and J. Schröder, "Observation of Eisenbud–Wigner–Smith states as principal modes in multimode fibre," *Nature Photonics,* vol. 9, no. 11, pp. 751-757, 2015.

[13] X. Shen, J. Kahn and M. Horowitz, "Compensation for multimode fiber dispersion by adaptive optics," *Optics letters,* vol. 30, no. 22, pp. 2985-2987, 2005.

[14] K. Cahoy, *et al.*, "Wavefront control in space with MEMS deformable mirrors for exoplanet direct imaging," *Journal of Micro/Nanolithography, MEMS, and MOEMS,* vol. 13, no. 1, pp. 011105-011105, 2014.

[15] I. Vellekoop, A. Lagendijk and A. Mosk, "Exploiting disorder for perfect focusing," *Nature photonics,* vol. 4, no. 5, pp. 320-322, 2010.

[16] E. Van Putten, D. Akbulut, J. Bertolotti, W. Vos, A. Lagendijk and A. Mosk, "Scattering lens resolves sub-100 nm structures with visible light," *Physical review letters,* vol. 106, no. 19, p. 193905, 2011.

[17] J. Park, *et al.*, "Subwavelength light focusing using random nanoparticles," *Nature photonics,* vol. 7, no. 6, pp. 454-458, 2013.

[18] O. Katz, E. Small and Y. Silberberg, " Looking around corners and through thin turbid layers in real time with scattered incoherent light," *Nature photonics,* vol. 6, no. 8, pp. 549-553, 2012.

[19] O. Katz, E. Small, Y. Guan and Y. Silberberg, "Noninvasive nonlinear focusing and

imaging through strongly scattering turbid layers," *Optica,* vol. 1, no. 3, pp. 170-174, 2014.

[20] M. Leonetti, C. Conti and C. Lopez, "Switching and amplification in disordered lasing resonators," *Nature communications,* vol. 4, p. 1740, 2013.

[21] S. A. Goorden, M. Horstmann, A. P. Mosk, B. Škorić and P. Pinkse, "Quantum-secure authentication of a physical unclonable key," *Optica,* vol. 1, no. 6, pp. 421-424, 2014.

[22] I. Freund, "Looking through walls and around corners," *Physica A: Statistical Mechanics and its Applications,* vol. 168, no. 1, pp. 49-65, 1990.

[23] P. Sheng, Introduction to wave scattering, localization and mesoscopic phenomena, New York: Springer Science & Business Media, 2006.

[24] M. van Albada, B. A. van Tiggelen, A. Lagendijk and A. Tip, "Speed of propagation of classical waves in strongly scattering media," *Physical review letters,* vol. 66, no. 24, p. 3132, 1991.

[25] J. Ripoll, Principles of diffuse light propagation, World Scientific, 2012.

[26] A. Lagendijk and B. Van Tiggelen, " Resonant multiple scattering of light," *Physics Reports,* vol. 270, no. 3, pp. 143-215, 1996.

[27] P. Garcia Fernandez, From Photonic Crystals to Photonic Glasses through disorder, (ICMM) Tesis ed., Madrid: DIGITAL.CSIC, 2010.

[28] N. Garcia, A. Genack and A. Lisyansky, "Measurement of the transport mean free path of diffusing photons," *Physical Review B,* vol. 46, no. 22, p. 14475, 1992.

[29] J. Goodman, Speckle phenomena in optics: theory and applications, Englewood: Roberts and Company Publishers, 2007.

[30] J. Bertolotti, "Multiple scattering: Unravelling the tangle," *Nature Physics,* 2015.

[31] S. Popoff, G. Lerosey, R. Carminati, M. Fink, A. Boccara and S. Gigan, "Measuring the transmission matrix in optics: an approach to the study and control of light propagation in disordered media," *Physical review letters,* vol. 104, no. 10, p. 100601, 2010.

[32] J. Bertolotti, G. E. van Putten, C. Blum, A. Lagendijk, W. Vos and A. Mosk, "Non-invasive imaging through opaque scattering layers," *Nature,* vol. 491, no. 7423, pp. 232-234, 2012.

[33] J. Dainty, Laser speckle and related phenomena, vol. 9, New York: Springer Science & Business Media, 2013.

[34] R. Horstmeyer, H. Ruan and C. Yang, "Guidestar-assisted wavefront-shaping methods," *Nature Photonics,* vol. 9, no. 9, pp. 563-571, 2015.

[35] J. V. Thompson, G. A. Throckmorton, B. H. Hokr and V. Yakovlev, "Wavefront shaping enhanced Raman scattering in a turbid medium," *Optics letters,* vol. 41, no. 8, pp. 1769-1772, 2016.

[36] I. Vellekoop, "Feedback-based wavefront shaping," *Optics express,* vol. 23, no. 9, pp. 12189-12206, 2015.

[37] I. M. Vellekoop and A. P. Mosk, " Phase control algorithms for focusing light through turbid media," *Optics communications,* vol. 281, no. 11, pp. 3071-3080, 2008.

[38] E. Abbe, "Beiträge zur Theorie des Mikroskops und der mikroskopischen Wahrnehmung," *Archiv für mikroskopische Anatomie,* vol. 9, no. 1, pp. 413-418, 1873.

[39] G. Fowles, Introduction to modern optics, New York: Courier Corporation, 2012.

[40] F. Zernike, "The concept of degree of coherence and its application to optical problems," *Physica,* vol. 5, no. 8, pp. 785-795, 1938.

[41] J. Goodman, Introduction to Fourier optics, New York: McGrawn-Hill, 1996.

[42] D. Di Battista, G. Zacharakis and M. Leonetti, "Enhanced adaptive focusing through semi-transparent media," *Scientific reports,* vol. 5, p. 17406, 2015.

[43] Z. Yaqoob, D. Psaltis, M. Feld and C. Yang, "Optical phase conjugation for turbidity suppression in biological samples," *Nature photonics,* vol. 2, no. 2, pp. 110-115, 2008.

[44] R. Kop, P. d. Vries, R. Sprik and A. Lagendijk, "Observation of anomalous transport of strongly multiple scattered light in thin disordered slabs," *Physical review letters,* vol. 79, no. 22, p. 4369, 1997.

[45] P. García, R. Sapienza and C. López, "Photonic glasses: a step beyond white paint," *Advanced materials,* vol. 22, no. 1, pp. 12-19, 2010.

[46] M. Davy, Z. Shi and A. Genack, "Focusing through random media: Eigenchannel participation number and intensity correlation," *Physical Review B,* vol. 85, no. 3, p. 035105, 2012.

[47] M. Mazilu, J. Baumgartl, S. Kosmeier and K. Dholakia, "Optical Eigenmodes; exploiting the quadratic nature of the energy flux and of scattering interactions," *Optics express,* vol. 19, no. 2, pp. 933-945, 2011.

[48] E. De Tommasi, *et al.*, "Biologically enabled sub-diffractive focusing," *Optics express,* vol. 22, no. 22, pp. 27214-27227, 2014.

[49] E. McDowell, M. Cui, I. M. Vellekoop, V. Senekerimyan, Z. Yaqoob and C. Yang, " Turbidity suppression from the ballistic to the diffusive regime in biological tissues using optical phase conjugation," *Journal of biomedical optics,* vol. 15, no. 2, pp. 025004-025004, 2010.

[50] P. García, R. Sapienza, C. Toninelli, C. López and D. Wiersma, "Photonic crystals with controlled disorder," *Physical Review A,* vol. 84, no. 2, p. 023813, 2011.

[51] D. Di Battista, D. Ancora, M. Leonetti and G. Zacharakis, "Tailoring non-diffractive beams from amorphous light speckles," *Applied Physics Letters,* vol. 109, no. 12, p. 121110, 2016.

[52] B. Huang, M. Bates and X. Zhuang, "Super resolution fluorescence microscopy," *Annual review of biochemistry,* vol. 78, p. 993, 2009.

[53] E. Betzig, A. Lewis, A. Harootunian, M. Isaacson and E. Kratschmer, "Near Field Scanning Optical Microscopy (NSOM): Development and Biophysical Applications," *Biophysical Journal,* vol. 49, no. 1, p. 269, 1986.

[54] R. Heintzmann and C. Cremer, " Laterally modulated excitation microscopy: improvement of resolution by using a diffraction grating," *BiOS Europe'98. SPIE,* vol. 3568, pp. 185-196, 1999.

[55] M. Gustafsson, "Surpassing the lateral resolution limit by a factor of two using structured illumination microscopy," *Journal of microscopy,* vol. 198, no. 2, pp. 82-87, 2000.

[56] C. R. Alves, A. J. Jesus-Silva and E. Fonseca, "Self-reconfiguration of a speckle pattern," *Optics letters,* vol. 39, no. 21, pp. 6320-6323, 2014.

[57] E. Mudry, *et al.*, "Structured illumination microscopy using unknown speckle patterns," *Nature Photonics,* vol. 6, no. 5, pp. 312-315, 2012.

[58] M. Kim, C. Park, C. Rodriguez, Y. Park and Y. Cho, "Superresolution imaging with optical fluctuation using speckle patterns illumination," *Scientific reports,* vol. 5, p. 16525, 2015.

[59] T. Chaigne, J. Gateau, M. Allain, O. Katz, S. Gigan, A. Sentenac and E. Bossy, "Super-resolution photoacoustic fluctuation imaging with multiple speckle illumination,"

*Optica,* vol. 3, no. 1, pp. 54-57, 2016.

[60] H. Yilmaz, E. van Putten, J. Bertolotti, A. Lagendijk, W. L. Vos and A. Mosk, "Speckle correlation resolution enhancement of wide-field fluorescence imaging," *Optica,* vol. 2, no. 5, pp. 424-429, 2015.

[61] T. Čižmár, M. Mazilu and K. Dholakia, "In situ wavefront correction and its application to micromanipulation," *Nature Photonics,* vol. 4, no. 6, pp. 388-394, 2010.

[62] R. Di Leonardo and S. Bianchi, " Hologram transmission through multi-mode optical fibers," *Optics express,* vol. 19, no. 1, pp. 247-254, 2011.

[63] H. Defienne, M. Barbieri, B. Chalopin, B. Chatel, I. Walmsley, B. J. Smith and S. Gigan, "Nonclassical light manipulation in a multiple-scattering medium," *Optics letters,* vol. 39, no. 21, pp. 6090-6093, 2014.

[64] S. Huisman, T. Huisman, T. Wolterink, A. Mosk and P. Pinkse, "Programmable multiport optical circuits in opaque scattering materials," *Optics express,* vol. 23, no. 3, pp. 3102-3116, 2015.

[65] Y. Guan, O. Katz, E. Small, J. Zhou and Y. Silberberg, "Polarization control of multiply scattered light through random media by wavefront shaping," *Optics letters,* vol. 37, no. 22, pp. 4663-4665, 2012.

[66] J. H. Park, C. Park, H. Yu, Y. H. Cho and Y. Park, "Dynamic active wave plate using random nanoparticles," *Optics Express,* vol. 20, no. 15, pp. 17010-17016, 2012.

[67] J. H. Park, C. Park, H. Yu, Y. H. Cho and Y. Park, "Active spectral filtering through turbid media," *Optics letters,* vol. 37, no. 15, pp. 3261-3263, 2012.

[68] I. Ouadghiri-Idrissi, *et al.*, "Arbitrary shaping of on-axis amplitude of femtosecond Bessel beams with a single phase-only spatial light modulator," *Optics Express,* vol. 24, no. 11, pp. 11495-11504, 2016.

[69] P. Steinvurzel, K. Tantiwanichapan, M. Goto and S. Ramachandran, "Fiber-based Bessel beams with controllable diffraction-resistant distance," *Optics letters,* vol. 36, no. 23, pp. 4671-4673, 2011.

[70] O. Brzobohatý, T. Čižmár and P. Zemánek, "High quality quasi-Bessel beam generated by round-tip axicon," *Optics Express,* vol. 16, no. 17, pp. 12688-12700, 2008.

[71] D. McGloin and K. Dholakia, "Bessel beams: Diffraction in a new light," *Contemporary Physics,* vol. 46, no. 1, pp. 15-28, 2005.

[72] T. Planchon, L. Gao, D. Milkie, M. Davidson, J. A. Galbraith, C. Galbraith and E. Betzig, "Rapid three-dimensional isotropic imaging of living cells using Bessel beam plane illumination," *Nature methods,* vol. 8, no. 5, pp. 417-423, 2011.

[73] Z. Bouchal, J. Wagner and M. Chlup, "Self-reconstruction of a distorted nondiffracting beam," *Optics Communications,* vol. 151, no. 4, pp. 207-211, 1998.

[74] C. Arnold, *et al.*, "Nonlinear Bessel vortex beams for applications," *Journal of Physics B: Atomic, Molecular and Optical Physics,* vol. 48, no. 9, p. 094006, 2015.

[75] D. Lorenser, C. Singe, A. Curatolo and D. Sampson, "Energy-efficient low-Fresnel-number Bessel beams and their application in optical coherence tomography," *Optics letters,* vol. 39, no. 3, pp. 548-551, 2014.

[76] M. Rechtsman, A. Szameit, F. Dreisow, M. Heinrich, R. Keil, S. Nolte and M. Segev, "Amorphous Photonic Lattices: Band Gaps, Effective Mass, and Suppressed Transport," *Physical review letters,* vol. 106, no. 19, p. 193904, 2011.

[77] P. Ni, P. Zhang, X. Qi, J. Yang, Z. Chen and W. Man, "Light localization and nonlinear beam transmission in specular amorphous photonic lattices," *Optics express,* vol. 24, no. 3, pp. 2420-2426, 2016.

[78] M. Boguslawski, S. Brake, J. Armijo, F. Diebel, P. Rose and C. Denz, "Analysis of transverse Anderson localization in refractive index structures with customized random potential," *Optics express,* vol. 21, no. 26, pp. 31713-31724, 2013.

[79] D. Wiersma, "Disordered photonics," *Nature Photonics,* vol. 7, no. 3, pp. 188-196, 2013.

[80] S. Torquato, "Hyperuniformity and its generalizations," *Physical Review E,* vol. 94, no. 2, p. 022122, 2016.

[81] H. Zhang, D. Di Battista, G. Zacharakis and S. Tzortzakis, "Robust authentication through stochastic femtosecond laser filament induced scattering surfaces, Erratum from [Appl. Phys. Lett. 108, 211107 (2016)]," *Applied Physics Letters,* vol. 109, no. 3, 2016.

[82] B. Gassend, D. Clarke, M. Van Dijk and S. Devadas, "Silicon physical random functions," *Proceedings of the 9th ACM conference on Computer and communications security. ACM,* pp. 148-160, 2002.

[83] B. Gassend, M. Dijk, D. Clarke, E. Torlak, S. Devadas and P. Tuyls, "Controlled physical random functions and applications," *ACM Transactions on Information and System Security (TISSEC),* vol. 10, no. 4, p. 3, 2008.

[84] J. Guajardo, S. Kumar, G. Schrijen and P. Tuyls, "FPGA intrinsic PUFs and their use for IP protection," in *International workshop on Cryptographic Hardware and Embedded Systems*, Springer Berlin Heidelberg, 2007.

[85] R. Pappu, B. Recht, J. Taylor and N. Gershenfeld, "Physical One-Way Functions," *Science,* vol. 297, no. 5589, pp. 2026-2030, 2002.

[86] J. D. Buchanan and e. al., "Forgery: 'Fingerprinting' documents and packaging," *Nature,* vol. 436, no. 7050, pp. 475-475, 2005.

[87] R. Horstmeyer, B. Judkewitz, I. Vellekoop, S. Assawaworrarit and C. Yang, "Physical key-protected one-time pad," *Scientific reports,* vol. 3, p. 3543, 2013.

[88] C. Yeh, P. Sung, C. Kuo and R. Yeh, "Robust laser speckle recognition system for," *Optics express,* vol. 20, no. 22, pp. 24382-24393, 2012.

[89] S. Gottardo, R. Sapienza, P. Garcia, A. Blanco, D. Wiersma and C. López, "Resonance-driven random lasing," *Nature Photonics,* vol. 2, no. 7, pp. 429-432, 2008.

[90] A. Joglekar, H. Liu, E. Meyhöfer, G. Mourou and A. Hunt, "Optics at critical intensity: Applications to nanomorphing," *Proceedings of the national academy of sciences of the United States of America,* vol. 101, no. 16, pp. 5856-5861, 2004.

[91] A. Couairon and A. Mysyrowicz, "Femtosecond filamentation in transparent media," *Physics reports,* vol. 441, no. 2, pp. 47-189, 2007.

[92] L. Sudrie, A. Couairon, M. Franco, B. Lamouroux, B. Prade, S. Tzortzakis and A. Mysyrowicz, "Femtosecond Laser-Induced Damage and Filamentary Propagation in Fused Silica," *Physical Review Letters,* vol. 89, no. 18, p. 186601, 2002.

[93] A. Borowiec, M. Mackenzie, G. Weatherly and H. Haugen, "Transmission and scanning electron microscopy studies of single femtosecond- laser-pulse ablation of silicon," *Applied Physics A,* vol. 76, no. 2, pp. 201-207, 2003.

[94] M. Ahsan, Y. Kim and M. Lee, "Formation mechanism of nanostructures in soda–lime glass using femtosecond laser," *Journal of Non-Crystalline Solids,* vol. 357, no. 3, pp. 851-857, 2011.

[95] P. Balling and J. Schou, "Femtosecond-laser ablation dynamics of dielectrics: basics and applications for thin films," *Reports on Progress in Physics,* vol. 76, no. 3, p. 036502, 2013.

[96]  A. Ben-Yakar and R. Byer, "Femtosecond laser ablation properties of borosilicate glass," *Journal of applied physics,* vol. 96, no. 9, pp. 5316-5323, 2004.

[97]  S. Feng, C. Kane, P. A. Lee and A. Stone, "Correlations and Fluctuations of Coherent Wave Transmission through Disordered Media," *Physical review letters,* vol. 61, no. 7, p. 834, 1988.

[98]  S. Schott, J. Bertolotti, J. Léger, L. Bourdieu and S. Gigan, "Characterization of the angular memory effect of scattered light in biological tissues," *Optics express,* vol. 23, no. 10, pp. 13505-13516, 2015.

[99]  K. Minoshima, A. Kowalevicz, E. Ippen and J. Fujimoto, "Fabrication of coupled mode photonic devices in glass by nonlinear femtosecond laser materials processing," *Optics Express,* vol. 10, no. 15, pp. 645-652, 2002.

[100] Y. Liao, Y. Shen, L. Qiao, D. Chen, Y. Cheng, K. Sugioka and K. Midorikawa, "Femtosecond laser nanostructuring in porous glass with sub-50 nm feature sizes," *Optics letters,* vol. 38, no. 2, pp. 187-189, 2013.

[101] R. Vázquez, S. Eaton, R. Ramponi, G. Cerullo and R. Osellame, "Fabrication of binary Fresnel lenses in PMMA by femtosecond laser surface ablation," *Optics express,* vol. 19, no. 12, pp. 11597-11604, 2011.

[102] D. Di Battista, D. Ancora, H. Zhang, K. Lemonaki, E. Marakis, E. Liapis, S. Tzortzakis and G. Zacharakis, "Tailored light sheets through opaque cylindrical lenses," *Optica,* vol. 3, no. 11, pp. 1237-1240, 2016.

[103] Y. Choi, *et al.*, "Overcoming the diffraction limit using multiple light scattering in a highly disordered medium," *Physical review letters,* vol. 107, no. 2, p. 023902, 2011.

[104] P. Keller, A. Schmidt, J. Wittbrodt and E. Stelzer, "Reconstruction of zebrafish early embryonic development by scanned light sheet microscopy," *science,* vol. 322, no. 5904, pp. 1065-1069, 2008.

[105] J. Huisken, J. Swoger, F. Del Bene, J. Wittbrodt and E. Stelzer, "Optical sectioning deep inside live embryos by selective plane illumination microscopy," *Science,* vol. 305, no. 5686, pp. 1007-1009, 2004.

[106] Z. Yang, H. Downie, E. Rozbicki, L. Dupuy and M. MacDonald, "Light Sheet Tomography (LST) for in situ imaging of plant roots," *Optics express,* vol. 21, no. 14, pp. 16239-1624721, 2013.

[107] D. Shotton, " Confocal scanning optical microscopy and its applications for biological specimens," *Journal of Cell Science,* vol. 94, no. 2, pp. 175-206, 1989.

[108] O. Olarte, *et al.*, " Image formation by linear and nonlinear digital scanned light-sheet fluorescence microscopy with Gaussian and Bessel beam profiles," *Biomedical optics express,* vol. 3, no. 7, pp. 1492-1505, 2012.

[109] T. Vettenburg, *et al.*, "Light-sheet microscopy using an Airy beam," *Nature methods,* vol. 11, no. 5, pp. 541-544, 2014.

[110] S. Saghafi, K. Becker, C. Hahn and H. Dodt, "3D-ultramicroscopy utilizing aspheric optics," *Journal of biophotonics,* vol. 7, no. 1-2, pp. 117-125, 2014.

[111] F. O. Fahrbach and A. Rohrbach, "A line scanned light-sheet microscope with phase shaped self-reconstructing beams," *Optics express,* vol. 18, no. 23, pp. 24229-24244, 2010.

[112] N. Yu and F. Capasso, "Flat optics with designer metasurfaces," *Nature materials,* vol. 13, no. 2, pp. 139-150, 2014.

[113] F. Aieta, M. A. Kats, P. Genevet and F. Capasso, "Multiwavelength achromatic metasurfaces by dispersive phase compensation," *Science,* vol. 347, no. 6228, pp. 1342-

1345, 2015.

[114] Q. Wang, E. Rogers, B. Gholipour, C. Wang, G. Yuan, J. Teng and N. Zheludev, "Optically reconfigurable metasurfaces and photonic devices based on phase change materials," *Nature Photonics,* vol. 10, no. 1, pp. 60-65, 2016.

[115] V. Parigi, *et al.*, " Near-field to far-field characterization of speckle patterns generated by disordered nanomaterials," *Optics express,* vol. 24, no. 7, pp. 7019-7027, 2016.

[116] D. Wilding, P. Pozzi, O. Soloviev, G. Vdovin, C. Sheppard and M. Verhaegen, "Pupil filters for extending the field-of-view in light-sheet microscopy," *Optics letters,* vol. 41, no. 6, pp. 1205-1208, 2016.

[117] A. E. Siegman, Lasers, Mill Valley, CA 37 : University Science Books, 1986.

[118] S. Psilodimitrakopoulos, I. Amat-Roldan, P. Loza-Alvarez and D. Artigas, "Effect of molecular organization on the image histograms of polarization SHG microscopy," *Biomedical optics express,* vol. 3, no. 10, pp. 2681-2693, 2012.

[119] M. Fehrenbach, Illustrated dental embryology, histology, and anatomy, New York, USA: Elsevier Health Sciences, 2015.

[120] A. Nanci, Ten cate's oral histology-pageburst on vitalsource: development, structure, and function, New York, USA: Elsevier Health Sciences, 2007.

[121] R. Garberoglio and M. Brännström, "Scanning electron microscopic investigation of human dentinal tubules," *Archives of Oral Biology,* vol. 21, no. 6, pp. 355-362, 1976.

[122] R. Fickler and R. W. Boyd, "Custom-Tailored Sorting of Structured Light by Controlled Scattering," in *Frontiers in Optics. Optical Society of America*, New York, USA, 2016.

[123] M. Leonetti, S. Karbasi, A. Mafi, E. DelRe and C. Conti, "Secure information transport by transverse localization of light," *Scientific Reports,* vol. 6, p. 29918, 2016.

[124] T. Tentrup, T. Hummel, T. Wolterink, R. Uppu, A. Mosk and P. Pinkse, "Transmitting more than 10 bit with a single photon," *arXiv preprint,* vol. 1609, no. 04200, 2016.

[125] L. Levi, Y. Krivolapov, S. Fishman and M. Segev, "Hyper-transport of light and stochastic acceleration by evolving disorder," *Nature Physics,* vol. 8, no. 12, pp. 912-917, 2012.

[126] S. Reddy, P. Chithrabhanu, P. Vaity, A. Aadhi, S. Prabhakar and R. Singh, "Non-diffracting speckles of a perfect vortex beam," *Journal of Optics,* vol. 18, no. 5, p. 055602, 2016.

[127] D. Phillips, "Non-diffractive computational ghost imaging," *Optics Express,* vol. 24, no. 13, pp. 14172-14182, 2016.

[128] L. Wan, Z. Chen, H. Huang and J. Pu, "Focusing light into desired patterns through turbid media by feedback-based wavefront shaping," vol. 122, no. 7, pp. 1-7, 2016.

[129] H. Yu, *et al.*, "In vivo deep tissue imaging using wavefront shaping optical coherence tomography," vol. 21, no. 10, pp. 101406-101406, 2016.

[130] J. Park, J. Cho, C. Park, K. Lee, H. Lee, Y. Cho and Y. Park, "Scattering optical elements: stand-alone optical elements exploiting multiple light scattering," *ACS NANO,* vol. 10, no. 7, pp. 6871-6876, 2016.

[131] S. Yu, X. Piao, J. Hong and N. Park, "Metadisorder for designer light in random systems," *Science Advances,* vol. 2, no. 10, p. e1501851, 2016.

[132] R. Bruck, *et al.*, "All-optical spatial light modulator for reconfigurable silicon photonic circuits," *Optica,* vol. 3, no. 4, pp. 396-402, 2016.

[133] D. Ancora, D. Di Battista, G. Giasafaki, S. Psycharakis, E. Liapis and G. Zacharakis,

"Phase-Retrieved Tomography enables imaging of a Tumor Spheroid in Mesoscopy Regime," vol. 1610, no. 06847, 2016.